\documentclass[11pt,a4paper]{article}
\pdfoutput=1
\usepackage[T1]{fontenc}
\usepackage[utf8]{inputenc}
\usepackage[top=2.5cm,bottom=2.5cm,left=2.5cm,right=2.5cm]{geometry}
\usepackage{amsmath,amssymb,amsfonts,mathrsfs,amsthm,mathtools}
\usepackage{graphicx,xcolor,booktabs,array,longtable,ragged2e,microtype}
\usepackage[colorlinks=true]{hyperref}
\definecolor{ABJMlink}{HTML}{176C68}
\definecolor{ABJMcite}{HTML}{8A4F6B}
\hypersetup{linkcolor=ABJMlink,urlcolor=ABJMlink,citecolor=ABJMcite,
 pdftitle={Constants and Nonperturbative Corrections in V(5,2) Theories},
 pdfauthor={Seyed Morteza Hosseini}}
\usepackage{caption,orcidlink,cite,enumitem}
\graphicspath{{figures/}}
\theoremstyle{plain}
\newtheorem{theorem}{Theorem}[section]
\newtheorem{proposition}[theorem]{Proposition}

\theoremstyle{definition}
\newtheorem{conjecture}[theorem]{Conjecture}

\numberwithin{equation}{section}
\DeclareMathOperator{\Tr}{Tr}
\DeclareMathOperator{\re}{Re}
\DeclareMathOperator{\im}{Im}
\DeclareMathOperator{\Li}{Li}

\newcommand{\ii}{\mathrm{i}}
\newcommand{\e}{\mathrm{e}}
\newcommand{\rd}{\mathrm{d}}
\newcommand{\cW}{\mathcal{W}}
\newcommand{\W}{\mathscr{W}}
\newcommand{\Wt}{\widetilde{\mathcal{W}}}
\newcommand{\cL}{\mathcal{L}}
\newcommand{\cO}{\mathcal{O}}
\newcommand{\Nh}{\widehat N}
\newcommand{\lamh}{\widehat\lambda}
\newcommand{\Cltwo}{\mathrm{Cl}_2}
\newcommand{\U}{\mathrm U}
\newcommand{\SU}{\mathrm{SU}}
\newcommand{\be}{\begin{equation}}
\newcommand{\ee}{\end{equation}}
\newcommand{\bea}{\begin{equation}\begin{aligned}}
\newcommand{\eea}{\end{aligned}\end{equation}}

\newcommand{\WI}{\W_{\mathrm I}}
\newcommand{\WII}{\W_{\mathrm{II}}}
\newcommand{\CI}{C_{\mathrm I}}
\newcommand{\CII}{C_{\mathrm{II}}}
\newcommand{\Etilde}{\widetilde E}
\DeclareMathOperator{\Log}{Log}

\usepackage{tikz}
\usetikzlibrary{arrows.meta}

\title{\vspace{-1.5cm}\bfseries Constants and Nonperturbative Corrections\\[.15em]
in $V^{5,2}$ Theories}
\author{Seyed Morteza Hosseini\,\orcidlink{0000-0001-8205-400X}\\[.5em]
\small Centre for Theoretical Physics, Department of Physics and Astronomy,\\
\small Queen Mary University of London, London E1 4NS, United Kingdom\\[.5em]}
\date{\today}
\begin{document}
\maketitle
\thispagestyle{empty}
\begin{abstract}
We study perturbative and exponentially small corrections to the on-shell effective twisted superpotentials of two three-dimensional $\mathcal N=2$ theories associated with free and fixed-locus cyclic quotients of the non-toric Sasaki--Einstein manifold $V^{5,2}$. These are the two-node Chern--Simons theory at levels $(k,-k)$ and the one-node flavored theory with $N_f$ fundamental pairs, respectively. At superconformal chemical potentials, we propose the Chern--Simons counterpart of the all-order perturbative expression for the flavored theory and obtain convergent integral representations for the rank-independent constants of both theories. At integer parameters, these constants admit explicit evaluations involving finite sums of Clausen functions and $\zeta(3)$, while their large-parameter expansions Borel sum to the defining integrals. At $k=1,3$ and $N_f=1,3$, we conjecture complete exponential remainders expressed through Eichler integrals of weight-four Eisenstein series. Supported by high-precision Bethe solutions, the resulting completions coincide at unit quotient order but differ at order three. We determine the minimal modular levels of the Eisenstein generators and establish convergence and natural boundaries for their primitives. Within the relative M-theory localization prescription, a complementary bulk analysis recovers the Chern--Simons shifted rank and constant in the $b\to0^+$ limit of the squashed-sphere partition function. We propose a generalization incorporating fixed-locus contributions that yields the corresponding flavored-theory result. The two theories therefore share a modular structure in their nonperturbative coefficient laws, but distinct perturbative shifts and exponential corrections resolve differences invisible in their common leading $N^{3/2}$ behavior.
\end{abstract}
\clearpage
\setcounter{page}{1}
\noindent\rule{\textwidth}{0.6pt}\vspace{-0.5cm}
\begingroup
\hypersetup{linktoc=page}
\fontsize{9}{10}\selectfont
\tableofcontents
\endgroup
\vspace{0.2cm}\noindent\rule{\textwidth}{0.6pt}
\clearpage

\section{Introduction}
\label{sec:intro}

The leading large-$N$ behavior of M2-brane theories is closely tied
to the geometry of their holographic backgrounds. For many
supersymmetric observables, the volume of the internal
Sasaki--Einstein manifold governs the coefficient of the
characteristic $N^{3/2}$ term. This geometric universality leaves
open how theories with the same leading behavior can be distinguished
beyond the continuum approximation. Subleading powers of $N$,
rank-independent constants, and exponentially small corrections
probe information not fixed by that volume. Here we study these
contributions to the on-shell effective twisted superpotentials of
two three-dimensional $\mathcal N=2$ theories associated with distinct
cyclic quotients of the non-toric Sasaki--Einstein manifold
$V^{5,2}$.

The common geometry is the Calabi--Yau quadric
\begin{equation}
 X=\{z_0^2+A_1B_1+A_2B_2=0\}\subset\mathbb C^5\,,
 \label{eq:intro-cone}
\end{equation}
whose link is $V^{5,2}=\mathrm{SO}(5)/\mathrm{SO}(3)$.
Two circle actions preserving the quadric define the quotients
associated with the theories~\cite{Martelli:2009ga,Jafferis:2009th}.
Rotating $A_1,A_2$ by a common phase and $B_1,B_2$ by its inverse
gives a free action on the link. Its $\mathbb Z_k$ quotient
corresponds to the two-node Chern--Simons theory with gauge group
$\U(N)_k\times \U(N)_{-k}$, which we call Model I. Rotating only
$A_1$ and $B_1$ by opposite phases instead fixes a complex surface
in the cone, whose intersection with the link is
$\mathbb{RP}^3$. The corresponding $\mathbb Z_{N_f}$ quotient
has a transverse $A_{N_f-1}$ singularity along this locus for
$N_f>1$. It is associated with Model II, a one-node $\U(N)$ theory
at vanishing Chern--Simons level with three adjoint chirals and
$N_f$ fundamental--antifundamental pairs.

At equal quotient order $\mathfrak k$, with $\mathfrak k=k$ in
Model I and $\mathfrak k=N_f$ in Model II, both volumes equal
$\operatorname{Vol}(V^{5,2})/\mathfrak k$: the fixed locus has zero
seven-dimensional measure. The two theories have the same leading
$N^{3/2}$ on-shell twisted superpotential at the superconformal
chemical potentials~\cite{Hosseini:2016ume}, consistently with this
volume dependence. The non-toric geometry admits volume
minimization, and sphere-matrix-model calculations reproduce the
leading free energies~\cite{Martelli:2006yb,Jafferis:2011zi}.
The consistent reduction to minimal four-dimensional
$\mathcal N=2$ gauged supergravity retains a common universal
sector~\cite{Gauntlett:2007ma}, but not the full invariant spectra
or additional fixed-locus fields. Comparing the two theories
therefore tests which corrections distinguish their quotient
actions despite the common leading normalization.

The effective twisted superpotential $\Wt$ determines the Bethe
vacua of a three-dimensional $\mathcal N=2$ theory compactified on
a circle~\cite{Nekrasov:2009uh}. Its on-shell values govern the
leading semiclassical behavior of holomorphic
blocks~\cite{Beem:2012mb} and enter the construction of
supersymmetric partition functions~\cite{Closset:2018ghr}.
For theories whose generic mass deformations yield isolated
massive vacua, the supersymmetric index on
$S^2\times S^1$~\cite{Bhattacharya:2008zy} admits the block
decomposition~\cite[Eq.~(1.3)]{Beem:2012mb}
\begin{equation}
 \mathcal I(\boldsymbol m,\boldsymbol y;\mathfrak q)
 =\sum_\alpha
 B^\alpha(\mathfrak q^{\boldsymbol m/2}\boldsymbol y;\mathfrak q)
 B^\alpha(\mathfrak q^{\boldsymbol m/2}\boldsymbol y^{-1};
          \mathfrak q^{-1})\,,
 \label{eq:intro-blocks}
\end{equation}
where $\boldsymbol m$ and $\boldsymbol y$ denote background
magnetic fluxes and flavor fugacities. The label $\alpha$
specifies the vacuum common to the two blocks, evaluated in
complementary chambers. Within a fixed chamber, taking
$\hbar=\log\mathfrak q\to0$ at fixed rank, flavor arguments,
and circle radius gives
\begin{equation}
 \log B^\alpha(\boldsymbol y;\mathfrak q)
 =\frac{\Wt_\alpha^{\mathrm{block}}(\boldsymbol y)}{\hbar}
  +o(\hbar^{-1})\,,
 \label{eq:intro-block-asymptotics}
\end{equation}
with $\Wt_\alpha^{\mathrm{block}}$ in the block
normalization~\cite[Eq.~(2.22)]{Beem:2012mb}.%
\footnote{The leading chiral contribution in Eq.~(3.18)
of~\cite{Beem:2012mb} is
$m_\phi^2/4+\Li_2(-\e^{-m_\phi})$.
Setting $m_\phi=\ii(\pi-z)$ gives
$\mathfrak w(z)-\pi^2/12$, with $\mathfrak w$ defined in
Eq.~\eqref{eq:ell}. This real additive difference leaves the
imaginary on-shell value unchanged. The gauge-variable change
$\sigma=-\ii u$ maps the quadratic Chern--Simons term and
exponentiated critical equation to our conventions;
Section~\ref{subsec:conventions} fixes the logarithmic lifts
and contact terms.}
Thus an exponentially small correction in $N$ still contributes
to the leading semiclassical block action at fixed rank.
The parameter $\mathfrak q$ is distinct from the rank-dependent
instanton nome $q$ used below.

The on-shell twisted superpotential does not determine the full
index, which also requires the block fluctuations and gluing sum.
For the topologically twisted index on
$\Sigma_{\mathfrak g}\times S^1$, flux operators and Jacobian
factors supply the additional weights in the Bethe-vacuum
sum~\cite{Benini:2015noa,Benini:2016hjo,Closset:2018ghr}.
Its leading large-$N$ extremization reproduces supersymmetric
$\mathrm{AdS}_4$ black-hole entropy~\cite{Benini:2015eyy}, and the
associated entropy functions admit a gravitational-block
formulation~\cite{Hosseini:2019iad}. Resolving the on-shell twisted
superpotential is therefore a necessary part of these comparisons,
not a reconstruction of the complete partition functions. We study
a specified Bethe branch without assuming its dominance in the
vacuum sums.

ABJM theory provides an important precedent for organizing
corrections beyond the leading geometric answer~\cite{Aharony:2008ug}.
Its sphere partition function has a Fermi-gas description in which
the perturbative rank expansion resums into an Airy function and
nonperturbative effects form membrane and worldsheet
sectors~\cite{Marino:2011eh,Hatsuda:2013oxa}. At special levels,
the grand partition function admits an exact theta-function
representation~\cite{Codesido:2014oua}. These results are specific
to the sphere observable: its Airy shift differs from the shifted
rank of the perturbative twisted index, even at superconformal
chemical potentials~\cite{Bobev:2022jte}. A corresponding spectral
description of the finite-rank Bethe problem remains to be
established.%
\footnote{An interacting-Fermi-gas description also motivates
Airy behavior for short-range one-node $\mathcal N=2$ sphere
matrix models, but does not fix their full constant and rank
shift~\cite[Section~3.2]{Marino:2012az}. It does not supply a
spectral problem for the Bethe observables studied here.}

High-precision studies of the ABJM and $\mathcal N=4$ ADHM
on-shell twisted superpotentials have revealed a different
organizing principle~\cite{Hosseini:2026dkj,Hosseini:2026lvj,Hosseini:ADHM2026}.
At special integer parameters, their reconstructed exponential
coefficients are consistent with cubic divisor laws generated by
Eichler integrals of weight-four Eisenstein series. Convergent
integrals for the rank-independent constants fix the normalization
needed to isolate these sectors~\cite{Hong:2026zul,Hosseini:ADHM2026,Hosseini:2026lvj}.
The arithmetic structure constrains infinitely many coefficients,
but its origin in the microscopic Bethe equations is not known.
Both theories have enhanced supersymmetry and geometries obtained
from quotients of flat space. The $V^{5,2}$ theories test whether
the same structure persists in a non-toric $\mathcal N=2$ setting,
and how its coefficient laws depend on the quotient action.

We work at the superconformal chemical potentials specified in
Section~\ref{sec:bethe} and write $\W=\im\Wt$ for the on-shell
value on the selected branch. Following the leading continuum
results for both models~\cite{Hosseini:2016ume} and the shifted-rank
perturbative proposal for the flavored
theory~\cite{Bobev:2023lkx,Bobev:2024mqw}, we propose the
Chern--Simons counterpart and determine integral representations
for both constant functions. The perturbative expressions are
\begin{equation}
 \begin{aligned}
 \WI^{\mathrm{pert}}(N,k)
 &=\frac{8\pi^2\sqrt{k}}{27}
   \left(N-\frac{k}{12}+\frac1{2k}\right)^{3/2}
   +\CI(k)\,,\\
 \WII^{\mathrm{pert}}(N,N_f)
 &=\frac{8\pi^2\sqrt{N_f}}{27}
   \left(N+\frac{N_f}{6}+\frac1{4N_f}\right)^{3/2}
   +\CII(N_f)\,.
 \end{aligned}
 \label{eq:intro-parents}
\end{equation}
The parentheses define the shifted ranks $\Nh_{\mathrm I}$ and
$\Nh_{\mathrm{II}}$; we take the positive square root on
$\Nh>0$. Here ``perturbative'' refers to the large-$N$ expansion
at fixed level or flavor number. For Model I, the shift and
constant integral are new proposals inferred from the numerical
Bethe solutions. For Model II, the shift is inherited from the
previous proposal, while we give an integral formula for its constant, previously
estimated numerically.

Both constant integrals converge absolutely for all real
$k,N_f>0$. At positive integer parameters, we evaluate them in
terms of $\zeta(3)$ and finite sums of Clausen functions. We also
derive their divergent large-parameter expansions and prove that
Borel summation in the positive inverse-parameter direction
recovers the defining integrals. These are exact statements about
the proposed constant functions; their identification with the
physical Bethe constants and the absence of further algebraic
corrections to Eq.~\eqref{eq:intro-parents} remain conjectural.
Fixing the constants is essential for the exponential analysis ---
an error in a rank-independent term eventually exceeds every
correction of order $\e^{-\mathrm{const.}\sqrt N}$.

For Model I at $k=1,3$ and Model II at $N_f=1,3$, we conjecture
the complete remainder
\begin{equation}
 \W^{\mathrm{np}}
 =\frac{\mathfrak k^2}{2\pi}\sum_{m\geq1}c_m
       \left(t+\frac1m\right)q^m\,,
 \qquad
 t=2\pi\sqrt{\frac{\Nh}{\mathfrak k}}\,,
 \qquad q=\e^{-t}\,,
 \label{eq:intro-series}
\end{equation}
where $\W^{\mathrm{np}}=\W-\W^{\mathrm{pert}}$ and the sequence
$c_m$ depends on the model and parameter. The quantities $m^2c_m$
are alternating rational combinations of $\sigma_3(m)$,
$\sigma_3(m/3)$, and $\sigma_3(m/9)$, where
$\sigma_3(n)=\sum_{d\mid n}d^3$ for positive integer $n$ and
vanishes otherwise. The coefficients of $tq^m$ and $q^m$ are
fitted independently. Their prefactor relation is a post-fit
test, not a constraint on the least-squares problem.

The divisor laws give a precise arithmetic description of the
proposed remainders. Define
\begin{equation*}
 \Phi(q)=\frac{\mathfrak k^2}{2\pi}
       \sum_{m\geq1}\frac{c_m}{m}q^m\,,
 \qquad D=q\partial_q\,.
\end{equation*}
The right-hand side of Eq.~\eqref{eq:intro-series} is then
$(1+tD)\Phi(q)|_{q=\e^{-t}}$, with $t$ held fixed during
differentiation. Three logarithmic derivatives recover cubic
divisor sums, whose alternating signs are absorbed by the
half-period shift $Q=-q$. Each $\Phi$ is consequently a
Fourier-normalized Eichler integral of a weight-four Eisenstein
combination with its constant mode
subtracted~\cite{Dorigoni:2020oon,Pasol:2013}.
We determine minimal congruence levels $3,9,9$ for the common
unit-parameter generator and the two parameter-three generators
in $Q$, and minimal integral levels $12,36,36$ in $q$.
We prove absolute convergence in the unit disk, establish its
boundary as a natural boundary for the primitives and their
first logarithmic derivatives, and derive explicit tail bounds
on the physical ray. Dilation and root-of-unity identities relate
the proposed functions. The modular transformation laws belong
to the Eisenstein generators, not to the completed on-shell twisted
superpotentials.

The two theories share this organization without having identical
completions. Their proposed perturbative terms and exponential
remainders coincide at unit quotient order. At order three, the
divisor laws differ even after the actions and shifted powers are
aligned, with opposite signs for the leading correction. At every
fixed common order $\mathfrak k>1$, the unequal shifts already
predict a separation at order $N^{1/2}$. High-precision comparisons
support the special-parameter completions at ranks outside the
reconstruction windows, testing the combined contribution of
sectors beyond those reconstructed individually. For Model II,
a strictly convex, coercive variational formulation also proves
existence and uniqueness of the selected imaginary-holonomy
branch. The Model-I branch is specified by symmetry, strip
conditions, and numerical continuation.

The special-parameter formulas do not determine a generic-coupling
completion. At parameter two, the reconstructed prefactors depart
from Eq.~\eqref{eq:intro-series} and involve higher and inverse
powers of $t$. The large-parameter limit requires separate care:
when $N/\mathfrak k$ remains fixed with positive shifted coupling,
$q$ approaches a nonzero value. The exponential rank corrections
can then contribute to ordinary genus coefficients. Borel
summation of the constant functions therefore does not determine
the full physical genus expansion, which requires control of the
remainder in the joint large-rank, large-parameter limit.

The perturbative terms admit a separate bulk characterization.
The relative M-theory localization prescription of
Refs.~\cite{BenettiGenolini:2026cyc,BenettiGenolini:2026kjy}
describes the sphere partition function in an ensemble with
chemical potential $\mu$ conjugate to the M2-brane number $N$,
identified with the gauge-theory rank. Its perturbative grand
potential takes the form
\begin{equation*}
 J_{\mathrm{pert}}(\mu)
 =\frac{\mathcal C}{3}\mu^3
  +\mathcal B\mu+\mathcal A\,.
\end{equation*}
The cubic and linear terms follow from the localized action,
including the relevant curvature and quotient contributions,
while the $\mu$-independent term $\mathcal A$ is supplied by
the proposed one-loop determinant computed from a relative
tangent--cotangent character. These coefficients depend on
the squashing and quotient parameters. An inverse Laplace
transform in $\mu$, with the sphere contour and measure,
then gives a perturbative Airy expression at fixed $N$.
Our comparison with the Bethe
observable uses the leading small-squashing coefficient of its
free energy, not an equality of the observables at finite
squashing. The distinction is also relevant to the flavored-theory
sphere Airy proposal of Ref.~\cite{Bobev:2025ltz}.

For Model I, we obtain the shifted rank by combining the
curvature contribution of the original quadric with the
gravitational charge of the isolated quotient singularity
at its apex. For $k>1$, a circle-preserving smoothing gives
two local $\mathbb C^4/\mathbb Z_k$ singularities, whose
gravitational charges determine the additional contribution
through a relative anomaly
prescription~\cite{Martelli:2009ga,Bergman:2009zh}.
The projected one-loop character independently yields
$\CI(k)/(2\pi)$ through an exact integral identity.
The resulting small-squashing limit agrees with the proposed
Model-I perturbative expression.

For Model II, we generalize the localization prescription to
include the contributions associated with the fixed locus.
A partial resolution of the transverse $A_{N_f-1}$ singularity
gives a curvature contribution that yields the Model-II shift.
The resolution also determines a candidate index for each
abelian gauge multiplet associated with its exceptional cycles.
We assume that the same index applies to every generator of
the $\SU(N_f)$ gauge symmetry restored when these cycles
collapse. Combining the resulting localized contribution with
the invariant parent character gives $\CII(N_f)/(2\pi)$.
A broader treatment of singular sectors in M-theory
localization, retaining the finite-squashing and
geometric-source dependence of the perturbative sphere
partition function, will be presented in
\cite{Hosseini:MTheoryLocalizationI}.

The character and curvature identities are exact within these
geometric constructions. Their physical interpretation assumes
the relative localization and anomaly prescriptions, together
with compatible boundary conditions and flux sectors.
In particular, the assignment of the abelian index to the
enhanced gauge sector still requires an independent derivation
from the gauge-fixed seven-dimensional theory.

The geometric and arithmetic results concern complementary
parts of the same observable. The bulk calculation characterizes
the perturbative terms, while the Eisenstein description
constrains the conjectured exponential remainder. These results
do not derive the Eisenstein structure from M-theory localization
or establish a modular action on the microscopic Bethe equations. Nor does the
unit-parameter agreement imply equivalence of the full theories;
their abelian chiral rings have additional sectors sensitive to
ordinary-superpotential deformations~\cite{Cremonesi:2016nbo}.
Independent fixed-rank expansions for the real Model-II branch
and the local large-level Model-I branch give a further comparison
directly from the Bethe equations. The common leading volume and recurring modular
structure therefore coexist with differences in the perturbative
shifts and exponential coefficient laws.

The paper is organised as follows.
Section~\ref{sec:bethe} introduces the theories, their quotient
geometries, and the selected Bethe branches.
Section~\ref{sec:constants} derives the continuum limits and
studies the shifted-rank proposals and constant functions.
Section~\ref{sec:sectors} reconstructs the exponential sectors
and compares the two theories at finite rank.
Section~\ref{sec:modular} develops the Eisenstein and Eichler-integral
description, including modular levels, convergence, natural
boundaries, and function identities.
Section~\ref{sec:genus} derives the large-parameter expansions and
Borel sums. Section~\ref{sec:localization} presents the classical-action
and one-loop calculations for both quotients, followed by the
discussion in Section~\ref{sec:discussion}. The appendices contain
the integral evaluations, numerical methods, fixed-rank asymptotics,
and detailed bulk character, curvature, and charge calculations.

\section{The two theories and their Bethe branches}
\label{sec:bethe}

\subsection{Common conventions}
\label{subsec:conventions}

All holonomies and chemical potentials are dimensionless.
A chiral multiplet with angular chemical potential $\Delta$
has fugacity $\e^{\ii\Delta}$. Following Ref.~\cite{Hosseini:ADHM2026},
we assign each chiral weight the contribution
\begin{equation}
 \mathfrak w(z)=\Li_2(\e^{\ii z})-\frac12g_2(z)\,,
 \qquad
 g_2(z)=\frac{z^2}{2}-\pi z+\frac{\pi^2}{3}\,,
 \label{eq:ell}
\end{equation}
where $z$ includes both the gauge and flavor arguments.
This prescription fixes the polynomial and additive terms
in the chiral contribution. We use the principal logarithm
and dilogarithm, with the dilogarithm cut along $[1,\infty)$.
For $0<\re z<2\pi$,
\begin{equation}
 \mathfrak w'(z)
 =-\ii\Log(1-\e^{\ii z})-\frac z2+\frac\pi2\,,
 \qquad
 \e^{\ii\mathfrak w'(z)}=2\sin\frac z2\,.
 \label{eq:ell-derivative}
\end{equation}
Dilogarithm inversion is performed in this strip before
continuation along the chosen branch. In the same strip,
$\overline{\mathfrak w(z)}=-\mathfrak w(\overline z)$.
Adjoint zero weights are included in the twisted superpotentials
and contribute to their critical values, although their
holonomy derivatives vanish.

Our Bethe-equation convention is
$\exp(\ii\partial_{h_A}\Wt_0)=1$, where $h_A$ denotes an
independent complex gauge holonomy and $\Wt_0$ is a chosen
representative before fixing the integer logarithmic data.
A logarithmic lift replaces it by
\begin{equation}
 \Wt=\Wt_0+2\pi\sum_A\nu_A h_A\,,
 \qquad
 \nu_A\in\mathbb Z\,,
 \qquad
 \frac{\partial\Wt}{\partial h_A}=0\,.
 \label{eq:logarithmic-lift}
\end{equation}
The integers do not alter the exponentiated equations, but
they specify the unwrapped equations solved below.
In this normalization, a node of Chern--Simons level $k_a$
contributes $-k_a\sum_i h_{a,i}^2/2$, and a topological
chemical potential $\Delta_{T,a}$ contributes
$-\Delta_{T,a}\sum_i h_{a,i}$~\cite{Hosseini:2016tor}.
A change of topological fugacity by a sign is therefore
distinct from an integer logarithmic lift. The displayed
linear terms include the phase conventions used for each
model; no further vector-multiplet term is to be added.

We use the angular abbreviations
\begin{equation}
 \theta=\frac{2\pi}{3}\,,
 \qquad
 \varphi=\frac{\pi}{3}\,.
 \label{eq:angles}
\end{equation}
Their assignment to the fields is specified separately
for each model. The superconformal chemical potentials
are $\Delta_\Phi=\pi R_\Phi$ for the corresponding
R-charge assignment. They do not require a choice of
magnetic fluxes. Since these charges are fractional,
the universal-twist fluxes are not integral on $S^2$;
compatible integral twists can be implemented on
$\Sigma_\mathfrak g$ at suitable genus~\cite{Hosseini:2016ume}.

Define
\begin{equation}
 K_s(z)=\mathfrak w'(s+z)-\mathfrak w'(s-z)\,,
 \qquad
 r_s(a,b)=\frac{b-\e^{\ii s}a}{a-\e^{\ii s}b}\,.
 \label{eq:kernel}
\end{equation}
For $0<s<\pi$ and $|\re z|<s$, the kernel and its
derivative can be evaluated as
\begin{equation}
 \begin{aligned}
 K_s(z)
 &=-\ii\left[
 \Log\sin\frac{s+z}{2}-\Log\sin\frac{s-z}{2}\right]\,,\\
 K_s'(z)
 &=-\frac{\ii}{2}\left[
 \cot\frac{s+z}{2}+\cot\frac{s-z}{2}\right]\,.
 \end{aligned}
 \label{eq:sine-kernel}
\end{equation}
Both logarithms are continued from $z=0$ within this
strip. For the real evaluations, we use
\begin{equation}
 D_s(d)=\im\Li_2(\e^{-d+\ii s})\,,
 \qquad d\geq0\,,
 \qquad
 \Cltwo(s)=D_s(0)
         =\sum_{m\geq1}\frac{\sin(ms)}{m^2}\,,
 \label{eq:D}
\end{equation}
with real $s$.

\subsection{Model I: the Chern--Simons theory}
\label{subsec:model-I}

Model I has gauge group $\U(N)_k\times\U(N)_{-k}$~\cite{Martelli:2009ga}.
The chiral multiplets $A_a$ transform in
$(\boldsymbol {\rm N},\overline{\boldsymbol {\rm N}})$ and $B_a$
in $(\overline{\boldsymbol {\rm N}},\boldsymbol {\rm N})$, with
$a=1,2$. There is an adjoint chiral $\phi_a$ at each
node. Figure~\ref{fig:V52-quivers}(a) displays these
representations. We use the ordinary superpotential
\begin{equation}
 W_{\mathrm I}
 =\Tr\left[
 \phi_1^3+\phi_2^3
 +\phi_1(A_1B_1+A_2B_2)
 +\phi_2(B_1A_1+B_2A_2)\right]\,.
 \label{eq:ordinary-I}
\end{equation}
The labels differ from those of
Ref.~\cite{Hosseini:2016ume} by interchanging $B_1$
and $B_2$. The cubic terms impose
$3\Delta_{\phi_1}=3\Delta_{\phi_2}=2\pi$ and
$\Delta_{\phi_r}+\Delta_{A_a}+\Delta_{B_a}=2\pi$
for $r,a=1,2$. We take the superconformal chemical
potentials
\begin{equation}
 \Delta_{\phi_1}=\Delta_{\phi_2}
 =\Delta_{A_a}=\Delta_{B_a}=\theta=\frac{2\pi}{3}\,,
 \qquad a=1,2\,.
 \label{eq:chemical-I}
\end{equation}
The finite-rank observables are defined at this fixed
assignment.\footnote{The terminology refers to the
superconformal assignment associated with the
$V^{5,2}$ construction, not to a proof of the exact
infrared R-symmetry at every finite rank. In particular,
the rank-one, level-one theory has an equivalent
Wess--Zumino description subject to the
accidental-symmetry constraints of
Ref.~\cite{Chester:2015qca}. These do not change the
definition of the twisted superpotential at the
specified chemical potentials.}

Let $u_i$ and $\widetilde u_i$ denote the holonomies associated
with the nodes at Chern--Simons levels $k$ and $-k$, respectively, with
all gauge indices running from $1$ to $N$. We fix
\begin{equation}
 n_i=\left\lfloor\frac N2\right\rfloor+1-i\,,
 \qquad
 \widetilde n_i=i-\left\lceil\frac N2\right\rceil\,.
 \label{eq:lifts}
\end{equation}
With the topological chemical potential set to zero in the
convention of Eq.~\eqref{eq:logarithmic-lift}, the lifted
twisted superpotential is
\bea
 \Wt_{\mathrm I}={}&
 \frac{k}{2}\sum_i(\widetilde u_i^2-u_i^2)
 +2\pi\sum_i(n_i u_i-\widetilde n_i\widetilde u_i)
 \\
 &+2\sum_{i,j}\bigl[
 \mathfrak w(\theta+u_i-\widetilde u_j)
 +\mathfrak w(\theta+\widetilde u_j-u_i)\bigr]
 \\
 &+\sum_{i,j}\bigl[
 \mathfrak w(\theta+u_i-u_j)
 +\mathfrak w(\theta+\widetilde u_i-\widetilde u_j)\bigr]\,.
 \label{eq:potential-I}
\eea
The factor of two in the bifundamental contribution
counts the two fields in each orientation.\footnote{
At this specialization, Eq.~(2.31) of
Ref.~\cite{Bobev:2024mqw} differs from
Eq.~\eqref{eq:potential-I} by the real constant
$-\pi^2N^2/3$. Our angular chemical potentials are
$\pi$ times the dimensionless ones used there.
In a convention retaining a separate factor
$(-1)^{N-1}$ in each node's Bethe equation, the
compensating topological chemical potential is $\pi(N-1)$
modulo $2\pi$. Such a compensation changes the
fugacity convention, not the integer data in
Eq.~\eqref{eq:lifts}.}

Differentiating Eq.~\eqref{eq:potential-I}
with respect to the independent complex holonomies
gives
\begin{equation}
 \begin{aligned}
 0={}&-ku_i+2\pi n_i
      +\sum_jK_\theta(u_i-u_j)
      +2\sum_jK_\theta(u_i-\widetilde u_j)\,,\\
 0={}&k\widetilde u_i-2\pi\widetilde n_i
      +\sum_jK_\theta(\widetilde u_i-\widetilde u_j)
      +2\sum_jK_\theta(\widetilde u_i-u_j)\,.
 \end{aligned}
 \label{eq:BAE-I}
\end{equation}
The diagonal adjoint terms vanish because $K_\theta(0)=0$.
For $x_i=\e^{\ii u_i}$ and
$\widetilde x_i=\e^{\ii\widetilde u_i}$,
Eq.~\eqref{eq:ell-derivative} implies
$\e^{\ii K_s(u_i-u_j)}=r_s(x_i,x_j)$.
The exponentiated equations are consequently
\begin{equation}
 \begin{aligned}
 x_i^k
 =\prod_{j\ne i}r_\theta(x_i,x_j)
   \prod_jr_\theta(x_i,\widetilde x_j)^2\,,\qquad
 \widetilde x_i^{-k}
 =\prod_{j\ne i}r_\theta(\widetilde x_i,\widetilde x_j)
   \prod_jr_\theta(\widetilde x_i,x_j)^2\,.
 \end{aligned}
 \label{eq:exp-I}
\end{equation}
These equations provide an independent consistency
check, but do not retain the logarithmic lifts.

To describe the branch used in the computation, let
$\epsilon_N=1$ for even $N$ and $0$ for odd $N$,
and remove the parity-dependent centre,
\begin{equation}
 \begin{aligned}
 u_i&=c_N+z_i\,,
 &\qquad \widetilde u_i&=c_N+\widetilde z_i\,,
 &\qquad c_N&=\frac{\pi\epsilon_N}{k}\,,\\
 \widetilde z_i&=-\overline z_i\,,
 & z_{N+1-i}&=-z_i\,.
 \end{aligned}
 \label{eq:symmetry-I}
\end{equation}
Writing $z_i=p_i+\ii y_i$, we order the imaginary
parts as $y_1<\cdots<y_N$ and impose
$-\pi/3<p_i<0$ whenever $y_i>0$.
Reflection fixes the other half and sets the central
$z_i$ to zero at odd rank. These inequalities keep
all chiral arguments in the prescribed strip.
Since
$2\pi n_i-kc_N=\pi(N+1-2i)$, the first-node
equations reduce to
\begin{equation}
 G_i=-kz_i+\pi(N+1-2i)
      +\sum_jK_\theta(z_i-z_j)
      +2\sum_jK_\theta(z_i+\overline z_j)=0\,.
 \label{eq:reduced-I}
\end{equation}
The second-node equations are their complex conjugates.
The reduction is imposed after differentiating the
holomorphic twisted superpotential; the resulting
equations are real analytic in $p_i,y_i$.
The symmetry and strip conditions specify the domain
used in numerical continuation. They do not establish
global uniqueness of the Model-I solution.

On this symmetry slice, conjugation pairs the chiral
contributions, and the real parts of the Chern--Simons
and linear terms cancel. The canonical twisted
superpotential is therefore purely imaginary. We write
$\WI(N,k)=\im\Wt_{\mathrm I}$ at the selected
critical point. For a numerically stable evaluation,
inversion in $0<s<2\pi$ gives
$\im[\mathfrak w(s+\ii d)+\mathfrak w(s-\ii d)]
=(\pi-s)d+2D_s(d)$.
Pairing the adjoint and bifundamental terms and
using $d_{ij}=|y_i-y_j|$, we obtain on the ordered slice
\begin{align}
 \im\Wt_{\mathrm I}={}&
 -2k\sum_i p_i y_i+2N\Cltwo(\theta)
 \notag\\
 &+2\sum_{i<j}\bigl[
 D_{\theta+p_i-p_j}(d_{ij})
 +D_{\theta-p_i+p_j}(d_{ij})\bigr]
 \notag\\
 &+4\sum_{i<j}\bigl[
 D_{\theta+p_i+p_j}(d_{ij})
 +D_{\theta-p_i-p_j}(d_{ij})\bigr]
 \notag\\
 &+2\sum_i\bigl[
 \Cltwo(\theta+2p_i)+\Cltwo(\theta-2p_i)\bigr]\,.
 \label{eq:stable-I}
\end{align}
The linear terms from inversion cancel the logarithmic
lifts because $6(\pi-\theta)=2\pi$ and
$\sum_{i<j}d_{ij}=\sum_i(2i-N-1)y_i$.
The Chern--Simons contribution
$-2k\sum_i p_i y_i$ remains.
All dilogarithm arguments in Eq.~\eqref{eq:stable-I}
have modulus at most one.

\subsection{Model II: the flavored theory}
\label{subsec:model-II}
\label{subsec:real}

\begin{figure}[tbp]
 \centering
 \begingroup
 \tikzset{
  v52 gauge/.style={
   circle,draw,line width=.65pt,minimum size=13mm,
   inner sep=1pt,font=\small
  },
  v52 flavor/.style={
   rectangle,draw,line width=.65pt,minimum size=10mm,
   inner sep=1pt,font=\small
  },
  v52 arrow/.style={
   -{Latex[length=1.6mm,width=1.0mm]},line width=.6pt
  },
  v52 label/.style={font=\small,inner sep=1pt}
 }
 \begin{minipage}[t]{.48\textwidth}
  \centering
  \begin{tikzpicture}[baseline=(current bounding box.north)]
   \path[use as bounding box] (-1.35,-2.35) rectangle (4.45,1.85);
   \node[v52 gauge] (L) at (0,0) {$\U(N)_k$};
   \node[v52 gauge] (R) at (3.1,0) {$\U(N)_{-k}$};

   \draw[v52 arrow]
    (R.140) to[out=140,in=40]
    node[v52 label,above] {$A_1$} (L.40);
   \draw[v52 arrow]
    (R.110) to[out=125,in=55]
    node[v52 label,above] {$A_2$} (L.70);

   \draw[v52 arrow]
    (L.320) to[out=-40,in=220]
    node[v52 label,below] {$B_1$} (R.220);
   \draw[v52 arrow]
    (L.290) to[out=-55,in=235]
    node[v52 label,below] {$B_2$} (R.250);

   \draw[v52 arrow]
    (L.145) to[out=145,in=215,looseness=4.2]
    node[v52 label,left] {$\phi_1$} (L.215);
   \draw[v52 arrow]
    (R.35) to[out=35,in=-35,looseness=4.2]
    node[v52 label,right] {$\phi_2$} (R.325);
   \node[font=\small] at (1.55,-2.12) {(a) Model I};
  \end{tikzpicture}
 \end{minipage}
 \hfill
 \begin{minipage}[t]{.48\textwidth}
  \centering
  \begin{tikzpicture}[baseline=(current bounding box.north)]
   \path[use as bounding box] (-1.75,-2.35) rectangle (4.05,1.85);
   \node[v52 gauge] (G) at (0,0) {$\U(N)_0$};
   \node[v52 flavor] (F) at (3.0,0) {$N_f$};

   \draw[v52 arrow]
    (G.145) to[out=145,in=215,looseness=4.2]
    node[v52 label,left] {$\Psi_1$} (G.215);
   \draw[v52 arrow]
    (G.55) to[out=55,in=125,looseness=4.2]
    node[v52 label,above] {$\Psi_2$} (G.125);
   \draw[v52 arrow]
    (G.235) to[out=235,in=305,looseness=4.2]
    node[v52 label,below] {$\Psi_3$} (G.305);

   \draw[v52 arrow]
    (F.north west) to[out=155,in=25]
    node[v52 label,above] {$\psi$} (G.25);
   \draw[v52 arrow]
    (G.335) to[out=-25,in=205]
    node[v52 label,below] {$\widetilde\psi$} (F.south west);
   \node[font=\small] at (1.15,-2.12) {(b) Model II};
  \end{tikzpicture}
 \end{minipage}
 \endgroup
 \caption{Matter representations of the two $V^{5,2}$
 theories. Circular nodes are gauged, with their
 Chern--Simons levels displayed as subscripts.
 An arrow from one node to another is antifundamental
 at its tail and fundamental at its head; an adjoint
 chiral is represented by a loop. The square node in
 Model II is ungauged and labels the $N_f$ flavor pairs.
 Each adjoint and each bifundamental chiral is drawn
 separately.}
 \label{fig:V52-quivers}
\end{figure}
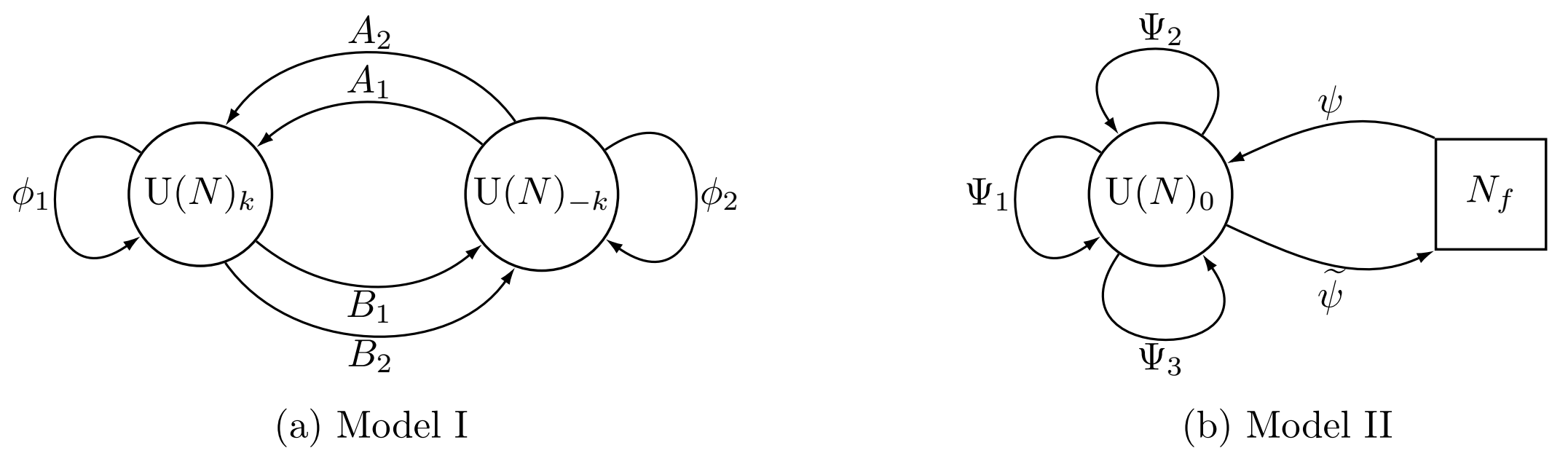

Model II has gauge group $\U(N)$ at vanishing
Chern--Simons level, three adjoint chiral multiplets
$\Psi_I$, $I=1,2,3$, and $N_f$ fundamental--antifundamental
pairs $(\psi_a,\widetilde\psi_a)$, with
$a=1,\ldots,N_f$~\cite{Jafferis:2009th}.
We take $\psi_a$ to be a fundamental column and
$\widetilde\psi_a$ an antifundamental row.
In the complex basis used here, the ordinary
superpotential is
\begin{equation}
 W_{\mathrm{II}}
 =\Tr\bigl(\Psi_3[\Psi_1,\Psi_2]\bigr)
 +\sum_{a=1}^{N_f}\widetilde\psi_a
   \bigl(\Psi_1\Psi_2+\Psi_2\Psi_1-\Psi_3^2\bigr)\psi_a\,.
 \label{eq:ordinary-II}
\end{equation}
Let $\Delta_I$ denote the adjoint chemical potentials,
and let $\Delta_q,\Delta_{\widetilde q}$ be the common
fundamental and antifundamental chemical potentials.
The interactions impose
\begin{equation}
 \begin{aligned}
 \Delta_1+\Delta_2+\Delta_3=2\pi\,,\qquad
 \Delta_q+\Delta_{\widetilde q}+\Delta_1+\Delta_2
 =\Delta_q+\Delta_{\widetilde q}+2\Delta_3=2\pi\,.
 \end{aligned}
 \label{eq:chemical-constraints-II}
\end{equation}
We use the superconformal chemical potentials
\begin{equation}
 \Delta_1=\Delta_2=\Delta_3=\theta=\frac{2\pi}{3}\,,
 \qquad
 \Delta_q=\Delta_{\widetilde q}=\varphi=\frac{\pi}{3}\,,
 \qquad \Delta_m=0\,,
 \label{eq:chemical-II}
\end{equation}
where $\Delta_m$ is the topological chemical potential.
The matter representations are shown in
Fig.~\ref{fig:V52-quivers}(b).

The holonomies are $v_i$, and the logarithmic data are
\begin{equation}
 n_i^{\mathrm{II}}
 =\left\lfloor\frac{N+1}{2}\right\rfloor-i\,,
 \qquad
 \Delta_T^{\mathrm{II}}=\Delta_m-\pi\epsilon_N\,.
 \label{eq:lifts-II}
\end{equation}
Thus
$2\pi n_i^{\mathrm{II}}-\Delta_T^{\mathrm{II}}
=\pi(N+1-2i)-\Delta_m$.
The chemical potential $\Delta_T^{\mathrm{II}}$ is the one entering
the linear term in the chosen localization convention.
Even at $\Delta_m=0$, its parity-dependent phase cannot
be absorbed into integer logarithmic lifts.\footnote{
Equation~(E.8) of Ref.~\cite{Bobev:2024mqw} gives this
dictionary: $u_{1,i}=v_i$,
$n_{1,i}=n_i^{\mathrm{II}}$ and
$\pi\Delta_{m,1}^{\mathrm{there}}
=\Delta_m-\pi\epsilon_N$.
At the chemical potentials in Eq.~\eqref{eq:chemical-II},
our canonical chiral prescription adds the real constant
$\pi^2N(3N-N_f)/18$ to the twisted superpotential in
Eq.~(3.9) of that reference. This change leaves its
derivatives and imaginary critical value unchanged.}
The twisted superpotential is
\begin{align}
 \Wt_{\mathrm{II}}=
 \sum_i(2\pi n_i^{\mathrm{II}}-\Delta_T^{\mathrm{II}})v_i
 +3\sum_{i,j}\mathfrak w(\theta+v_i-v_j)
 +N_f\sum_i\bigl[
 \mathfrak w(\varphi+v_i)+\mathfrak w(\varphi-v_i)\bigr]\,.
 \label{eq:potential-II}
\end{align}
Differentiation gives
\begin{equation}
 0=2\pi n_i^{\mathrm{II}}-\Delta_T^{\mathrm{II}}
   +3\sum_jK_\theta(v_i-v_j)+N_fK_\varphi(v_i)\,.
 \label{eq:BAE-II}
\end{equation}
At $\Delta_m=0$, its multiplicative form, with
$X_i=\e^{\ii v_i}$, can be written as
\begin{equation}
 \left(\frac{X_i-\e^{\ii\varphi}}
              {1-\e^{\ii\varphi}X_i}\right)^{N_f}
 \prod_{j\ne i}
 \left(\frac{X_i-\e^{\ii\theta}X_j}
              {X_i-\e^{-\ii\theta}X_j}\right)^3=1\,.
 \label{eq:exp-II}
\end{equation}
The parity phase in Eq.~\eqref{eq:lifts-II} cancels the
factors introduced when rewriting the adjoint kernels
in this form. The self-factor is excluded: at $j=i$
its cube would be $-1$.

We consider imaginary holonomies $v_i=\ii y_i$ on the
specified logarithmic branch. For $0<s<\pi$,
\begin{equation}
 K_s(\ii y)=f_s(y)
 =2\arctan\left(\cot\frac s2\tanh\frac y2\right)\,,
 \qquad
 f_s'(y)=\frac{\sin s}{\cosh y-\cos s}>0\,,
 \label{eq:f}
\end{equation}
where the inverse tangent takes its real principal
value. The logarithmic equations become
\begin{equation}
 B_i(\boldsymbol y)
 =N_f f_\varphi(y_i)
  +3\sum_j f_\theta(y_i-y_j)
  -\pi(2i-N-1)=0\,.
 \label{eq:real-II}
\end{equation}
The left-hand side of Eq.~\eqref{eq:exp-II} is
$\e^{-\ii B_i(\boldsymbol y)}$.
Imposing $B_i=0$ therefore retains the specified lift,
which the condition $B_i\in2\pi\mathbb Z$ would lose.

The real logarithmic problem has the variational function
\begin{equation}
 \begin{aligned}
 H_s(y)&=\int_0^y f_s(v)\,\rd v\,,\\
 \mathcal Y_{N_f}(\boldsymbol y)
 &=N_f\sum_iH_\varphi(y_i)
   +3\sum_{i<j}H_\theta(y_i-y_j)
   -\pi\sum_i(2i-N-1)y_i\,,
 \end{aligned}
 \label{eq:convex-functional}
\end{equation}
whose gradient is $\boldsymbol B$.
Its coercivity depends on the superconformal value
$3(\pi-\theta)=\pi$.

\begin{theorem}[The real logarithmic branch]
\label{thm:branch}
For every integer $N\geq1$ and real $N_f>0$,
Eq.~\eqref{eq:real-II} has exactly one real solution
$\boldsymbol y^*$. It satisfies
$y_1^*<\cdots<y_N^*$ and $y_{N+1-i}^*=-y_i^*$,
and depends real analytically on $N_f$.
\end{theorem}

\begin{proof}
The Hessian $J=\nabla^2\mathcal Y_{N_f}$ satisfies
\begin{equation}
 \boldsymbol a^{\mathsf T}J\boldsymbol a
 =N_f\sum_i f_\varphi'(y_i)a_i^2
 +3\sum_{i<j}f_\theta'(y_i-y_j)(a_i-a_j)^2>0
 \qquad(\boldsymbol a\ne0)\,.
 \label{eq:positive-J}
\end{equation}
Thus $\mathcal Y_{N_f}$ is strictly convex.
To establish existence, integrate Eq.~\eqref{eq:f} to obtain
\begin{equation}
 H_s(y)=(\pi-s)|y|+2D_s(|y|)-2\Cltwo(s)\,.
 \label{eq:H-explicit}
\end{equation}
For $0<s<\pi$, the function
$\im\Li_2(r\e^{\ii s})$ vanishes at $r=0$ and has
derivative $-\arg(1-r\e^{\ii s})/r>0$ for $0<r\leq1$.
Hence $D_s(d)>0$ for finite $d\geq0$.
Using
$\sum_i(2i-N-1)y_i
=\sum_{i<j}(y_j-y_i)\leq\sum_{i<j}|y_j-y_i|$
and $3(\pi-\theta)=\pi$ gives
\begin{equation}
 \mathcal Y_{N_f}(\boldsymbol y)
 \geq\frac{2\pi N_f}{3}\sum_i|y_i|
 -2NN_f\Cltwo(\varphi)
 -6\binom N2\Cltwo(\theta)\,.
 \label{eq:coercive}
\end{equation}
The function is therefore coercive and attains a unique
global minimum, which solves $\boldsymbol B=0$.

At a solution, the function
$T(u)=N_f f_\varphi(u)+3\sum_jf_\theta(u-y_j^*)$
is strictly increasing and obeys
$T(y_i^*)=\pi(2i-N-1)$.
Since
$T(y_{i+1}^*)-T(y_i^*)=2\pi>0$,
strict monotonicity implies
$y_{i+1}^*>y_i^*$ for every $i$.
Hence $y_1^*<\cdots<y_N^*$.
Reflection combined with reversal of the labels preserves
the equations, so uniqueness gives
$y_{N+1-i}^*=-y_i^*$.
Finally, the equations are real analytic in
$(\boldsymbol y,N_f)$ and their Jacobian is positive
definite. The analytic implicit-function theorem
therefore gives real-analytic dependence on $N_f$.
\end{proof}

This argument adapts the convex construction of
Ref.~\cite[Section~2.3]{Hosseini:ADHM2026}.
The continuation to real $N_f>0$ is a property of the
logarithmic equations --- the physical flavor number is
integer. The theorem does not classify complex Bethe
vacua outside the imaginary slice.

The canonical $\Wt_{\mathrm{II}}$ is purely imaginary
on this slice. Its imaginary part, evaluated on ordered
inputs, has the decaying-argument expression
\begin{equation}
 \im\Wt_{\mathrm{II}}(\ii\boldsymbol y)
 =3N\Cltwo(\theta)+6\sum_{i<j}D_\theta(d_{ij})
 +N_f\sum_i\left[
   2D_\varphi(|y_i|)+\frac{2\pi}{3}|y_i|\right]\,,
 \label{eq:stable-II}
\end{equation}
where $d_{ij}=|y_i-y_j|$. As in Model I, the adjoint
linear terms from inversion cancel the logarithmic
lifts. The fundamental linear terms remain.
Writing $\WII(N,N_f)$ for the critical value, the
off-shell identity
$\im\Wt_{\mathrm{II}}(\ii\boldsymbol y)
=\mathcal Y_{N_f}(\boldsymbol y)
 +3N^2\Cltwo(\theta)+2NN_f\Cltwo(\varphi)$ gives
\begin{equation}
 \WII
 =\min_{\boldsymbol y}\mathcal Y_{N_f}(\boldsymbol y)
  +3N^2\Cltwo(\theta)+2NN_f\Cltwo(\varphi)\,.
 \label{eq:value-minimum}
\end{equation}
This determines the flavor dependence independently
of any large-rank proposal. The envelope theorem and
differentiation of the critical equations yield
\begin{equation}
 \begin{aligned}
 \frac{\partial\WII}{\partial N_f}
 &=\sum_i\left[
   2D_\varphi(|y_i^*|)+\frac{2\pi}{3}|y_i^*|\right]>0\,,\\
 \frac{\partial^2\WII}{\partial N_f^2}
 &=-\boldsymbol f_\varphi^{\mathsf T}
    J^{-1}\boldsymbol f_\varphi\leq0\,,
 \qquad
 \boldsymbol f_\varphi
   =\bigl(f_\varphi(y_1^*),\ldots,f_\varphi(y_N^*)\bigr)^{\mathsf T}\,,
 \end{aligned}
 \label{eq:monotonicity}
\end{equation}
with $J$ evaluated at the critical point.
For $N>1$, strict ordering makes
$\boldsymbol f_\varphi\ne0$, so the second inequality
is strict. The rank-one value is linear in $N_f$.

\subsection{Free and fixed-locus quotients}
\label{subsec:geometry}

The two models use different circle actions on the
quadric in Eq.~\eqref{eq:intro-cone}. Their finite
subgroups have weights
\begin{equation}
 \begin{array}{c|rrrrr}
 &A_1&A_2&B_1&B_2&z_0\\
 \hline
 \mathbb Z_k\text{ in Model I}&1&1&-1&-1&0\\
 \mathbb Z_{N_f}\text{ in Model II}&1&0&-1&0&0
 \end{array}\,.
 \label{eq:quotient-weights}
\end{equation}
A coordinate of weight $w$ is multiplied by
$\e^{2\pi\ii w/\mathfrak k}$ under the cyclic
generator, with $\mathfrak k=k$ or $N_f$ in the
corresponding row~\cite{Martelli:2009ga,Jafferis:2009th}.
Both actions preserve the defining equation.

In Model I, invariance under any nonidentity circle
element forces all four charged coordinates to vanish,
since their weights are $\pm1$. The quadric then sets
$z_0=0$. The only fixed point on the cone is its apex,
which is absent from the link. The circle and each of
its finite subgroups therefore act freely on $V^{5,2}$.
In Model II, the circle instead fixes
\begin{equation}
 S=\{A_1=B_1=0,\ z_0^2+A_2B_2=0\}\,.
 \label{eq:fixed-locus}
\end{equation}
For $N_f>1$, every nonidentity element of the finite
group has the same fixed set. The map
$(u,v)\mapsto(A_2,B_2,z_0)=(u^2,v^2,\ii uv)$
identifies $S$ with $\mathbb C^2/\mathbb Z_2$:
the invariants of $(u,v)\mapsto(-u,-v)$ are generated
by $u^2,v^2,uv$ with their single quadratic relation.
Thus the fixed surface is an $A_1$ cone, and its
three-dimensional link is diffeomorphic to
$S^3/\mathbb Z_2\simeq\mathbb{RP}^3$.

At any nonzero point of $S$, the differential
$2z_0\,\rd z_0+B_2\,\rd A_2+A_2\,\rd B_2$
is nonzero. The ambient quadric is smooth there,
and $A_1,B_1$ are local normal coordinates with
weights $+1,-1$. Their invariant monomials are
generated by $U=A_1^{N_f}$, $V=B_1^{N_f}$ and
$T=A_1B_1$, satisfying $UV=T^{N_f}$.
The transverse quotient is therefore of type
$A_{N_f-1}$. This singularity along the fixed link
for $N_f>1$ is distinct from the intrinsic $A_1$
singularity of $S$ at the cone apex.

Reduction along the Model-II circle produces
$N_f$ coincident D6-branes at its degeneration
locus. D2--D6 strings supply the fundamental--antifundamental
pairs~\cite[Section~2]{Benini:2009qs}, giving the
flavored realization of the non-toric cone described
in Ref.~\cite[Section~5.2]{Jafferis:2009th}.
For $N_f=1$ the finite quotient is trivial and the
transverse space is smooth, but the circle still
degenerates and its reduction contains one D6-brane.
The Model-I circle is free on the link. Quotienting
its fibers by $\mathbb Z_k$ multiplies the first
Chern class by $k$, producing $k$ units of
Ramond--Ramond two-form flux and the levels $(k,-k)$,
without a D6-brane locus on the
link~\cite[Sections~3.2 and~3.5]{Martelli:2009ga}.
The unequal fixed sets distinguish the embeddings
at every common order $k=N_f>1$.

At equal quotient order $\mathfrak k$, both maps are
$\mathfrak k$-to-one almost everywhere, since the fixed
locus of Model II has measure zero. The induced
Sasaki--Einstein metrics therefore give
\begin{equation}
 \operatorname{Vol}(V^{5,2}/\mathbb Z_{\mathfrak k})
 =\frac{\operatorname{Vol}(V^{5,2})}{\mathfrak k}\,,
 \qquad
 \operatorname{Vol}(V^{5,2})=\frac{27\pi^4}{128}\,,
 \label{eq:volume}
\end{equation}
in the normalization $\operatorname{Ric}=6g$ of
Ref.~\cite[Eq.~(3.22)]{Martelli:2009ga}, with the
Model-II volume understood in the orbifold sense
for $N_f>1$.
The inverse-square-root dependence on this volume
accounts for the common leading
$\sqrt{\mathfrak k}\,N^{3/2}$ normalization of the
Bethe critical values. The spectrum surviving the
quotient and the fields localized at its fixed locus
depend on the group embedding and are not determined
by the volume.

The charge and discrete-flux sectors used in the bulk comparison
are specified in Section~\ref{subsec:relative-localization}.

\subsection{Critical-value data and numerical continuation}
\label{subsec:numerics}

At rank one, the selected solutions are $z_1=0$ in
Model I and $y_1^*=0$ in Model II.
Equations~\eqref{eq:stable-I} and~\eqref{eq:stable-II}
then give the exact normalizations
\begin{equation}
 \WI(1,k)=4\Cltwo(\pi/3)\,,
 \qquad
 \WII(1,N_f)=2(N_f+1)\Cltwo(\pi/3)\,,
 \label{eq:rank-one}
\end{equation}
where $\Cltwo(2\pi/3)=\tfrac23\Cltwo(\pi/3)$.
These are finite-rank checks of the chosen
twisted-superpotential representatives.

The numerical construction solves the reduced equations
with analytic Jacobians and increasing-precision Newton
iteration. At each fixed integer parameter, ranks are
followed in ascending order. The first solution is
initialized from the continuum profile. Subsequent
predictors interpolate the ordered roots at the new
rank quantiles and rescale their imaginary coordinates
by $\sqrt{N/(N-1)}$. For Model I, the parity centre
is removed before interpolation and restored afterwards.
Only the positive half is solved; reflection supplies
the other half and the central zero at odd rank.
Backtracking preserves ordering and the Model-I strip
and sign conditions. The Newton displacement relative
to the preceding root spacing supplies a continuity
diagnostic, not a global uniqueness proof for Model I.
Related interpolation, $\sqrt N$ rescaling and damped
Newton refinement were used for the sphere saddle
equations of chiral Chern--Simons--matter theories
in Ref.~\cite{Hosseini:2025jxb}.

The critical-value data cover $2\leq N\leq300$ and
$1\leq k,N_f\leq30$. The target residual precision is
$P=800$ decimal digits through $N=100$ and $P=200$
at higher rank. The working precision $d$ starts at
$40$ digits and increases by
$d\mapsto\min(P+120,2d-10)$, with stage threshold
$10^{-(d-12)}$.
Both logarithmic and exponentiated equations are
evaluated. Root neighborhoods are propagated through
Eqs.~\eqref{eq:stable-I} and~\eqref{eq:stable-II}
with midpoint-radius arithmetic~\cite{Johansson:2017}.
The computed critical-value enclosures are denoted $\bar w\pm r$,
with midpoint $\bar w$ and absolute radius $r$, including decimal export error.
Appendix~\ref{app:numerics} specifies the branch
controls, local enclosure conditions, fitting windows
and series-evaluation prescriptions.
All fits use these critical-value midpoints in arbitrary-precision
decimal arithmetic without conversion to binary
machine precision.

\section{The continuum limit, shifted ranks and constant functions}
\label{sec:constants}

\subsection{Continuum densities and the leading critical values}
\label{subsec:continuum}

We consider $N\to\infty$ at fixed $k$ or $N_f$ and
assume the continuum scaling
$y_i=\sqrt N\,\xi_i$, with $\xi_i=O(1)$ uniformly in $i$.
Together with the local-sampling assumption stated below,
this scaling extracts the leading critical values from
the decaying-argument expressions in
Eqs.~\eqref{eq:stable-I} and~\eqref{eq:stable-II}.
The resulting profiles reproduce the superconformal
specialization of the solutions in
Ref.~\cite[Section~5.1]{Hosseini:2016ume}.

Assume that the empirical densities converge weakly,
\begin{equation}
 \rho_N(\xi)=\frac1N\sum_{i=1}^N\delta(\xi-\xi_i)
 \rightharpoonup\rho(\xi)\,,
 \qquad
 \rho(\xi)\geq0\,,
 \qquad
 \int_{\mathbb R}\rho(\xi)\,\rd\xi=1\,,
 \label{eq:continuum-density-definition}
\end{equation}
where $\rho$ is bounded, piecewise smooth and compactly
supported. In Model I, the symmetry
$\widetilde z_i=-\overline z_i$ identifies the imaginary
coordinates at the two nodes, so the common density
is normalized to one at each node. After removing
the parity centre, we also assume
$p_i=p(\xi_i)+o(1)$ locally uniformly in the interior
of the support, with a smooth bulk profile $p$.

The kernel $D_s(d)$ defined in Eq.~\eqref{eq:D} obeys
the uniform bound
$|D_s(d)|\leq\zeta(2)\e^{-d}$ for real $s$ and $d\geq0$.
Its integral is
\begin{equation}
 \mathcal S_3(s)
 =\int_0^\infty D_s(d)\,\rd d
 =\sum_{m\geq1}\frac{\sin(ms)}{m^3}
 =\frac{s(s-\pi)(s-2\pi)}{12}\,,
 \qquad 0\leq s\leq2\pi\,.
 \label{eq:continuum-kernel-moment}
\end{equation}
Termwise integration is justified by
$\sum_{m\geq1}m^{-2}\int_0^\infty\e^{-md}\,\rd d
=\zeta(3)$.
The polynomial follows by integrating the uniformly
convergent Fourier series
$\sum_{m\geq1}\cos(ms)/m^2
=\pi^2/6-\pi s/2+s^2/4$ from zero to $s$.
In particular, $\mathcal S_3(\theta)=2\pi^3/81$.

The exponential decay localizes the interaction on
separations $d_{ij}=|y_i-y_j|=O(1)$, corresponding to
$|\xi_i-\xi_j|=O(N^{-1/2})$.
We assume that the roots sample $\rho$ locally on this
shrinking scale; weak convergence alone does not imply
this property.
For effective pair angles with a smooth local limit
$s_{ij}=s(\xi_i)+o(1)$, locally uniformly in the bulk
for bounded $d_{ij}$, the local-density replacement gives
\begin{equation}
 \begin{aligned}
 \sum_{i<j}D_{s_{ij}}(|y_i-y_j|)
 &\simeq N^2\int_{\mathbb R}\rd\xi\,\rho(\xi)
       \int_\xi^\infty\rd\eta\,\rho(\eta)
       D_{s(\xi)}\!\left(\sqrt N(\eta-\xi)\right)\\
 &\simeq N^{3/2}
       \int_{\mathbb R}\rho(\xi)^2
       \mathcal S_3\!\left(s(\xi)\right)\,\rd\xi\,.
 \end{aligned}
 \label{eq:continuum-local-pairs}
\end{equation}
The change of variables $\eta=\xi+d/\sqrt N$ supplies
the factor $N^{-1/2}$.
At leading order, the nearby density and effective
angle are replaced by $\rho(\xi)$ and $s(\xi)$.
The separation integral is then
$\int_0^\infty D_{s(\xi)}(d)\,\rd d
=\mathcal S_3(s(\xi))$, with $\xi$ held fixed.
For ordered roots, the region $\eta>\xi$ counts each
pair once, so no additional factor of $1/2$ appears.
Here $\simeq$ denotes the leading local-density
approximation.

Single gauge-index sums with uniformly bounded summands
contribute at most $O(N)$ at fixed parameter.
By contrast, the Chern--Simons term proportional to
$\sum_i p_i y_i$ and the fundamental term proportional
to $\sum_i|y_i|$ contribute at order $N^{3/2}$ and must
be retained.

\paragraph{Model I.}
In the local limit, the adjoint angles
$\theta\pm(p_i-p_j)$ approach $\theta$, whereas the
bifundamental angles $\theta\pm(p_i+p_j)$ approach
$\theta\pm2p(\xi)$.
Including the multiplicities in Eq.~\eqref{eq:stable-I},
the combined coefficient of $\rho^2$ is
\begin{equation*}
 4\mathcal S_3(\theta)
 +4\bigl[
   \mathcal S_3(\theta+2p)
  +\mathcal S_3(\theta-2p)\bigr]
 =\frac{8\pi^3}{27}-\frac{8\pi}{3}p^2\,.
\end{equation*}
Together with the Chern--Simons contribution, this gives
\begin{equation}
 \frac{\im\Wt_{\mathrm I}}{N^{3/2}}
 \simeq\mathcal V_{\mathrm I}[\rho,p]
 =\int_{\mathbb R}
 \left[
 -2k\xi p(\xi)\rho(\xi)
 +\left(
   \frac{8\pi^3}{27}-\frac{8\pi}{3}p(\xi)^2
  \right)\rho(\xi)^2
 \right]\rd\xi\,.
 \label{eq:continuum-functional-I}
\end{equation}
We use the branch $|p|<\pi/3$ in the interior of the
support, for which $0<\theta\pm2p<2\pi$ and the
polynomial representation in
Eq.~\eqref{eq:continuum-kernel-moment} applies.

Introduce a Lagrange multiplier $\mu_{\mathrm I}$ for
$\int\rho\,\rd\xi=1$ and impose stationarity of
$\mathcal V_{\mathrm I}
-\mu_{\mathrm I}(\int\rho\,\rd\xi-1)$.
At points where $\rho>0$, variation with respect to
$p$ gives
$-2k\xi\rho-(16\pi/3)p\rho^2=0$.
Substitution into the density equation cancels the
$p$-dependent terms and yields
\begin{equation}
 p(\xi)=-\frac{3k\xi}{8\pi\rho(\xi)}\,,
 \qquad
 \rho(\xi)=\frac{27\mu_{\mathrm I}}{16\pi^3}\,.
 \label{eq:continuum-stationarity-I}
\end{equation}
The density is therefore constant on the interior
of the support, and $p$ is linear in $\xi$.
At fixed $\rho$, the functional is concave in $p$,
and the stationary profile is a saddle point.

The interior equations do not determine the support
without an endpoint condition.
At the superconformal chemical potentials, the inner
and outer endpoints of the general Model-I solution
coincide, so no tail interval of nonzero width
remains~\cite[Eqs.~(5.7)--(5.11)]{Hosseini:2016ume}.%
\footnote{The continuum coordinate $t$ and relative real
part $\delta v$ in that reference correspond to
$\xi$ and $-2p(\xi)$, respectively.}
For this branch, the reflection-symmetric support is
$[-L_{\mathrm I},L_{\mathrm I}]$, with
$p(\pm L_{\mathrm I})=\mp\pi/3$ understood as one-sided
limits. Combining this endpoint condition with
$2L_{\mathrm I}\rho_{\mathrm I}=1$ gives
\begin{equation}
 \begin{aligned}
 \rho_{\mathrm I}(\xi)
 &=\frac{3\sqrt{k}}{4\pi}\,,
 &\qquad
 p(\xi)&=-\frac{\sqrt{k}}2\,\xi\,,
 \qquad |\xi|<L_{\mathrm I}\,,\\
 L_{\mathrm I}
 &=\frac{2\pi}{3\sqrt{k}}\,,
 &
 \mu_{\mathrm I}&=\frac{4\pi^2\sqrt{k}}9\,.
 \end{aligned}
 \label{eq:continuum-solution-I}
\end{equation}
The density vanishes outside this interval.
On this profile, the Chern--Simons and combined
interaction terms in
Eq.~\eqref{eq:continuum-functional-I} each contribute
$4\pi^2\sqrt{k}/27$. Hence
\begin{equation}
 \WI(N,k)
 \sim\frac23\mu_{\mathrm I}N^{3/2}
 =\frac{8\pi^2\sqrt{k}}{27}N^{3/2}\,,
 \label{eq:continuum-critical-I}
\end{equation}
in agreement with Eq.~(5.17) of
Ref.~\cite{Hosseini:2016ume}.
Saturation of $|p|=\pi/3$ at the continuum endpoints
does not relax the strict strip conditions imposed
on finite-rank configurations in
Section~\ref{sec:bethe}.

\paragraph{Model II.}
In Eq.~\eqref{eq:stable-II}, the adjoint pair angle is
constant, $s_{ij}=\theta$.
Equation~\eqref{eq:continuum-local-pairs} therefore gives
the coefficient
$6\mathcal S_3(\theta)=4\pi^3/27$ for the adjoint
interaction. Including the fundamental linear term
gives
\begin{equation}
 \frac{\im\Wt_{\mathrm{II}}}{N^{3/2}}
 \simeq\mathcal V_{\mathrm{II}}[\rho]
 =\frac{4\pi^3}{27}
       \int_{\mathbb R}\rho(\xi)^2\,\rd\xi
  +\frac{2\pi N_f}{3}
       \int_{\mathbb R}|\xi|\rho(\xi)\,\rd\xi\,.
 \label{eq:continuum-functional-II}
\end{equation}
We minimize this strictly convex functional over
nonnegative unit-mass densities for which it is finite.
Introducing a Lagrange multiplier $\mu_{\mathrm{II}}$,
stationarity where $\rho>0$ gives
\begin{equation}
 \frac{8\pi^3}{27}\rho(\xi)
 +\frac{2\pi N_f}{3}|\xi|=\mu_{\mathrm{II}}\,.
 \label{eq:continuum-stationarity-II}
\end{equation}
Where $\rho=0$, the nonnegativity constraint instead
requires $(2\pi N_f/3)|\xi|\geq\mu_{\mathrm{II}}$.
These conditions fix the support to
$[-L_{\mathrm{II}},L_{\mathrm{II}}]$, with
$L_{\mathrm{II}}=3\mu_{\mathrm{II}}/(2\pi N_f)$.
Normalization determines the triangular profile
and its multiplier,
\begin{equation}
 \begin{aligned}
 \rho_{\mathrm{II}}(\xi)
 &=\frac{3\sqrt{N_f}}{2\pi}
   -\frac{9N_f}{4\pi^2}|\xi|\,,
 &\qquad |\xi|&\leq L_{\mathrm{II}}\,,\\
 L_{\mathrm{II}}
 &=\frac{2\pi}{3\sqrt{N_f}}\,,
 &
 \mu_{\mathrm{II}}&=\frac{4\pi^2\sqrt{N_f}}9\,.
 \end{aligned}
 \label{eq:continuum-solution-II}
\end{equation}
The density vanishes outside this interval.
Strict convexity makes this profile the unique
minimizer of the continuum functional.
It agrees with the superconformal specialization
of Eqs.~(5.25)--(5.27) of
Ref.~\cite{Hosseini:2016ume}, where $k$ denotes the
number of fundamental pairs.

The moments
$\int\rho_{\mathrm{II}}^2\,\rd\xi=2/(3L_{\mathrm{II}})$
and
$\int|\xi|\rho_{\mathrm{II}}\,\rd\xi=L_{\mathrm{II}}/3$
give equal adjoint and fundamental contributions,
each $4\pi^2\sqrt{N_f}/27$, to the continuum critical
value. Hence
\begin{equation}
 \WII(N,N_f)
 \sim\frac23\mu_{\mathrm{II}}N^{3/2}
 =\frac{8\pi^2\sqrt{N_f}}{27}N^{3/2}\,.
 \label{eq:continuum-critical-II}
\end{equation}

At equal quotient order, the two continuum solutions
have the same support length and leading critical
value, with a constant density in Model I and a
triangular density in Model II.
The shifted ranks, constants and exponentially small sectors
require information beyond this leading approximation.

\subsection{The shifted-rank perturbative proposal}
\label{subsec:parents}

At fixed $k$ or $N_f$, we propose that all algebraic
corrections to the leading critical values are contained
in a shifted power and a rank-independent constant,
\begin{equation}
 \begin{aligned}
 \WI^{\mathrm{pert}}(N,k)
 &=\frac{8\pi^2\sqrt{k}}{27}\Nh_{\mathrm I}^{3/2}
   +\CI(k)\,,
 &\qquad
 \Nh_{\mathrm I}&=N+s_{\mathrm I}(k)\,,\\
 \WII^{\mathrm{pert}}(N,N_f)
 &=\frac{8\pi^2\sqrt{N_f}}{27}\Nh_{\mathrm{II}}^{3/2}
   +\CII(N_f)\,,
 &\qquad
 \Nh_{\mathrm{II}}&=N+s_{\mathrm{II}}(N_f)\,,
 \end{aligned}
 \label{eq:parents}
\end{equation}
where
\begin{equation}
 s_{\mathrm I}(k)=-\frac{k}{12}+\frac1{2k}\,,
 \qquad
 s_{\mathrm{II}}(N_f)=\frac{N_f}{6}+\frac1{4N_f}\,.
 \label{eq:rank-shifts}
\end{equation}
We use these functions on the domain $\Nh>0$,
taking the positive square root. This condition holds
throughout the physical Model-II parameter range.
For Model I it requires $N>k/12-1/(2k)$ and excludes
some small-rank, large-$k$ points from the comparisons.
It restricts the proposed perturbative expression,
not the existence of a Bethe solution.

The Model-II shift is the superconformal specialization
of the proposal in
Refs.~\cite{Bobev:2023lkx,Bobev:2024mqw}.
With the normalization of Section~\ref{sec:bethe},
Eq.~(3.16) of Ref.~\cite{Bobev:2024mqw} gives
$\CII(N_f)=2\pi\widehat g_0$ at these chemical potentials.
The Model-I shift is inferred independently from its
critical-value data.

For Model I, adjacent-rank differences eliminate any
rank-independent contribution. With the leading
coefficient fixed by the continuum calculation,
define $s_{\mathrm{eff}}(N,k)$ implicitly by
\begin{equation}
 \begin{aligned}
 \WI(N+1,k)-\WI(N,k)
 &=a(k)\left[
 (N+1+s_{\mathrm{eff}})^{3/2}
 -(N+s_{\mathrm{eff}})^{3/2}\right]\,,\\
 a(k)&=\frac{8\pi^2\sqrt{k}}{27}\,.
 \end{aligned}
 \label{eq:shift-fit}
\end{equation}
As a function of a trial shift $s>-N$, the right-hand
side has derivative
$3a(k)/[2(\sqrt{N+1+s}+\sqrt{N+s})]>0$
and ranges over $(a(k),\infty)$.
Thus a unique real effective shift exists precisely
when the adjacent-rank difference exceeds $a(k)$.

The large-rank estimates support the Model-I shift
in Eq.~\eqref{eq:rank-shifts}.
Restricting its level dependence to
$s=\alpha k+\beta/k$, the estimates from the rank pair
$(299,300)$ at $k=1,2$ give the rational candidates
$\alpha=-1/12$ and $\beta=1/2$.
The remaining levels test this parameter dependence
without readjusting the coefficients.
Fits with the leading coefficient, shift and constant
all free use different three-rank subsets to assess
their sensitivity to the choice of ranks.
Appendix~\ref{app:numerics} gives the reconstruction
rule, rank selections and precision checks.
Adjacent differencing removes the constant but retains
the difference of every rank-dependent remainder.
The effective shifts therefore need not equal the
asymptotic shift at finite rank, even when the
critical values are known to high precision.

At fixed parameter, the binomial expansion of each
shifted power fixes the coefficients of
$N^{3/2-j}$ for every integer $j\geq0$.
The functions $\CI$ and $\CII$ specified below supply
the separate $N^0$ terms.
Their identification with the physical constants remains conjectural.

\subsection{Integral representations of the constants}
\label{subsec:constant-integrals}

We define the proposed constant functions by
\begin{align}
 \CI(k)={}&\frac{k^2\zeta(3)}{18\pi}
 +\frac{k}{\pi}\int_0^\infty\log(1-\e^{-kx})
 \left[
 2\log(1+2\cosh x)
 -\frac43\log\left(2\cosh\frac x2\right)
 \right]\,\rd x\,,
 \label{eq:constant-I}\\
 \CII(N_f)={}&\frac{2N_f^2}{9}
 \left[
 \frac{\zeta(3)}{\pi}-\Cltwo\left(\frac\pi3\right)
 \right]
 +\frac{N_f}{\pi}\int_0^\infty
 \log(1-\e^{-N_fx})\log(1+2\cosh x)\,\rd x\,.
 \label{eq:constant-II}
\end{align}
Let $\mathcal K_{\mathrm I}(x)$ and
$\mathcal K_{\mathrm{II}}(x)$ denote the respective
kernels multiplying $\log(1-\e^{-kx})$ and
$\log(1-\e^{-N_fx})$.
Both kernels are even and analytic at the origin,
and their large-$x$ behavior is
$\mathcal K_{\mathrm I}(x)=4x/3+O(\e^{-x})$ and
$\mathcal K_{\mathrm{II}}(x)=x+O(\e^{-x})$.
The integrands are therefore $O(|\log x|)$ as
$x\to0^+$ and decay as
$O(x\e^{-kx})$ or $O(x\e^{-N_fx})$ at infinity.
The integrals converge absolutely for every real
$k,N_f>0$, defining real continuations beyond the
physical integer parameters.

The retrospective Model-I reconstruction within this kernel family
varies the coefficient
of $k^2\zeta(3)/\pi$ and the two kernel coefficients
in Eq.~\eqref{eq:constant-I}, keeping the common
integral prefactor $k/\pi$ fixed.
Subtracting the shifted power from the critical
values at $N=300$ and $k=1,2,3$ gives the constant
estimates used to infer the rational candidates
$1/18$, $2$ and $-4/3$.
The remaining levels test this expression without
refitting. For Model II, the inherited shift gives
constant estimates at successive ranks for comparison
with Eq.~\eqref{eq:constant-II}.
Appendix~\ref{app:numerics} specifies these tests.
The finite-rank estimates retain an exponentially
small bias, and the kernel family is an assumption
of the reconstruction.

The common kernel $\log(1+2\cosh x)$ relates these
functions to the ABJM Bethe constant
\begin{equation}
 C_{\mathrm B}(K)
 =-\frac{K^2\zeta(3)}{16\pi}
 +\frac{2K}{\pi}\int_0^\infty
       \log(2\cosh x)\log(1-\e^{-Kx})\,\rd x\,,
 \label{eq:ABJM-constant}
\end{equation}
in the normalization of
Refs.~\cite{Hosseini:2026dkj,Hosseini:2026lvj}.
At $K=2k$, the change of variables $x=u/2$ gives
\begin{equation*}
 C_{\mathrm B}(2k)
 =-\frac{k^2\zeta(3)}{4\pi}
 +\frac{2k}{\pi}\int_0^\infty
       \log\left(2\cosh\frac u2\right)
       \log(1-\e^{-ku})\,\rd u\,.
\end{equation*}
The common kernels cancel in $\CI(k)-2\CII(k)$,
and the remaining integral is $-2/3$ times the
integral contribution to $C_{\mathrm B}(2k)$.
Combining the explicit quadratic terms yields
\begin{equation}
 \CI(k)=2\CII(k)-\frac23C_{\mathrm B}(2k)
 +\frac{k^2}{9}
 \left[
 4\Cltwo\left(\frac\pi3\right)
 -\frac{5\zeta(3)}{\pi}
 \right]\,.
 \label{eq:constant-relation}
\end{equation}
Absolute convergence justifies the substitution,
so this identity holds for the defined functions
at every real $k>0$, independently of their
identification with physical Bethe constants.

After separating the explicit quadratic terms,
the large-parameter expansions are determined by
the even Taylor coefficients of the kernels at
the origin. Section~\ref{sec:genus} justifies the
finite-order expansion under the integral and
derives its positive-direction Borel sum.
At integer parameter, the finite Clausen formulas
obtained next evaluate the constants without
numerical quadrature. Sections~\ref{subsec:bulk-constant}
and~\ref{subsec:bulk-constant-II} also derive
$\CI(k)/(2\pi)$ and $\CII(N_f)/(2\pi)$ from
the small-squashing limits of the corresponding
bulk one-loop characters.
The Model-II calculation includes the fixed-locus
contribution specified in
Section~\ref{subsec:localized-index}.

\subsection{Finite Clausen evaluations}
\label{subsec:finite}

For $a>0$, define $L_a(x)=\log(1-\e^{-ax})$.
At $x>0$, the kernel factorizations are
\begin{equation}
 \begin{aligned}
 \log(1+2\cosh x)
 =x+L_3(x)-L_1(x)\,,\qquad
 \log\left(2\cosh\frac x2\right)
 =\frac x2+L_2(x)-L_1(x)\,.
 \end{aligned}
 \label{eq:kernel-factorization}
\end{equation}
Expanding $L_a$ into exponentials and integrating
termwise gives
$\int_0^\infty xL_a(x)\,\rd x=-\zeta(3)/a^2$.
The remaining terms involve the symmetric integral
\begin{equation}
 D(a,b)
 =\int_0^\infty L_a(x)L_b(x)\,\rd x\,,
 \label{eq:D-integral}
\end{equation}
which converges absolutely for $a,b>0$.
Substitution into the constant functions yields
\begin{align}
 \CI(k)={}&
 \frac{\zeta(3)}{\pi}
 \left(\frac{k^2}{18}-\frac{4}{3k}\right)
 +\frac{k}{\pi}
 \left[
 2D(k,3)-\frac43D(k,2)
                   -\frac23D(k,1)
 \right]\,,
 \label{eq:constant-I-log-integrals}\\
 \CII(N_f)={}&
 \frac{2N_f^2}{9}
 \left[\frac{\zeta(3)}{\pi}-\Cltwo\left(\frac\pi3\right)\right]
 -\frac{\zeta(3)}{\pi N_f}
 +\frac{N_f}{\pi}
 \left[D(N_f,3)-D(N_f,1)\right]\,.
 \label{eq:constant-II-log-integrals}
\end{align}
These identities hold for all real $k,N_f>0$.

At positive integer arguments, $D(a,b)$ admits
a finite evaluation. With $g=\gcd(a,b)$,
\begin{equation}
 \begin{aligned}
 D(a,b)={}
 \frac{\zeta(3)}2
 \left(\frac a{b^2}+\frac b{a^2}
                   +\frac{2g^3}{a^2b^2}\right)
 +\frac{\pi}{ab}
 \left[
 \sum_{j=1}^{a-1}j\Cltwo\left(\frac{2\pi bj}{a}\right)
 +\sum_{j=1}^{b-1}j\Cltwo\left(\frac{2\pi aj}{b}\right)
 \right]\,.
 \end{aligned}
 \label{eq:D-finite}
\end{equation}
Appendix~\ref{app:clausen} proves this identity by
boundary integration and root-of-unity filtering.
The endpoint contributions and logarithmic jumps
are retained before applying the root-of-unity
filter that produces the $g^3$ term.

For integer $n$, let $\chi_{-3}$ denote the primitive real odd Dirichlet character
modulo three, defined by $\chi_{-3}(n)=0,1,-1$ for $n\equiv0,1,2\pmod3$.
Periodicity and oddness of the Clausen function,
together with
$\Cltwo(2\pi/3)=\frac23\Cltwo(\pi/3)$, give
\begin{equation}
 \sum_{j=1}^{2}j\Cltwo\left(\frac{2\pi nj}{3}\right)
 =-\frac23\chi_{-3}(n)\Cltwo\left(\frac\pi3\right)\,,
 \qquad
 \Cltwo(\pi n)=0\,.
 \label{eq:short-Clausen-sums}
\end{equation}
These identities evaluate the short sums associated
with the fixed arguments $2$ and $3$.
Inserting Eq.~\eqref{eq:D-finite} into
Eqs.~\eqref{eq:constant-I-log-integrals}
and~\eqref{eq:constant-II-log-integrals} therefore
gives, at integer $k,N_f$,
\begin{align}
 \CI(k)={}&
 -\frac49\chi_{-3}(k)\Cltwo\left(\frac\pi3\right)
 +\left[
 -\frac{k^2}{3}
 +\frac{2\gcd(k,3)^3-3\gcd(k,2)^3-6}{9k}
 \right]\frac{\zeta(3)}{\pi}
 \notag\\
 &+\frac23\sum_{j=1}^{k-1}j
 \left[
 \Cltwo\left(\frac{6\pi j}{k}\right)
 -\Cltwo\left(\frac{4\pi j}{k}\right)
 -\Cltwo\left(\frac{2\pi j}{k}\right)
 \right]\,,
 \label{eq:CI-finite}\\
 \CII(N_f)={}&
 -\frac29\bigl[N_f^2+\chi_{-3}(N_f)\bigr]
          \Cltwo\left(\frac\pi3\right)
 +\left[
 -\frac{2N_f^2}{9}
 +\frac{\gcd(N_f,3)^3-9}{9N_f}
 \right]\frac{\zeta(3)}{\pi}
 \notag\\
 &+\sum_{j=1}^{N_f-1}j
 \left[
 \frac13\Cltwo\left(\frac{6\pi j}{N_f}\right)
 -\Cltwo\left(\frac{2\pi j}{N_f}\right)
 \right]\,.
 \label{eq:CII-finite}
\end{align}
Empty sums are understood to vanish.
Table~\ref{tab:constants} lists the values at parameters
one, two and three, together with the reductions at
four and six. The same Clausen identities reduce these
entries to $\Cltwo(\pi/3)$, Catalan's constant
$G=\Cltwo(\pi/2)$ and $\zeta(3)/\pi$.

\begin{table}[tbp]
 \centering\small
 \renewcommand{\arraystretch}{1.5}
 \begin{tabular}{@{}rll@{}}
  \toprule
  $\mathfrak k$ & $\CI(\mathfrak k)$ & $\CII(\mathfrak k)$ \\
  \midrule
  1
  & $-\dfrac49\Cltwo(\pi/3)-\dfrac{10\zeta(3)}{9\pi}$
  & $-\dfrac49\Cltwo(\pi/3)-\dfrac{10\zeta(3)}{9\pi}$
  \\[.35em]
  2
  & $\dfrac49\Cltwo(\pi/3)-\dfrac{26\zeta(3)}{9\pi}$
  & $-\dfrac23\Cltwo(\pi/3)-\dfrac{4\zeta(3)}{3\pi}$
  \\[.35em]
  3
  & $-\dfrac{4\zeta(3)}{3\pi}$
  & $-\dfrac43\Cltwo(\pi/3)-\dfrac{4\zeta(3)}{3\pi}$
  \\[.35em]
  4
  & $-\dfrac49\Cltwo(\pi/3)+\dfrac83G
      -\dfrac{55\zeta(3)}{9\pi}$
  & $-\dfrac{34}{9}\Cltwo(\pi/3)+\dfrac83G
      -\dfrac{34\zeta(3)}{9\pi}$
  \\[.35em]
  6
  & $\dfrac{40}{9}\Cltwo(\pi/3)
      -\dfrac{104\zeta(3)}{9\pi}$
  & $-\dfrac83\Cltwo(\pi/3)-\dfrac{23\zeta(3)}{3\pi}$
  \\
  \bottomrule
 \end{tabular}
 \caption{Selected values of the integral-defined
 constant functions, evaluated using
 Eqs.~\eqref{eq:CI-finite} and~\eqref{eq:CII-finite}.
 Here $G=\Cltwo(\pi/2)$ is Catalan's constant.}
 \label{tab:constants}
\end{table}

The quadrature comparisons described in
Appendix~\ref{app:numerics} check the numerical
implementations of these identities.
The finite evaluations supply the constants used
to isolate the exponential remainder from the
critical-value data in Section~\ref{sec:sectors}.

\section{Exponentially suppressed sectors}\label{sec:sectors}

\subsection{Finite-rank reconstruction}
\label{subsec:reconstruction}

For either model, define the remainder after subtracting
the proposed perturbative expression,
\begin{equation}
 \begin{aligned}
 \W^{\mathrm{np}}(N,\mathfrak k)
 &=\W(N,\mathfrak k)-\W^{\mathrm{pert}}(N,\mathfrak k)\,,\\
 t_{\mathrm I}
 &=2\pi\sqrt{\frac{\Nh_{\mathrm I}}{k}}\,,
 \qquad
 t_{\mathrm{II}}
 =2\pi\sqrt{\frac{\Nh_{\mathrm{II}}}{N_f}}\,,
 \qquad q=\e^{-t}\,.
 \end{aligned}
 \label{eq:actions}
\end{equation}
Here $t$ denotes the corresponding action on the
domain $\Nh>0$.
At $k=1,3$ and $N_f=1,3$, the observed exponential
scale motivates a fit in integer multiples of $t$,
\begin{equation}
 \frac{2\pi}{\mathfrak k^2}\W^{\mathrm{np}}(N,\mathfrak k)
 \simeq\sum_{m=1}^{M}(A_m t+B_m)q^m\,.
 \label{eq:free-fit}
\end{equation}
The shifted powers and the constants in
Table~\ref{tab:constants} are held fixed.
All $2M$ coefficients are free, with no divisor law
or relation between $A_m$ and $B_m$ imposed in the
least-squares problem. The QR solve uses the data
and model divided by $q$, with the column normalization
specified in Eq.~\eqref{eq:QR-columns}.

For Model I, the final fits use every rank
in the disjoint windows $10\leq N\leq54$ and
$56\leq N\leq100$, with $M=16$ in each.
For Model II, one fit uses the even ranks in
$20\leq N\leq100$ with $M=12$, and the other uses
the even ranks in $10\leq N\leq100$ with $M=16$.
Odd ranks are withheld from both Model-II fits.
The overlapping Model-II windows test sensitivity
to the rank window and sector cutoff; they are
not disjoint training sets. All four fitting
configurations use $850$ working decimal digits.

Continued fractions assign rational candidates
with denominator at most $10^5$ separately to
the fitted $A_m$ and $B_m$. The prefactor relation
\begin{equation}
 B_m=\frac{A_m}{m}
 \label{eq:prefactor-relation}
\end{equation}
is then tested as an identity between the candidate
fractions.
In Model I, the candidates agree across the disjoint
windows and satisfy Eq.~\eqref{eq:prefactor-relation}
through the twelve reported sectors at $k=1,3$.
For Model II, acceptance requires agreement of
both rational candidates across the windows,
Eq.~\eqref{eq:prefactor-relation}, and the numerical
deviation threshold specified in
Appendix~\ref{app:numerics}. Eleven sectors pass
these criteria at $N_f=1,3$.
Thus the Model-II identity is a post-fit selection
condition, although the underlying coefficients
are fitted independently.

Let $c_m$ denote the retained rational candidate
for $A_m$. Table~\ref{tab:coefficients} lists these
values; the corresponding candidate for $B_m$ is
$c_m/m$. The unit-parameter candidates agree between
the models through their common reconstructed range.
The fitted decimals depend on the input uncertainty
and on the sectors omitted from the finite ansatz.
Agreement across windows and bounded-denominator
approximation do not provide an interval enclosure
of the exact physical coefficients.

\begin{table}[t]
 \centering\small
 \begin{tabular}{rcccc}
\toprule
& \multicolumn{2}{c}{Model I} & \multicolumn{2}{c}{Model II} \\
\cmidrule(lr){2-3}\cmidrule(lr){4-5}
$m$ & $k=1$ & $k=3$ & $N_f=1$ & $N_f=3$ \\
\midrule
1 & $\dfrac{20}{3}$ & $\dfrac{2}{3}$ & $\dfrac{20}{3}$ & $-\dfrac{4}{3}$ \\
\addlinespace[.55em]
2 & $-15$ & $-\dfrac{3}{2}$ & $-15$ & $3$ \\
\addlinespace[.55em]
3 & $\dfrac{380}{27}$ & $\dfrac{56}{27}$ & $\dfrac{380}{27}$ & $\dfrac{56}{27}$ \\
\addlinespace[.55em]
4 & $-\dfrac{365}{12}$ & $-\dfrac{73}{24}$ & $-\dfrac{365}{12}$ & $\dfrac{73}{12}$ \\
\addlinespace[.55em]
5 & $\dfrac{168}{5}$ & $\dfrac{84}{25}$ & $\dfrac{168}{5}$ & $-\dfrac{168}{25}$ \\
\addlinespace[.55em]
6 & $-\dfrac{95}{3}$ & $-\dfrac{14}{3}$ & $-\dfrac{95}{3}$ & $-\dfrac{14}{3}$ \\
\addlinespace[.55em]
7 & $\dfrac{6880}{147}$ & $\dfrac{688}{147}$ & $\dfrac{6880}{147}$ & $-\dfrac{1376}{147}$ \\
\addlinespace[.55em]
8 & $-\dfrac{975}{16}$ & $-\dfrac{195}{32}$ & $-\dfrac{975}{16}$ & $\dfrac{195}{16}$ \\
\addlinespace[.55em]
9 & $\dfrac{10100}{243}$ & $\dfrac{1352}{243}$ & $\dfrac{10100}{243}$ & $\dfrac{1352}{243}$ \\
\addlinespace[.55em]
10 & $-\dfrac{378}{5}$ & $-\dfrac{189}{25}$ & $-\dfrac{378}{5}$ & $\dfrac{378}{25}$ \\
\addlinespace[.55em]
11 & $\dfrac{8880}{121}$ & $\dfrac{888}{121}$ & $\dfrac{8880}{121}$ & $-\dfrac{1776}{121}$ \\
\addlinespace[.55em]
12 & $-\dfrac{6935}{108}$ & $-\dfrac{511}{54}$ & \textemdash & \textemdash \\
\bottomrule
\end{tabular}

 \caption{Rational candidates $c_m$ for the coefficient
 $A_m$ in Eq.~\eqref{eq:free-fit}.
 The separately reconstructed candidate for $B_m$
 is $c_m/m$.
 Model I is reported through twelve sectors;
 eleven sectors pass the Model-II acceptance rule.
 The dash at $m=12$ in Model II denotes failure
 to meet that rule, not a vanishing coefficient.
 The proposed divisor laws reproduce every displayed
 fraction exactly.}
 \label{tab:coefficients}
\end{table}

\begin{figure}[t]
 \centering
 \includegraphics[width=.85\textwidth]{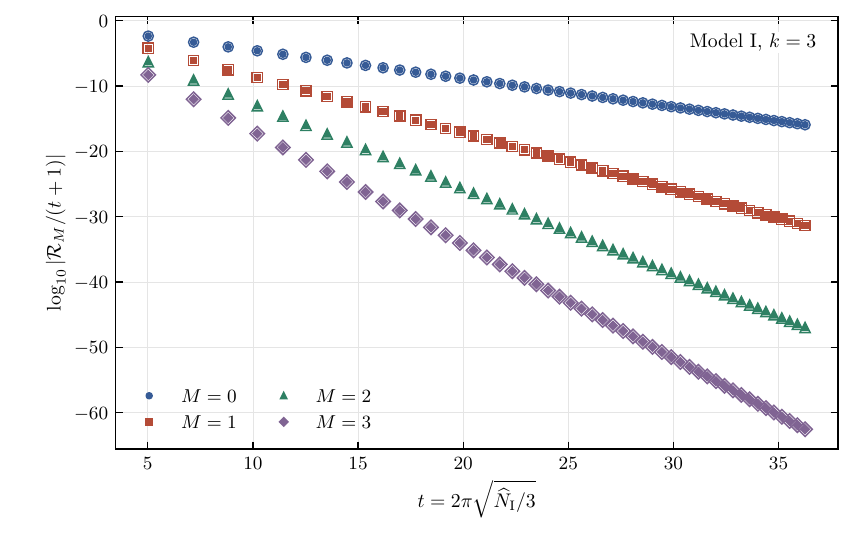}
 \caption{Successive exponential subtractions in
 Model I at $k=3$, evaluated at even ranks
 $2\leq N\leq100$.
 Filled markers show $\log_{10}|\mathcal R_M/(t+1)|$,
 where
 $\mathcal R_M=(2\pi/9)(\WI-\WI^{\mathrm{pert}})
 -\sum_{m=1}^{M}c_{m,\mathrm I}^{(3)}
 (t+1/m)\e^{-mt}$.
 Open markers show
 $\log_{10}|c_{M+1,\mathrm I}^{(3)}
 (t+1/(M+1))\e^{-(M+1)t}/(t+1)|$
 at the same ranks, for $M=0,1,2,3$.
 The empty subtraction at $M=0$ leaves the normalized
 remainder $(2\pi/9)(\WI-\WI^{\mathrm{pert}})$.
 The plotted discrepancies are evaluated from
 critical-value midpoints.}
 \label{fig:hierarchy}
\end{figure}

Figure~\ref{fig:hierarchy} compares successive
subtractions in Model I at $k=3$ with the first
omitted terms. The separation of the remainders
into successive exponential scales tests the
first few reconstructed sectors.

To determine the divisor structure, apply exact
arithmetic to the retained candidates and set
$a_m=(-1)^{m+1}m^2c_m$.
The convolution with the cubic divisor sum can be
inverted recursively,
\begin{equation}
 \begin{aligned}
 a_m=\sum_{d\mid m}b_d\sigma_3(m/d)\,,\qquad
 b_m=a_m-
 \sum_{\substack{d\mid m\\d<m}}b_d\sigma_3(m/d)\,.
 \end{aligned}
 \label{eq:deconvolution}
\end{equation}
Here $\sigma_3(m)=\sum_{d\mid m}d^3$, with
$\sigma_3(x)=0$ unless $x$ is a positive integer.
Since $\sigma_3(1)=1$, the recursion determines
$b_m$ uniquely through the reconstructed order
for any initial sequence $a_m$.
For the candidates in Table~\ref{tab:coefficients},
all weights in this range vanish except those
listed below:
\begin{equation}
 \begin{array}{c|rrr}
  &b_1&b_3&b_9\\
 \hline
 k=N_f=1
  &20/3&-60&0\\
 \text{Model I},\ k=3
  &2/3&0&-54\\
 \text{Model II},\ N_f=3
  &-4/3&56&-108
 \end{array}\,.
 \label{eq:dilation-weights}
\end{equation}
These weights reproduce the twelve reported
Model-I coefficients and the eleven accepted
Model-II coefficients.
Taking them to be the complete support of the
convolution, and extending
Eq.~\eqref{eq:prefactor-relation} to every sector,
gives the following all-order conjecture.

\begin{conjecture}[Special-parameter completion]
\label{conj:completion}
For Model I at $k=1,3$ and Model II at $N_f=1,3$,
the critical values on the branches of
Section~\ref{sec:bethe}, at the specified superconformal
chemical potentials, satisfy
\begin{equation}
 \W(N,\mathfrak k)=\W^{\mathrm{pert}}(N,\mathfrak k)
       +\frac{\mathfrak k^2}{2\pi}\sum_{m\geq1}c_m
           \left(t+\frac1m\right)q^m\,,
 \label{eq:completion}
\end{equation}
for every integer $N\geq1$.
The action $t$ and nome $q$ are defined in Eq.~\eqref{eq:actions}.
The perturbative term is given by Eq.~\eqref{eq:parents},
with the constants in Eqs.~\eqref{eq:constant-I}
and~\eqref{eq:constant-II}.
The common unit-parameter sequence and the two
parameter-three sequences are
\begin{align}
 c_m^{(1)}
 &=\frac{20(-1)^{m+1}}{3m^2}
   \bigl[\sigma_3(m)-9\sigma_3(m/3)\bigr]\,,
 \label{eq:divisor-unit}\\
 c_{m,\mathrm I}^{(3)}
 &=\frac{2(-1)^{m+1}}{3m^2}
   \bigl[\sigma_3(m)-81\sigma_3(m/9)\bigr]\,,
 \label{eq:divisor-I3}\\
 c_{m,\mathrm{II}}^{(3)}
 &=-\frac{4(-1)^{m+1}}{3m^2}
   \bigl[\sigma_3(m)-42\sigma_3(m/3)
                    +81\sigma_3(m/9)\bigr]\,.
 \label{eq:divisor-II3}
\end{align}
\end{conjecture}

At these parameters, the shifted ranks are positive
for every $N\geq1$, so $0<q<1$.
Section~\ref{sec:modular} proves absolute convergence
of the explicit series and the tail bound
in Eq.~\eqref{eq:tail}, independently of their
identification with the Bethe critical values.

With the constants and divisor laws fixed, the
complete expressions have been compared with the
computed critical values at every
rank $2\leq N\leq300$ in all four model--parameter
cases. For each comparison, an interval enclosing
the finite sum, including the perturbative term
and arithmetic errors, is enlarged by the analytic
tail bound. The resulting full-series interval
lies strictly inside the recorded critical-value
interval, as required by
Eq.~\eqref{eq:scalar-compatibility}.
The comparisons include ranks outside the fitting
windows and the odd ranks withheld from the
Model-II fits.

\begin{figure}[t]
 \centering
 \includegraphics[width=\textwidth]{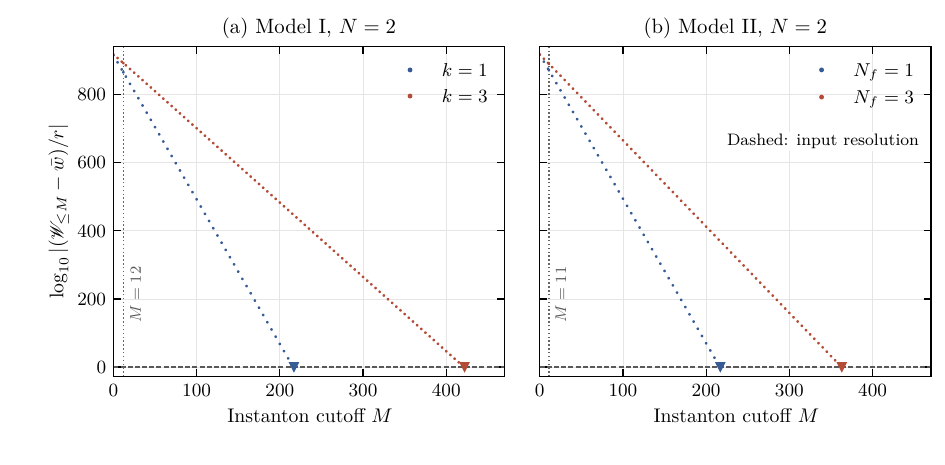}
 \caption{Partial sums of Eq.~\eqref{eq:completion}
 at $N=2$, for quotient orders one and three.
 Model I is shown on the left and Model II on the right.
 The ordinate is
 $\log_{10}|(\W_{\leq M}-\bar w)/r|$, where
 $\W_{\leq M}=\W^{\mathrm{pert}}
 +\W^{\mathrm{np}}_{\leq M}$ and $\bar w\pm r$ is the
 computed critical-value enclosure.
 Every integer cutoff is evaluated. Filled markers
 sample every fifth cutoff with
 $|\W_{\leq M}-\bar w|>r$, together with the reported
 reconstruction boundary and the last cutoff above
 this threshold.
 Downward triangles are placed on the horizontal
 dashed line at the first cutoffs satisfying
 $|\W_{\leq M}-\bar w|\leq r$:
 $M=217,422$ in Model I and $M=217,363$ in Model II,
 for parameters one and three, respectively.
 Their vertical position marks the resolution
 threshold, not the computed ordinate.
 The vertical dotted lines mark the reported
 coefficient ranges, $M=12$ in Model I and $M=11$
 in Model II, not the fitted sector cutoffs.
 These point-arithmetic partial-sum diagnostics
 exclude the omitted tail and are distinct from
 the full-series interval comparisons.}
 \label{fig:convergence}
\end{figure}

At small rank, the partial sums require many more
terms to reach the recorded uncertainty than were
recovered from the finite-sector fits.
Figure~\ref{fig:convergence} illustrates this
difference at $N=2$. The additional coefficients
are supplied by the proposed divisor laws, so
agreement tests their combined contribution,
not the accuracy of each coefficient separately.
At $N=1$, the proposed sums also agree numerically
with the exact branch values in
Eq.~\eqref{eq:rank-one}. The exact evaluation of
these infinite sums remains part of the conjecture.

\subsection{Parameter two and the limits of the linear ansatz}
\label{subsec:two}

At $k=2$ and $N_f=2$, we retain the model-dependent
actions in Eq.~\eqref{eq:actions} but enlarge the
prefactor basis to Laurent polynomials in $t$.
With $q=\e^{-t}$, the finite-sector candidates are
\begin{align}
 \frac\pi2\WI^{\mathrm{np}}(N,2)
 &\simeq\frac43(t+2)q+\mathcal P_2(t)q^2+\cdots\,,
 \label{eq:I2-sectors}\\
 \mathcal P_2(t)&=-\frac{16t^3}{9\pi^2}-\frac{32t^2}{3\pi^2}
       -\left(\frac{43}{9}+\frac{64}{3\pi^2}\right)t
       +\frac{491}{54}-\frac{128}{9\pi^2}+\frac4t\,,
 \label{eq:I2-prefactor}\\
 \frac\pi2\WII^{\mathrm{np}}(N,2)
 &\simeq-\frac{2\pi}{3}q+
 \left(-\frac{23t}{3}-\frac{19}{6}+\frac{\pi^2}{t}\right)q^2\notag\\
 &\hspace{1.5em}+
 \left(\frac{866\pi}{27}-\frac{4\pi}{t}
                    -\frac{3\pi^3}{t^2}-\frac{\pi^3}{t^3}\right)q^3+\cdots\,.
 \label{eq:II2-sectors}
\end{align}
The factor $\pi/2$ is the normalization $2\pi/\mathfrak k^2$
at $\mathfrak k=2$. Here $\simeq$ denotes finite-sector numerical
reconstruction, described in
Appendix~\ref{app:numerics}, with no asymptotic
error bound assigned to the unshown terms.

At parameter two, the unresolved structure lies within
the exponential sectors themselves: the reconstructed
Laurent prefactors may terminate or may be the first
terms of longer expansions in $1/t$.
Any further inverse-action correction multiplying
$\e^{-mt}$ would eventually dominate the displayed
terms at order $\e^{-(m+1)t}$.
An all-order description must therefore determine
the complete prefactors together with their relations
across sectors. The proposed completions at parameters
one and three specify both through a single Eisenstein
primitive. Explaining this simplification requires
a microscopic account of the action dependence,
as well as the origin of the divisor laws.

\subsection{Rank and parameter dependence of the critical values}
\label{subsec:finite-comparison}
\label{sec:comparison}

\begin{figure}[t]
 \centering
 \includegraphics[width=.85\textwidth]{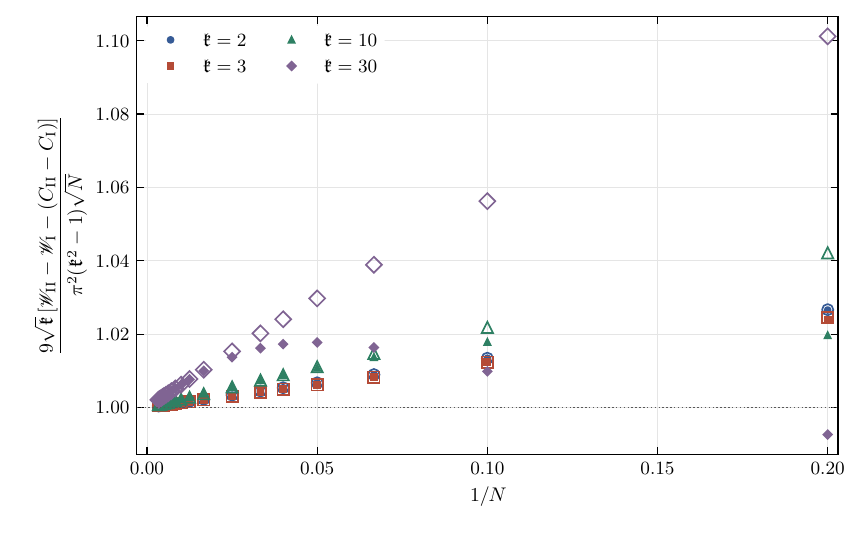}
 \caption{Separation of the critical values at equal
 quotient order $\mathfrak k=2,3,10,30$, plotted against
 $1/N$. Filled markers show
 $\frac{9\sqrt{\mathfrak k}}
 {\pi^2(\mathfrak k^2-1)\sqrt N}
 [\bar w_{\mathrm{II}}-\bar w_{\mathrm I}
  -\CII(\mathfrak k)+\CI(\mathfrak k)]$,
 where $\bar w_{\mathrm I}$ and $\bar w_{\mathrm{II}}$
 are the archived critical-value midpoints.
 Open markers use the difference of the perturbative
 functions with the same subtraction and normalization,
 retaining the shifted powers but no exponential terms.
 The horizontal line marks the large-rank limit, one.
 All displayed ranks have $\Nh_{\mathrm I}>0$.}
 \label{fig:quotients}
\end{figure}

At equal quotient order $k=N_f=\mathfrak k$, the
proposed shifts differ by
\begin{equation}
 s_{\mathrm{II}}(\mathfrak k)-s_{\mathrm I}(\mathfrak k)
 =\frac14 \left(\mathfrak k-\frac1{\mathfrak k}\right)\,.
 \label{eq:shift-difference}
\end{equation}
This shift difference predicts a quantitative separation
beyond the common leading coefficient. At fixed
$\mathfrak k$, the perturbative expressions give
\begin{equation}
 \begin{aligned}
 \WII^{\mathrm{pert}}(N,\mathfrak k)
 -\WI^{\mathrm{pert}}(N,\mathfrak k)
 =\frac{\pi^2(\mathfrak k^2-1)}{9\sqrt{\mathfrak k}}
      \sqrt N
 +\CII(\mathfrak k)-\CI(\mathfrak k)+O(N^{-1/2})\,.
 \end{aligned}
 \label{eq:large-rank-difference}
\end{equation}
Figure~\ref{fig:quotients} compares the Bethe data
with the leading $\sqrt N$ separation and the
unexpanded difference of the shifted powers.
The numerical intervals resolve the strict ordering
$\WI<\WII$ throughout the tested range
$2\leq N\leq300$, $2\leq\mathfrak k\leq30$.

At parameter three, the translation $N\mapsto N-2/3$
in Model II gives
$\Nh_{\mathrm{II}}(N-2/3,3)=\Nh_{\mathrm I}(N,3)=N-1/12$,
matching both the shifted powers and the instanton
actions. With the common action
$t=2\pi\sqrt{(N-1/12)/3}$, defined for $N>1/12$,
the reconstructed completions give
\begin{equation}
 \WI(N,3)-\WII(N-2/3,3)
 =\frac43\Cltwo\left(\frac\pi3\right)
  +\frac9\pi(t+1)\e^{-t}+O(t\e^{-2t})\,,
 \label{eq:aligned-difference}
\end{equation}
as $t\to\infty$. The remainder is bounded by
Eq.~\eqref{eq:tail} with $M=1$.
Subtracting the constant offset therefore leaves
a difference at the leading exponential order,
which distinguishes the two completions even
when their perturbative contributions are matched.

At fixed $N$, expanding the Bethe equations about
vanishing centred holonomies gives the large-parameter
limits
\begin{align}
 \WII(N,N_f)
 &=2N(N_f+N)\Cltwo\left(\frac\pi3\right)
 -\frac{\pi^2N(N^2-1)}{6\sqrt3(N_f+N)}
 +O(N_f^{-3})\,,
 \label{eq:fixed-rank-II}\\
 \WI(N,k)
 &=4N^2\Cltwo\left(\frac\pi3\right)
 -\frac{\pi^2N^2(N^2-1)}{\sqrt3(k^2+3N^2)}
 +O(k^{-4})\,.
 \label{eq:fixed-rank-I}
\end{align}
The flavored critical value grows linearly with
$N_f$ on the unique real branch of
Theorem~\ref{thm:branch}, while the Model-I value
approaches $4N^2\Cltwo(\pi/3)$ on the local
large-$k$ branch.
The derivation in Appendix~\ref{app:fixed-rank}
retains the denominators obtained from the quadratic
stationary equations, with the first omitted
quartic interactions contributing at the displayed
remainder orders.

At $N=1$ and $k=N_f=1$, both theories contain
the quadric cone as a geometric branch, but their
full chiral rings differ. In Model I, the additional
nonzero classes are $\Sigma=\phi_1+\phi_2$ and
$\Sigma^2$, with $\Sigma^3=0$.
These nilpotent operators have vanishing expectation
values in every vacuum, so they introduce no new
geometric direction. In Model II, the meson
$\mathcal M=\widetilde\psi\psi$ instead has
unrestricted powers and parametrizes an additional
complex line meeting the cone at its apex.

Adding $\mathcal M^3$ to the Model-II ordinary
superpotential leaves the geometric branch unchanged,
since the fundamental fields vanish there.
The modified $F$-term equations give
\begin{equation*}
 \mathcal M\Psi_I=0\quad(I=1,2,3)\,,
 \qquad
 \mathcal M^3=0\,.
\end{equation*}
The mesonic line is thus replaced by the nonzero
nilpotent classes $\mathcal M$ and $\mathcal M^2$,
matching the additional operator content of
Model I~\cite[Section~6.2.4]{Cremonesi:2016nbo}.
This deformation is absent from
Eq.~\eqref{eq:ordinary-II}, so the agreement
of Bethe critical values coexists with distinct
full chiral rings.

At $N=k=1$, the combinations
$\Sigma=\phi_1+\phi_2$ and
$z_0=\frac{\sqrt3}{2}(\phi_1-\phi_2)$
put the ordinary superpotential
in Eq.~\eqref{eq:ordinary-I} into the form
\begin{equation*}
 W_{\mathrm I}
 =\Sigma\bigl(z_0^2+A_1B_1+A_2B_2\bigr)
  +\frac14\Sigma^3\,.
\end{equation*}
The quadratic expression is the defining polynomial
of the cone in Eq.~\eqref{eq:intro-cone}.
In the Wess--Zumino description, its five variables
$(z_0,A_1,A_2,B_1,B_2)$ form an $\mathrm{SO}(5)$
vector in a complex basis, while $\Sigma$ is a
singlet. A superconformal fixed point retaining
both interactions would require all six fields
to have R-charge $2/3$.

The combined localization and bootstrap analysis
of Ref.~\cite{Chester:2015qca} excludes this
candidate and supports an infrared limit in
which the singlet cubic interaction becomes
irrelevant. Without that interaction, the
superpotential preserves an additional $\U(1)$
flavor symmetry assigning charge $1$ to the five
vector components and $-2$ to $\Sigma$.
Mixing with this symmetry allows their R-charges
to vary subject to
$R(\Sigma)+2R(z_0)=2$, with a common R-charge
for the vector components.
The infrared superconformal assignment must
then be determined by $F$-maximization including
this mixing. It need not coincide with the
fixed chemical potentials $\Delta=2\pi/3$
used to define the Bethe critical values.

\section{Eisenstein series and Eichler integrals}\label{sec:modular}

\subsection{Eichler primitives of the divisor series}

The prefactor $t+1/m$ in Eq.~\eqref{eq:completion}
allows the proposed exponential correction to be
generated from a single holomorphic function.
For each of the three divisor sequences, define
\begin{equation*}
 \Phi(q)=\frac{\mathfrak k^2}{2\pi}
          \sum_{m\geq1}\frac{c_m}{m}q^m\,,
 \qquad D=q\partial_q\,,
 \qquad |q|<1\,.
\end{equation*}
The divisor estimate $\sigma_3(m)\leq\zeta(3)m^3$
implies $c_m=O(m)$, so this series and its
logarithmic derivatives converge locally uniformly
in the unit disk. At fixed $t$,
\begin{equation}
 \left.(1+tD)\Phi(q)\right|_{q=\e^{-t}}
 =\frac{\mathfrak k^2}{2\pi}
  \sum_{m\geq1}c_m
       \left(t+\frac1m\right)\e^{-mt}\,.
 \label{eq:operator}
\end{equation}
Along the physical ray, the same operation is
$(1-t\partial_t)\Phi(\e^{-t})$.
Equation~\eqref{eq:operator} is an identity for
the explicit series. Its equality to the Bethe
remainder $\W^{\mathrm{np}}$ is the assertion
of Conjecture~\ref{conj:completion}.

Three logarithmic derivatives remove the inverse
powers of $m$ in $\Phi$ and recover the cubic
divisor sums. Their alternating signs are
absorbed by the half-period shift
\begin{equation}
 q=\e^{2\pi\ii\tau}\,,
 \qquad Q=-q=\e^{2\pi\ii\tau_2}\,,
 \qquad \tau_2=\tau+\frac12\,,
 \qquad \im\tau>0\,.
 \label{eq:coordinates}
\end{equation}
Thus $D=Q\partial_Q=(2\pi\ii)^{-1}\partial_{\tau_2}$,
and $\tau=\ii t/(2\pi)$ on the physical ray.
Using the normalization
$E_4(Q)=1+240\sum_{m\geq1}\sigma_3(m)Q^m$,
with the argument denoting a nome, introduce
the full Eisenstein combinations
\begin{equation}
 \begin{aligned}
 h_1^{\mathrm{sh}}(\tau_2)
 &=E_4(Q)-9E_4(Q^3)\,,\\
 h_{\mathrm I,3}^{\mathrm{sh}}(\tau_2)
 &=E_4(Q)-81E_4(Q^9)\,,\\
 h_{\mathrm{II},3}^{\mathrm{sh}}(\tau_2)
 &=E_4(Q)-42E_4(Q^3)+81E_4(Q^9)\,.
 \end{aligned}
 \label{eq:full-forms}
\end{equation}
Their constant Fourier modes are $-8$, $-80$
and $40$, respectively. Since
$Q^m=(-1)^mq^m$, the divisor laws give
\begin{equation}
 \begin{aligned}
 D^3\Phi_1(q)
 &=-\frac{h_1^{\mathrm{sh}}(\tau_2)+8}{72\pi}\,,\\
 D^3\Phi_{\mathrm I,3}(q)
 &=-\frac{h_{\mathrm I,3}^{\mathrm{sh}}(\tau_2)+80}
          {80\pi}\,,\\
 D^3\Phi_{\mathrm{II},3}(q)
 &=\frac{h_{\mathrm{II},3}^{\mathrm{sh}}(\tau_2)-40}
         {40\pi}\,,
 \end{aligned}
 \label{eq:derivative-forms}
\end{equation}
where $\Phi_1=\Phi_{\mathrm I,1}=\Phi_{\mathrm{II},1}$.
The subtraction of the constant modes is required
because $D^3$ annihilates the Fourier constant
of any function holomorphic at the origin.

These equations identify the generating functions
as normalized Eichler integrals. For a weight-four
form $h(\tau_2)=a_0+\sum_{m\geq1}a_mQ^m$, our
inverse-derivative convention is
\begin{equation}
 \begin{aligned}
 D^{-3}(h-a_0)
 =\sum_{m\geq1}\frac{a_m}{m^3}Q^m
 =-\frac{(2\pi\ii)^3}{2}
   \int_{\tau_2}^{\ii\infty}
   (z-\tau_2)^2\bigl[h(z)-a_0\bigr]\,\rd z\,.
 \end{aligned}
 \label{eq:Eichler-integral}
\end{equation}
The contour is vertical. Subtraction of $a_0$
ensures exponential decay at its upper endpoint,
and integration of each Fourier mode gives
the factor $m^{-3}$.
The primitive is uniquely fixed by holomorphy
at $Q=0$ and its vanishing value there.
These conditions exclude the polynomial of
degree at most two in $\log Q$ allowed by
indefinite integration.
This is the noncuspidal Eichler prescription
of Ref.~\cite[Eq.~(8.1)]{Pasol:2013}, with the
normalization used in the ABJM and ADHM
analyses~\cite{Hosseini:2026lvj,Hosseini:ADHM2026}.

The finite dilation structure of the generators
gives explicit Lambert representations.
Define
\begin{equation}
 \cL_3(Q)
 =\sum_{r\geq1}\Li_3(Q^r)
 =\sum_{m\geq1}\frac{\sigma_3(m)}{m^3}Q^m\,.
 \label{eq:Lambert}
\end{equation}
Expanding the trilogarithm and grouping terms
with $m=rn$ gives the coefficient
$\sum_{n\mid m}n^{-3}=\sigma_3(m)/m^3$.
The rearrangement is justified by
$\sum_{r,n\geq1}|Q|^{rn}/n^3
\leq\zeta(3)|Q|/(1-|Q|)$.
For a positive integer $L$,
$D^3\cL_3(Q^L)=L^3[E_4(Q^L)-1]/240$,
so the normalized combination
\begin{equation}
 \begin{aligned}
 \Etilde_L(Q)
 &:= D^{-3}\bigl[
 E_4(Q)-L^2E_4(Q^L)-(1-L^2)\bigr]\\
 &=240\bigl[\cL_3(Q)-L^{-1}\cL_3(Q^L)\bigr]
 \end{aligned}
 \label{eq:Etilde}
\end{equation}
has vanishing Fourier constant. In this notation,
\begin{equation}
 \begin{aligned}
 \Phi_1(q)
 &=-\frac{\Etilde_3(-q)}{72\pi}\,,\\
 \Phi_{\mathrm I,3}(q)
 &=-\frac{\Etilde_9(-q)}{80\pi}\,,\\
 \Phi_{\mathrm{II},3}(q)
 &=\frac6\pi\left[
 \cL_3(-q)-\frac{14}{9}\cL_3(-q^3)
          +\frac19\cL_3(-q^9)\right]\,.
 \end{aligned}
 \label{eq:primitives}
\end{equation}
Here $Q^d=-q^d$ for the odd dilations $d=1,3,9$.
The factors $d^{-3}$ from the inverse derivatives
account for the coefficients of the dilated
Lambert series, so these expressions have
precisely the Fourier coefficients
$\mathfrak k^2c_m/(2\pi m)$ specified above.

The homogeneous weight-four transformation laws
belong to the full generators in
Eq.~\eqref{eq:full-forms}, whose minimal levels
are determined in Section~\ref{subsec:levels}.
Their nonzero constant modes contribute to the
inhomogeneous transformation laws of the
Eichler primitives~\cite[Section~8]{Pasol:2013}.
The additional action dependence in
Eq.~\eqref{eq:operator} must also be retained
when transforming the proposed correction.
Its consequence for the physical expression
is made explicit in
Eq.~\eqref{eq:correction-monodromy}.

\subsection{Half-period rewriting and minimal levels}
\label{subsec:levels}

For a positive integer $L$, define
\begin{equation}
 \begin{aligned}
 \Gamma_0(L)
 &=\left\{\gamma=
 \begin{pmatrix}
 \mathsf a&\mathsf b\\
 \mathsf c&\mathsf d
 \end{pmatrix}
 \in\mathrm{SL}_2(\mathbb Z):L\mid\mathsf c\right\}\,,\\
 (h|_4\gamma)(\tau)
 &=(\mathsf c\tau+\mathsf d)^{-4}h(\gamma\tau)\,,
 \qquad
 \gamma\tau=\frac{\mathsf a\tau+\mathsf b}
                  {\mathsf c\tau+\mathsf d}\,.
 \end{aligned}
 \label{eq:Gamma0}
\end{equation}
The space $M_4(\Gamma_0(L))$ consists of functions
holomorphic on the upper half-plane and at every
cusp, invariant under this slash action.
All characters are trivial.

For a positive integer dilation $d$ and
$\gamma\in\Gamma_0(d)$, the matrix
$\big(\begin{smallmatrix}
 \mathsf a&d\mathsf b\\
 \mathsf c/d&\mathsf d
 \end{smallmatrix}\big)
 \in\mathrm{SL}_2(\mathbb Z)$
maps $d\tau$ to $d\gamma\tau$ with automorphy
factor $\mathsf c\tau+\mathsf d$.
The level-one transformation law of $E_4$,
together with holomorphy at the cusps under the
degeneracy construction, gives
$E_4(\e^{2\pi\ii d\tau})\in
M_4(\Gamma_0(d))$~\cite[Theorem~5.8]{Stein:2007}.
A linear combination with all dilations dividing
$L$ therefore belongs to $M_4(\Gamma_0(L))$.
In particular, the shifted forms in
Eq.~\eqref{eq:full-forms} belong to levels
$3,9,9$. These memberships do not yet establish
minimality.

To express the same functions in the physical
coordinate $\tau$, recall that
$q=\e^{2\pi\ii\tau}$ and
$Q=-q=\e^{2\pi\ii\tau_2}$, with
$\tau_2=\tau+1/2$.
The Lambert identity in
Ref.~\cite[Eq.~(2.52)]{Dorigoni:2020oon},
specialized to $s=3$, reads%
\footnote{For $|z|<1$, absolute convergence gives
\begin{equation*}
 \Li_3(z)+\Li_3(-z)
 =2\sum_{n\geq1}\frac{z^{2n}}{(2n)^3}
 =\frac14\Li_3(z^2)\,.
\end{equation*}
In $\cL_3(q)+\cL_3(-q)$, the even Lambert
indices $r$ satisfy $(-q)^r=q^r$ and contribute
$2\sum_{j\geq1}\Li_3(q^{2j})=2\cL_3(q^2)$.
For odd $r$, the trilogarithm identity gives
\begin{equation*}
 \begin{aligned}
 \sum_{\substack{r\geq1\\r\ \mathrm{odd}}}
 \bigl[\Li_3(q^r)+\Li_3(-q^r)\bigr]
 =\frac14
   \sum_{\substack{r\geq1\\r\ \mathrm{odd}}}
   \Li_3(q^{2r})
 =\frac14\bigl[\cL_3(q^2)-\cL_3(q^4)\bigr]\,.
 \end{aligned}
\end{equation*}
The last equality removes the even-$r$ terms
from $\sum_{r\geq1}\Li_3(q^{2r})$.
All rearrangements are justified by
$\sum_{r,n\geq1}|q|^{rn}/n^3
\leq\zeta(3)|q|/(1-|q|)$.
Adding the two contributions proves
Eq.~\eqref{eq:half-period-primitive} directly
for the Fourier-normalized functions.
If obtained instead by three inverse logarithmic
derivatives, the possible difference would be a
polynomial of degree at most two in $\log q$.
Holomorphy at $q=0$ excludes the logarithmic
terms, and the vanishing Fourier constant
excludes the remaining constant.}
\begin{equation}
 \cL_3(-q)
 =-\cL_3(q)+\frac94\cL_3(q^2)
          -\frac14\cL_3(q^4)\,,
 \qquad |q|<1\,.
 \label{eq:half-period-primitive}
\end{equation}
Applying $D^3$, with $D=q\partial_q$ and
$D^3\cL_3(q^d)=d^3[E_4(q^d)-1]/240$, gives
\begin{equation}
 E_4(-q)
 =-E_4(q)+18E_4(q^2)-16E_4(q^4)\,.
 \label{eq:half-period}
\end{equation}
The constant terms agree because
$-1+18-16=1$.
This is the normalized weight-four identity
of Ref.~\cite[Eq.~(2.56)]{Dorigoni:2020oon},
also used in
Ref.~\cite[Eq.~(3.29)]{Hosseini:2026lvj}.

For each shifted form, write
$h(\tau)=h^{\mathrm{sh}}(\tau_2)$.
All dilations in Eq.~\eqref{eq:full-forms}
are odd, so $Q^d=-q^d$ and
Eq.~\eqref{eq:half-period} gives
\begin{align}
 h_1(\tau)={}&
 -E_4(q)+18E_4(q^2)-16E_4(q^4)
 +9E_4(q^3)-162E_4(q^6)
                  +144E_4(q^{12})\,,
 \label{eq:physical-unit}\\
 h_{\mathrm I,3}(\tau)={}&
 -E_4(q)+18E_4(q^2)-16E_4(q^4)
 +81E_4(q^9)-1458E_4(q^{18})
                  +1296E_4(q^{36})\,,
 \label{eq:physical-I3}\\
 h_{\mathrm{II},3}(\tau)={}&
 -E_4(q)+18E_4(q^2)-16E_4(q^4)
 +42E_4(q^3)-756E_4(q^6)
                  +672E_4(q^{12})
 \notag\\
 &-81E_4(q^9)+1458E_4(q^{18})
                  -1296E_4(q^{36})\,.
 \label{eq:physical-II3}
\end{align}
The dilations establish membership at levels
$12,36,36$, respectively.
The constant Fourier modes remain $-8,-80,40$,
and Eq.~\eqref{eq:half-period-primitive} rewrites
the corresponding primitives without changing
their zero-mode prescription.

\begin{proposition}\label{prop:levels}
The minimal levels of
$h_1^{\mathrm{sh}}$, $h_{\mathrm I,3}^{\mathrm{sh}}$
and $h_{\mathrm{II},3}^{\mathrm{sh}}$ in the
coordinate $\tau_2$ are $3,9,9$.
The minimal levels of their pullbacks
$h_1$, $h_{\mathrm I,3}$ and $h_{\mathrm{II},3}$
in $\tau$ are $12,36,36$.
Minimality is taken among the spaces
$M_4(\Gamma_0(L))$ with positive integer $L$
and trivial character.
\end{proposition}
\begin{proof}
Consider a finite sum
$h(\tau)=\sum_d a_dE_4(\e^{2\pi\ii d\tau})$,
with constant Fourier mode $a_0=\sum_d a_d$.
For any positive integer $M$, the cusp-constant
formula proved in
Ref.~\cite[Section~5.4, Eq.~(5.14)]{Hosseini:ADHM2026}
gives%
\footnote{For one dilation, put $g=\gcd(d,M)$,
$\mathsf a=d/g$ and $\mathsf c=M/g$,
and choose integers $\mathsf b,\mathsf d$ with
$\mathsf a\mathsf d-\mathsf b\mathsf c=1$.
The associated matrix maps
$w=(g\tau-\mathsf b)/\mathsf a$ to $d\gamma_M\tau$.
The level-one transformation law then gives
\begin{equation*}
 (M\tau+1)^{-4}
 E_4(\e^{2\pi\ii d\gamma_M\tau})
 =\left(\frac gd\right)^4
   E_4(\e^{2\pi\ii w})\,.
\end{equation*}
Since $\im w=(g/\mathsf a)\im\tau\to\infty$,
the limit is $(g/d)^4$.
Summing over the finitely many dilations gives
Eq.~\eqref{eq:cusp-test}.}
\begin{equation}
 \lim_{\im\tau\to\infty}(h|_4\gamma_M)(\tau)
 =\sum_d a_d\frac{\gcd(d,M)^4}{d^4}\,,
 \qquad
 \gamma_M=\begin{pmatrix}1&0\\M&1\end{pmatrix}\,.
 \label{eq:cusp-test}
\end{equation}
If $h\in M_4(\Gamma_0(M))$, then
$\gamma_M$ belongs to its modular group and
$h|_4\gamma_M=h$.
The cusp constant in Eq.~\eqref{eq:cusp-test}
must therefore equal $a_0$.
This is a necessary condition for every candidate
level $M$, not only for divisors of a level
already known to give membership.

For the shifted forms, the cusp sums depend
only on whether $M$ is divisible by $3$ or $9$:
\begin{equation}
 \begin{array}{c|rrr|r}
 &3\nmid M&3\mid M,\ 9\nmid M&9\mid M&a_0\\
 \hline
 h_1^{\mathrm{sh}}
 &8/9&-8&-8&-8\\
 h_{\mathrm I,3}^{\mathrm{sh}}
 &80/81&0&-80&-80\\
 h_{\mathrm{II},3}^{\mathrm{sh}}
 &40/81&-40&40&40
 \end{array}\,.
 \label{eq:cusp-table}
\end{equation}
For example, the last row is obtained from
\begin{equation*}
 1-\frac{42}{3^4}\gcd(3,M)^4
   +\frac{81}{9^4}\gcd(9,M)^4\,.
\end{equation*}
Comparison with $a_0$ requires $3\mid M$ for
$h_1^{\mathrm{sh}}$ and $9\mid M$ for the
other two shifted forms.

For the physical forms, each odd-dilation term
is replaced by three terms with relative
coefficients $-1,18,-16$ and dilations $1,2,4$.
For odd $d$,
$\gcd(2^rd,M)=\gcd(2^r,M)\gcd(d,M)$.
Thus the physical cusp sum is the corresponding
entry of Eq.~\eqref{eq:cusp-table} multiplied
by the factor
\begin{equation}
 \begin{aligned}
 U_2(M)
 &:=-1+\frac{18}{16}\gcd(2,M)^4
       -\frac{16}{256}\gcd(4,M)^4\\
 &=
 \begin{cases}
  1/16&\text{if $M$ is odd}\,,\\
  16&\text{if $M\equiv2\pmod4$}\,,\\
  1&\text{if $4\mid M$}\,.
 \end{cases}
 \end{aligned}
 \label{eq:two-cusp-factor}
\end{equation}
This factor is positive.
For $h_1$, the positive cusp value when $3\nmid M$
cannot match the negative constant $a_0=-8$.
For $h_{\mathrm I,3}$, the nonnegative values
when $9\nmid M$ cannot match $a_0=-80$.
Hence the physical forms still require
$3\mid M$ and $9\mid M$, respectively.
For $h_{\mathrm{II},3}$, the cases $3\nmid M$
and $3\mid M$, $9\nmid M$ would require
$U_2(M)=81$ and $U_2(M)=-1$, respectively.
Neither is possible, so this form also requires
$9\mid M$.

Once the required power of three divides $M$,
the shifted cusp value equals the nonzero
constant $a_0$ in every case.
The physical cusp value can then equal $a_0$
only if $U_2(M)=1$, or $4\mid M$.
Every admissible physical level must therefore
be divisible by $12$, $36$ and $36$,
respectively.
The memberships established above attain these
bounds, proving all six minimal-level statements.
\end{proof}

The half-period translation also identifies a
group under which every pulled-back form is
modular. Set
$T_{1/2}=
\big(\begin{smallmatrix}1&1/2\\0&1\end{smallmatrix}\big)$.
If $h^{\mathrm{sh}}$ is modular on $\Gamma_0(L)$
in $\tau_2$, then $h(\tau)=h^{\mathrm{sh}}(T_{1/2}\tau)$
is modular on
$T_{1/2}^{-1}\Gamma_0(L)T_{1/2}$.
For an integral matrix
$\gamma=
\big(\begin{smallmatrix}
\mathsf a&\mathsf b\\\mathsf c&\mathsf d
\end{smallmatrix}\big)$
of determinant one,
\begin{equation}
 T_{1/2}\gamma T_{1/2}^{-1}
 =
 \begin{pmatrix}
 \mathsf a+\mathsf c/2&
 \mathsf b+(\mathsf d-\mathsf a)/2-\mathsf c/4\\
 \mathsf c&\mathsf d-\mathsf c/2
 \end{pmatrix}\,.
 \label{eq:conjugation}
\end{equation}
Integrality of the diagonal entries requires
$\mathsf c$ to be even.
The determinant condition then makes
$\mathsf a$ and $\mathsf d$ odd, so the
upper-right entry is integral precisely when
$4\mid\mathsf c$.
Membership in $\Gamma_0(L)$ additionally
requires $L\mid\mathsf c$.
Since $L=3,9$ is odd,
\begin{equation}
 \bigl(T_{1/2}^{-1}\Gamma_0(L)T_{1/2}\bigr)
 \cap\mathrm{SL}_2(\mathbb Z)
 =\Gamma_0(4L)\,.
 \label{eq:group-intersection}
\end{equation}
This calculation establishes an integral group
of modular transformations for every pullback.
A particular linear combination could have
additional symmetries, so the group intersection
alone does not prove minimality.
The cusp test excludes lower $\Gamma_0$ levels
for the three forms considered here.
Neither argument classifies their full
stabilizers in $\mathrm{SL}_2(\mathbb R)$.

\paragraph{The nonperturbative correction.}
The homogeneous weight-four laws concern the
full Eisenstein generators.
Their constant-subtracted Eichler primitives
have the inhomogeneous transformation laws
appropriate to noncuspidal
forms~\cite[Section~8]{Pasol:2013}.
The explicit action dependence of the proposed correction
is retained in its continuation to the upper half-plane,
\begin{equation}
 \mathscr F(\tau)
 =(1-\tau\partial_\tau)
   \Phi(\e^{2\pi\ii\tau})\,.
 \label{eq:correction-continuation}
\end{equation}
On $\tau=\ii t/(2\pi)$, this is the explicit
exponential series in Eq.~\eqref{eq:operator}.
The Fourier-normalized primitive and its
derivative are periodic under
$T:\tau\mapsto\tau+1$, but the factor of $\tau$
in Eq.~\eqref{eq:correction-continuation} gives
\begin{equation}
 \mathscr F(\tau+1)-\mathscr F(\tau)
 =-2\pi\ii\,D\Phi(q)\not\equiv0\,.
 \label{eq:correction-monodromy}
\end{equation}
This defect is nonzero because each primitive has a
nonzero first Fourier coefficient. It records the
logarithmic monodromy of the action: $T$ leaves $q$
unchanged but sends $t=-2\pi\ii\tau$ to $t-2\pi\ii$.

Every $\Gamma_0(L)$ contains $T$, whose
automorphy factor is one at any weight.
A scalar modular form with trivial character
must therefore be periodic, so $\mathscr F$
cannot be such a form.
Adding the perturbative expression does not
restore this property. Its shifted-power term
is $\mathfrak k^2t^3/(27\pi)$, whose translation
defect is a quadratic polynomial in $\tau$;
the constant contributes no defect.
This polynomial cannot cancel the nonconstant
Fourier series in
Eq.~\eqref{eq:correction-monodromy}, which
decays as $\im\tau\to\infty$.

The modular structure therefore constrains the
divisor coefficients through the Eisenstein
generators and their normalized primitives.
It does not make the proposed completed critical
value a scalar modular form on the groups
considered here, or establish a modular action
on the microscopic Bethe equations.

\subsection{Convergence, natural boundaries and truncation}
\label{subsec:convergence}

For the explicit divisor laws, the estimate
$\sigma_3(m)=m^3\sum_{d\mid m}d^{-3}\leq\zeta(3)m^3$
gives
\begin{equation*}
 \frac{\mathfrak k^2}{2\pi}|c_m|
 \leq\frac{\mathfrak k^2\zeta(3)}{2\pi}\,
 m\sum_{d\in\{1,3,9\}}\frac{|b_d|}{d^3}\,.
\end{equation*}
With the dilation weights in
Eq.~\eqref{eq:dilation-weights}, this is
$(\mathfrak k^2/2\pi)|c_m|\leq Km$, where
\begin{equation}
 K_1=\frac{40\zeta(3)}{9\pi}\,,
 \qquad
 K_{\mathrm I,3}=\frac{10\zeta(3)}{3\pi}\,,
 \qquad
 K_{\mathrm{II},3}=\frac{16\zeta(3)}{\pi}\,.
 \label{eq:coefficient-bounds}
\end{equation}
The constant $K_1$ applies to both unit-parameter
series. It follows that $\sum_{m\geq1}c_mq^m$
and $\sum_{m\geq1}c_mq^m/m$ converge absolutely
and locally uniformly for $|q|<1$.

To determine their radii of convergence, consider
a prime $\mathfrak p>3$. The terms with dilation
$3$ or $9$ vanish, and the alternating sign is
positive, giving
\begin{equation}
 c_{\mathfrak p}
 =\kappa\left(\mathfrak p+\mathfrak p^{-2}\right)\,,
 \qquad
 \kappa=\frac{20}{3},\ \frac23,\ -\frac43\,,
 \label{eq:prime-subsequence}
\end{equation}
for the common unit-parameter, Model-I
parameter-three and Model-II parameter-three
sequences, respectively.
Together with the polynomial upper bound, this
subsequence gives
$\limsup_m|c_m|^{1/m}
=\limsup_m|c_m/m|^{1/m}=1$.
The Cauchy--Hadamard formula therefore fixes both
radii to one, by the same argument used in
Ref.~\cite[Eqs.~(3.35)--(3.38)]{Hosseini:2026lvj}.

On $|q|=1$, the terms of neither series tend to
zero along the prime subsequence, so both series
diverge at every boundary point.
This does not by itself exclude analytic
continuation of the functions defined inside
the disk. The stronger natural-boundary
statement follows from radial singularities
of their Eisenstein generators at a dense set
of roots of unity.

\begin{proposition}\label{prop:natural-boundary}
For each primitive $\Phi$ in Eq.~\eqref{eq:primitives},
both $\Phi$ and $D\Phi$ have the unit circle
as a natural boundary. Neither admits a
holomorphic continuation through any point
of $|q|=1$.
\end{proposition}
\begin{proof}
Fix a reduced fraction $a/c$, with $c>0$ and
$\gcd(c,3)=1$, and approach it along
$\tau_2=a/c+\ii y$, $y\downarrow0$.
In the physical nome, this path is
\begin{equation*}
 q_y=q_*\,\e^{-2\pi y}\,,
 \qquad
 q_*=-\e^{2\pi\ii a/c}\,,
\end{equation*}
so it approaches $q_*$ radially from inside
the unit disk.

For each dilation $d=1,3,9$, the integers
$da$ and $c$ are coprime. Choose integers
$\mathsf b,\mathsf d$ satisfying
$da\,\mathsf d-c\,\mathsf b=1$ and write
\begin{equation*}
 \gamma_d=
 \begin{pmatrix}
 \mathsf a&\mathsf b\\
 \mathsf c&\mathsf d
 \end{pmatrix}
 \in\mathrm{SL}_2(\mathbb Z)\,,
 \qquad
 \mathsf a=da\,,
 \qquad
 \mathsf c=c\,.
\end{equation*}
With
$w=-\mathsf d/c+\ii/(c^2dy)$, the determinant
condition gives
\begin{equation*}
 \gamma_d w
 =\frac{\mathsf a w+\mathsf b}
       {\mathsf c w+\mathsf d}
 =d\tau_2\,,
 \qquad
 \mathsf c w+\mathsf d=\frac{\ii}{cdy}\,.
\end{equation*}
The level-one transformation law therefore yields
\begin{equation}
 E_4\!\left(\e^{2\pi\ii d(a/c+\ii y)}\right)
 =\frac{E_4(\e^{2\pi\ii w})}{c^4d^4y^4}
 \sim\frac1{c^4d^4y^4}\,,
 \qquad y\downarrow0\,.
 \label{eq:cusp-singularity}
\end{equation}
Indeed, $\im w=1/(c^2dy)\to\infty$, so the
Fourier expansion gives
$E_4(\e^{2\pi\ii w})\to1$.
The positive prefactor follows from $\ii^4=1$.

For the three shifted forms in
Eq.~\eqref{eq:full-forms}, the coefficients of
$(c^4y^4)^{-1}$ are
\begin{equation*}
 1-\frac9{3^4}=\frac89\,,
 \qquad
 1-\frac{81}{9^4}=\frac{80}{81}\,,
 \qquad
 1-\frac{42}{3^4}+\frac{81}{9^4}
 =\frac{40}{81}\,.
\end{equation*}
None vanishes, and subtracting the constant
Fourier modes changes only bounded terms.
Including the normalizations in
Eq.~\eqref{eq:derivative-forms} gives
\begin{equation*}
 \lim_{y\downarrow0}c^4y^4D^3\Phi(q_y)
 =
 \begin{cases}
 -\dfrac1{81\pi}\,,
   &\Phi=\Phi_1\ \text{or}\ \Phi_{\mathrm I,3}\,,\\[.4em]
 \phantom{-}\dfrac1{81\pi}\,,
   &\Phi=\Phi_{\mathrm{II},3}\,.
 \end{cases}
\end{equation*}
Thus $D^3\Phi$ is unbounded along the radial
approach to every such $q_*$.
A holomorphic extension to a neighborhood of
$q_*$ would have a finite limit there, so each
of these points is a nonremovable boundary
singularity of $D^3\Phi$.

These explicitly identified singularities are
dense on the unit circle.
For any real $x$, the dyadic rational
$\lfloor2^n x\rfloor/2^n$ differs from $x$
by less than $2^{-n}$. Reducing this fraction
leaves a denominator that is a power of two
and hence coprime to three.
The reduced fractions allowed above are
therefore dense on the real line.
Every nonempty open arc of the unit circle
has a preimage containing a real interval
under $x\mapsto-\e^{2\pi\ii x}$, so it contains
one of their images $q_*$.

Now let $q_0$ be an arbitrary point of the unit
circle and suppose that $\Phi$ has a holomorphic
continuation $F$ to an open disk $U$ centred
at $q_0$, agreeing with $\Phi$ on
$U\cap\{|q|<1\}$.
The operator $D=q\partial_q$ has holomorphic
coefficients, so $D^3F$ is holomorphic on $U$
and agrees there with $D^3\Phi$ inside the
unit disk.
By density, choose a cusp point $q_*\in U$
from the set above. For sufficiently small
$y>0$, its radial approach $q_y$ lies in
$U\cap\{|q|<1\}$.
Holomorphy would then imply
\begin{equation*}
 D^3\Phi(q_y)=D^3F(q_y)
 \longrightarrow D^3F(q_*)\,,
\end{equation*}
a finite limit, contradicting the radial
divergence.

If $D\Phi$ instead admitted a holomorphic
continuation $G$ to such a disk, then
$D^2G$ would extend $D^3\Phi$ and give the
same contradiction.
Neither function can therefore be continued
through any boundary point.
The argument uses the explicitly established
singularities only on a dense subset; it
does not require a cusp asymptotic at every
point of the circle.
\end{proof}

The cusp divergences are compatible with
holomorphy of the full modular forms at their
cusps, which is defined after the weight-four
slash transformation.
The proposition concerns the holomorphic
functions $\Phi$ and $D\Phi$.
The proposed correction additionally sets
$t=-\log q$, giving $(1-\log q\,D)\Phi(q)$
on a chosen logarithmic branch.
This expression is not holomorphic at $q=0$,
and its translation defect is given in
Eq.~\eqref{eq:correction-monodromy}.

For $t>0$, $q=\e^{-t}$ and an integer cutoff
$M\geq0$, define the omitted part of the
proposed series by
\begin{equation*}
 \mathrm{tail}_{>M}
 =\frac{\mathfrak k^2}{2\pi}
  \sum_{m=M+1}^{\infty}c_m
       \left(t+\frac1m\right)q^m\,.
\end{equation*}
The geometric series
$\sum_{m>M}q^m=q^{M+1}/(1-q)$ and its
logarithmic derivative give
\begin{equation}
 \begin{aligned}
 |\mathrm{tail}_{>M}|
 &\leq K\sum_{m=M+1}^{\infty}(tm+1)q^m\\
 &=Kq^{M+1}
 \left[
 t\left(\frac{M+1}{1-q}
        +\frac{q}{(1-q)^2}\right)
 +\frac1{1-q}
 \right]\,.
 \end{aligned}
 \label{eq:tail}
\end{equation}
This bound controls truncation of the explicit
series, independently of its identification
with the Bethe remainder.
The full-series comparisons add the resulting
symmetric tail interval to the numerical
enclosure of the partial sum, including the
perturbative term.
The cutoff selection and arithmetic and export
radii are specified in
Appendix~\ref{app:numerics}.

\subsection{Dilation and root-of-unity relations}
\label{subsec:function-identities}

The right-hand side of Eq.~\eqref{eq:completion} defines
an analytic function of real shifted rank $\Nh>0$, with
the positive square root fixing its action. In this
subsection, $\W$ denotes that function. Its identification
with the Bethe critical value remains conditional on
Conjecture~\ref{conj:completion}. The identities between
primitives hold throughout $|Q|<1$, with $Q=-q$.

The definition of $\Etilde_L$ gives, for positive
integers $a,b$,
\begin{equation}
 \Etilde_{ab}(Q)
 =\Etilde_a(Q)+\frac1a\Etilde_b(Q^a)\,.
 \label{eq:dilation}
\end{equation}
Indeed, the two terms proportional to $\cL_3(Q^a)$
on the right cancel, leaving
$240[\cL_3(Q)-(ab)^{-1}\cL_3(Q^{ab})]$.
Taking $a=b=3$ and using the primitive normalizations
in Eq.~\eqref{eq:primitives} yields
\begin{equation}
 \Phi_{\mathrm I,3}(q)
 =\frac9{10}\Phi_1(q)+\frac3{10}\Phi_1(q^3)\,.
 \label{eq:primitive-dilation}
\end{equation}
Here $(-q)^3=-q^3$, so no additional phase is introduced.
The logarithmic derivative acts on the dilated argument
according to
\begin{equation*}
 (1+tD)\Phi_1(q^d)
 =\left.(1+dtz\partial_z)\Phi_1(z)\right|_{z=q^d}\,.
\end{equation*}
Thus a dilation by $d$ replaces the action by $dt$
in the corresponding exponential correction.

At level three, $\Nh_{\mathrm I}(N,3)=N-1/12$.
The unit-level rank arguments required by actions
$t$ and $3t$ are fixed by
\begin{equation*}
 \begin{aligned}
 \Nh_{\mathrm I}\left(\frac N3-\frac49,1\right)
 =\frac13\Nh_{\mathrm I}(N,3)\,,\qquad
 \Nh_{\mathrm I}\left(3N-\frac23,1\right)
 =3\Nh_{\mathrm I}(N,3)\,.
 \end{aligned}
\end{equation*}
The shifted-power term equals $k^2t^3/(27\pi)$,
so the coefficients in Eq.~\eqref{eq:primitive-dilation}
match it as well, since $9/10+(3/10)3^3=9$.
The remaining constant difference is
$\CI(3)-\frac65\CI(1)=\frac8{15}\Cltwo(\pi/3)$.
Combining these contributions gives
\begin{equation}
 \WI(N,3)
 =\frac9{10}\WI\left(\frac N3-\frac49,1\right)
  +\frac3{10}\WI\left(3N-\frac23,1\right)
  +\frac8{15}\Cltwo\left(\frac\pi3\right)\,.
 \label{eq:rank-dilation}
\end{equation}
This identity holds for real $N>1/12$, where all three
shifted ranks are positive. At integer $N$, both transformed
arguments are noninteger and refer to the analytic
continuations of the unit-level expression.

For a prime $\mathfrak p$, the finite phase sum selects
Fourier modes divisible by $\mathfrak p$:
\begin{equation}
 \begin{aligned}
 \sum_{j=0}^{\mathfrak p-1}
 \cL_3(\e^{2\pi\ii j/\mathfrak p}Q)
 &=\frac1{\mathfrak p^2}
   \sum_{m\geq1}\frac{\sigma_3(\mathfrak p m)}{m^3}
                   Q^{\mathfrak p m}\\
 &=(\mathfrak p+\mathfrak p^{-2})\cL_3(Q^{\mathfrak p})
   -\mathfrak p^{-2}\cL_3(Q^{\mathfrak p^2})\,.
 \end{aligned}
 \label{eq:root-filter}
\end{equation}
Absolute convergence justifies the exchange with the
Fourier sum. The second equality uses
$\sigma_3(\mathfrak p m)=(1+\mathfrak p^3)\sigma_3(m)
-\mathfrak p^3\sigma_3(m/\mathfrak p)$, obtained by
separating the prime-power factor of $m$.
In the final term, only $m=\mathfrak p n$ contributes;
the factor $m^{-3}$ then supplies $\mathfrak p^{-3}$.

With $\zeta_3=\e^{2\pi\ii/3}$, the cubic argument
in $\Phi_1$ is unchanged by these phases. Equation
\eqref{eq:root-filter} consequently gives
\begin{equation*}
 \sum_{j=0}^{2}\Phi_1(\zeta_3^jq)
 =-\frac{10}{27\pi}
   \bigl[19\cL_3(Q^3)-\cL_3(Q^9)\bigr]\,.
\end{equation*}
Combining this with the three Lambert representations
in Eq.~\eqref{eq:primitives} proves
\begin{equation}
 \Phi_{\mathrm{II},3}(q)
 =\Phi_{\mathrm I,3}(q)-\frac95\Phi_1(q)
  +\frac9{10}
   \bigl[\Phi_1(\zeta_3q)+\Phi_1(\zeta_3^2q)\bigr]\,.
 \label{eq:root-primitive}
\end{equation}
For $0<q<1$, the real Fourier coefficients of $\Phi_1$
make the phase-rotated terms complex conjugates, so their
sum is real.

Applying $1+tD$ to Eq.~\eqref{eq:root-primitive}
keeps the same external $t$ in every term. A phase
rotation introduces no chain-rule multiplier:
$D[\Phi(\zeta_3^jq)]=(D\Phi)(\zeta_3^jq)$.
To compare with completions evaluated at their own
complex actions, choose
$q_j=\zeta_3^jq=\e^{-t_j}$ and
$t_j=t-2\pi\ii j/3$, $j=\pm1$. Then
\begin{equation}
 (1+tD)\Phi(q_j)
 =(1+t_jD)\Phi(q_j)+(t-t_j)D\Phi(q_j)\,.
 \label{eq:phase-operator}
\end{equation}
For conjugate phases, the sum of the additional terms is
$-(4\pi/3)\im[(D\Phi)(\zeta_3q)]$, which is real
and need not vanish. A complex-rank interpretation must retain
this correction and also match the shifted powers
and constants. Reality of the phase sum alone does
not give an identity of completed critical values.

\section{Large-parameter expansion and Borel summation}
\label{sec:genus}

The integral-defined constants~\eqref{eq:constant-I}--\eqref{eq:constant-II}
admit asymptotic
expansions at large $k$ or $N_f$. Combined with
the shifted powers, these determine the genus
expansion of the proposed perturbative expressions
at fixed 't~Hooft coupling.

We normalize the Bernoulli numbers by
\begin{equation*}
 \frac{x}{\e^x-1}
 =\sum_{n\geq0}B_n\frac{x^n}{n!}\,,
 \qquad |x|<2\pi\,.
\end{equation*}
For $a>0$, the identity
$L_a'(x)=a/(\e^{ax}-1)$ and the condition
$L_a(x)-\log(ax)\to0$ as $x\to0^+$ give
\begin{equation*}
 L_a(x)
 =\log(ax)-\frac{ax}{2}
 +\sum_{j\geq1}
  \frac{B_{2j}}{2j(2j)!}(ax)^{2j}\,,
 \qquad 0<x<\frac{2\pi}{a}\,.
\end{equation*}
Substitution into
Eq.~\eqref{eq:kernel-factorization} cancels
the logarithmic and linear terms. The resulting
even Taylor expansions are
\begin{equation}
 \begin{aligned}
 \mathcal K_{\mathrm I}(x)
 &=\frac23\log\frac{27}{4}
   +\sum_{j\geq1}
    \frac{\mathcal A_jB_{2j}}{2j(2j)!}x^{2j}\,,\qquad
    \mathcal A_j
    =2\cdot3^{2j}-\frac43\,2^{2j}-\frac23\,,\\
 \mathcal K_{\mathrm{II}}(x)
 &=\log3
   +\sum_{j\geq1}
    \frac{(3^{2j}-1)B_{2j}}{2j(2j)!}x^{2j}\,.
 \end{aligned}
 \label{eq:kernel-Taylor}
\end{equation}
The nearest complex zeros of $1+2\cosh x$ are
the simple zeros $x=\pm2\pi\ii/3$.
They produce logarithmic singularities in both
kernels. Since $2\cosh(x/2)=1$ at these points,
the additional logarithm in Model I cannot
cancel them. Both Taylor series therefore
have radius $2\pi/3$.

For every integer $j\geq0$ and real $a>0$,
the logarithmic moment is
\begin{equation}
 \begin{aligned}
 \int_0^\infty
 x^{2j}\log(1-\e^{-ax})\,\rd x
 =-\sum_{n\geq1}\frac1n
   \int_0^\infty x^{2j}\e^{-anx}\,\rd x
 =-\frac{(2j)!\zeta(2j+2)}{a^{2j+1}}\,.
 \end{aligned}
 \label{eq:moment}
\end{equation}
The sum of the absolute integrals is
$(2j)!\zeta(2j+2)/a^{2j+1}<\infty$, which
justifies termwise integration.

Only finite Taylor truncations can be used over the
full integration range. For either kernel $\mathcal K$,
fix $J\geq0$ and $0<\delta<2\pi/3$, and write
\begin{equation*}
 \mathcal K(x)
 =\sum_{j=0}^{J}\kappa_{2j}x^{2j}+R_J(x)\,,
 \qquad
 |R_J(x)|\leq C_{J,\delta}x^{2J+2}
 \quad(0\leq x\leq\delta)\,,
\end{equation*}
where $\kappa_{2j}$ are the Taylor coefficients
of the chosen kernel.
Equation~\eqref{eq:moment} bounds the local
remainder by
\begin{equation*}
 \left|
 \frac{\mathfrak k}{\pi}\int_0^\delta
 \log(1-\e^{-\mathfrak kx})R_J(x)\,\rd x
 \right|
 \leq
 \frac{C_{J,\delta}(2J+2)!\zeta(2J+4)}
      {\pi\mathfrak k^{2J+2}}\,.
\end{equation*}
For $x\geq\delta$,
\begin{equation*}
 |\log(1-\e^{-\mathfrak kx})|
 \leq
 \frac{\e^{-\mathfrak kx}}
      {1-\e^{-\mathfrak k\delta}}\,.
\end{equation*}
The exact kernel grows at most linearly, while
the retained Taylor polynomial has fixed degree.
Setting $x=\delta+u/\mathfrak k$ therefore bounds
both tail contributions, including the prefactor
$\mathfrak k/\pi$, by
$O(\e^{-\mathfrak k\delta})$ as
$\mathfrak k\to+\infty$.

Integrating the finite Taylor polynomials gives
\begin{align}
 \CI(k)
 &\sim\frac{k^2\zeta(3)}{18\pi}
 -\frac\pi9\log\frac{27}{4}
 -\frac1\pi\sum_{j\geq1}
  \frac{\mathcal A_jB_{2j}\zeta(2j+2)}
       {2j\,k^{2j}}\,,
 \label{eq:CI-asymptotic}\\
 \CII(N_f)
 &\sim\frac{2N_f^2}{9}
  \left[
   \frac{\zeta(3)}{\pi}
   -\Cltwo\left(\frac\pi3\right)
  \right]
 -\frac\pi6\log3
 -\frac1\pi\sum_{j\geq1}
  \frac{(3^{2j}-1)B_{2j}\zeta(2j+2)}
       {2j\,N_f^{2j}}\,.
 \label{eq:CII-asymptotic}
\end{align}
The parameter-independent terms are
$-\pi\mathcal K(0)/6$, from the $j=0$ moment.
For every fixed $J$, truncation after $j=J$
leaves an error $O(\mathfrak k^{-2J-2})$.
Equations~\eqref{eq:CI-asymptotic}
and~\eqref{eq:CII-asymptotic} are therefore
Poincar\'e asymptotic expansions of the defined
constant functions, with remainder constants that depend
on the fixed order $J$. Uniform control for growing $J$
is not supplied by this estimate.

For the joint large-rank, large-parameter limit,
set
$\lambda=N/\mathfrak k$,
$\lamh_{\mathrm I}=\lambda-1/12$ and
$\lamh_{\mathrm{II}}=\lambda+1/6$.
The perturbative expressions become
\begin{equation*}
 \begin{aligned}
 \WI^{\mathrm{pert}}
 =\frac{8\pi^2}{27}k^2
   \left(\lamh_{\mathrm I}+\frac1{2k^2}\right)^{3/2}
   +\CI(k)\,,\qquad
 \WII^{\mathrm{pert}}
 =\frac{8\pi^2}{27}N_f^2
   \left(\lamh_{\mathrm{II}}+\frac1{4N_f^2}\right)^{3/2}
   +\CII(N_f)\,.
 \end{aligned}
\end{equation*}
We hold $\lambda>1/12$ fixed in Model I and
$\lambda>0$ fixed in Model II, so the shifted
couplings remain positive.
Suppressing the model label when the convention
is common, define the genus coefficients by
\begin{equation}
 \W^{\mathrm{pert}}
 \sim-\sum_{h\geq0}
 (2\pi\ii\lambda)^{2h-2}
 \cW_h(\lambda)N^{2-2h}\,.
 \label{eq:genus-convention}
\end{equation}
The relation $N=\mathfrak k\lambda$ gives
\begin{equation*}
 -(2\pi\ii\lambda)^{2h-2}N^{2-2h}
 =(-1)^h(2\pi)^{2h-2}\mathfrak k^{2-2h}\,.
\end{equation*}
Thus a coefficient at order
$\mathfrak k^{2-2h}$ must be multiplied by
$(-1)^h(2\pi)^{2-2h}$ to obtain $\cW_h$.

The binomial expansion of the shifted power
contributes
\begin{equation*}
 \frac{8\pi^2}{27}\binom{3/2}{h}
 \begin{cases}
  2^{-h}\lamh_{\mathrm I}^{3/2-h}\,,
       &\text{Model I}\,,\\
  4^{-h}\lamh_{\mathrm{II}}^{3/2-h}\,,
       &\text{Model II}\,,
 \end{cases}
\end{equation*}
to the coefficient of $\mathfrak k^{2-2h}$.
Combining these terms with the quadratic and
constant terms of
Eqs.~\eqref{eq:CI-asymptotic}
and~\eqref{eq:CII-asymptotic} yields
\begin{equation}
 \begin{aligned}
 \cW_{\mathrm I,0}
 &=\frac{32\pi^4}{27}\lamh_{\mathrm I}^{3/2}
   +\frac{2\pi\zeta(3)}9\,,\\
 \cW_{\mathrm I,1}
 &=-\frac{2\pi^2}{9}\sqrt{\lamh_{\mathrm I}}
   +\frac\pi9\log\frac{27}{4}\,,\\
 \cW_{\mathrm{II},0}
 &=\frac{32\pi^4}{27}\lamh_{\mathrm{II}}^{3/2}
   +\frac{8\pi\zeta(3)}9
   -\frac{8\pi^2}{9}\Cltwo\left(\frac\pi3\right)\,,\\
 \cW_{\mathrm{II},1}
 &=-\frac{\pi^2}{9}\sqrt{\lamh_{\mathrm{II}}}
   +\frac\pi6\log3\,.
 \end{aligned}
 \label{eq:genus-low}
\end{equation}
For $h=j+1\geq2$, the constant contribution
comes from the inverse power
$\mathfrak k^{-2j}$.
Using
\begin{equation*}
 B_{2j}=(-1)^{j+1}|B_{2j}|\,,
 \qquad
 \zeta(2j+2)
 =\frac{(2\pi)^{2j+2}|B_{2j+2}|}
        {2(2j+2)!}\,,
\end{equation*}
the same normalization gives
\begin{equation}
 \begin{aligned}
 \cW_{\mathrm I,j+1}
 &=\frac{(-1)^{j+1}2^{2-3j}}
          {27\pi^{2j-2}}
   \binom{3/2}{j+1}\lamh_{\mathrm I}^{1/2-j}
 -\frac{\pi\mathcal A_j|B_{2j}B_{2j+2}|}
        {j(2j+2)!}\,,\\
 \cW_{\mathrm{II},j+1}
 &=\frac{(-1)^{j+1}2^{1-4j}}
          {27\pi^{2j-2}}
   \binom{3/2}{j+1}\lamh_{\mathrm{II}}^{1/2-j}
 -\frac{\pi(3^{2j}-1)|B_{2j}B_{2j+2}|}
        {j(2j+2)!}\,.
 \end{aligned}
 \label{eq:genus-all}
\end{equation}
In particular, the alternating Bernoulli signs
combine with the genus convention to give the
negative, $\lambda$-independent terms displayed
here. These coefficients are exact consequences
of the specified perturbative functions.

Let $a_{2j}$ denote the coefficient of
$\mathfrak k^{-2j}$ in the corresponding
constant-function expansion.
For $j\geq1$, these coefficients alternate
in sign. The identity
$|B_{2j}|=2(2j)!\zeta(2j)/(2\pi)^{2j}$,
together with
$\mathcal A_j\sim2\cdot3^{2j}$ and
$\zeta(2j)\to1$, gives
\begin{equation}
 |a_{2j}|
 \sim(2j-1)!\left(\frac3{2\pi}\right)^{2j}
 \begin{cases}
  4/\pi,&\text{Model I}\,,\\
  2/\pi,&\text{Model II}\,.
 \end{cases}
 \label{eq:factorial-growth}
\end{equation}
Consequently
$\limsup_{j\to\infty}|a_{2j}|^{1/j}=\infty$,
so the inverse-parameter series has zero radius
of convergence in $1/\mathfrak k^2$.
The binomial parts converge for
$|1/(2k^2\lamh_{\mathrm I})|<1$ and
$|1/(4N_f^2\lamh_{\mathrm{II}})|<1$.
Their coefficients grow at most geometrically
at fixed positive shifted coupling and cannot
cancel the factorial growth.
After separating the leading quadratic term,
the genus expansion of the full perturbative
expression therefore also has zero radius
of convergence.

The ratio of successive constant-function
terms is asymptotically
\begin{equation*}
 \frac{|a_{2j+2}|\mathfrak k^{-2j-2}}
      {|a_{2j}|\mathfrak k^{-2j}}
 \sim
 \frac{(2j)(2j+1)}{\mathfrak k^2}
 \left(\frac3{2\pi}\right)^2\,.
\end{equation*}
The least-magnitude term occurs near
$j=\pi\mathfrak k/3+O(1)$ for large $\mathfrak k$.
This estimates a useful truncation scale, but a bound
on the optimally truncated error would require control
uniform at $J=O(\mathfrak k)$.

\begin{figure}[t]
 \centering
 \includegraphics[width=\textwidth]{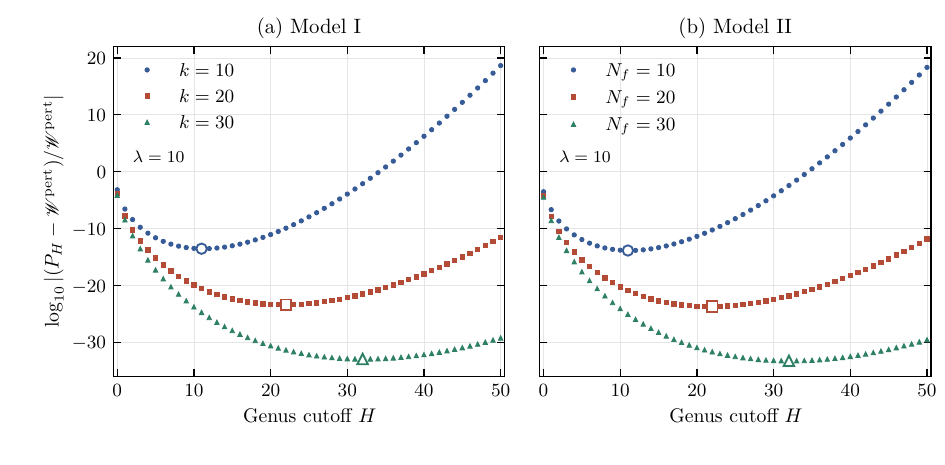}
\caption{Genus truncation of the proposed perturbative
expressions at $\lambda=10$, with Model I on the left
and Model II on the right.
The parameters $\mathfrak k=10,20,30$ correspond to
$N=100,200,300$, respectively.
Filled markers show
$\log_{10}|(P_H-\W^{\mathrm{pert}})/
\W^{\mathrm{pert}}|$ as a function of the genus cutoff $H$;
 open markers identify the smallest discrepancy
 in the displayed range.
 The reference values use the finite Clausen
 evaluations of the constants and the unexpanded
 shifted powers.
 These comparisons test the asymptotic expansions
 of the defined perturbative functions, not the
 full fixed-$\lambda$ Bethe critical values.}
 \label{fig:genus-truncation}
\end{figure}

For a finite genus cutoff $H$, define
\begin{equation}
 P_H(N,\mathfrak k)
 =-\sum_{h=0}^{H}
 (2\pi\ii\lambda)^{2h-2}
 \cW_h(\lambda)N^{2-2h}\,,
 \qquad \lambda=N/\mathfrak k\,.
 \label{eq:genus-truncation}
\end{equation}
At fixed $H\geq1$ and fixed allowed $\lambda$,
the preceding estimates give
$\W^{\mathrm{pert}}-P_H=O(\mathfrak k^{-2H})$.
Increasing $H$ at fixed parameter is a different
limit. For the parameters displayed in
Figure~\ref{fig:genus-truncation}, the discrepancy
first decreases and then grows.
The marked minima are numerical properties of
these partial sums, not exact consequences of
the least-term estimate.
The divisor-series truncations in
Figure~\ref{fig:convergence}, in contrast,
converge as their cutoff increases.

The divergent constant-function expansions
admit an exact Borel sum in the positive
$1/\mathfrak k$ direction.
For a formal series
$\sum_{n\geq0}a_n\mathfrak k^{-n}$, we use
the coefficient map
\begin{equation*}
 \sum_{n\geq0}a_n\mathfrak k^{-n}
 \ \longmapsto\
 \sum_{n\geq0}\frac{a_n}{n!}\xi^n\,,
\end{equation*}
with inverse Laplace transform
$\mathfrak k\int_0^\infty
\e^{-\mathfrak k\xi}(\cdots)\,\rd\xi$.
The constant term is included in this convention.
For the expansions above, $a_{2j+1}=0$.

\begin{proposition}\label{prop:Borel}
For either model, write
\begin{equation*}
 C(\mathfrak k)
 =c_{\mathrm{quad}}\mathfrak k^2
 +\frac{\mathfrak k}{\pi}
  \int_0^\infty
  \log(1-\e^{-\mathfrak kx})\mathcal K(x)\,\rd x\,,
\end{equation*}
where $c_{\mathrm{quad}}$ is the coefficient
of the explicit quadratic term.
The Borel transform of the asymptotic expansion
of $C(\mathfrak k)-c_{\mathrm{quad}}\mathfrak k^2$
continues analytically along the positive axis,
and its Laplace sum is
\begin{equation}
 \begin{aligned}
 \mathscr B(\xi)
 &=-\frac1\pi\sum_{n\geq1}
       \frac{\mathcal K(\xi/n)}{n^2}\,,\\
 C(\mathfrak k)-c_{\mathrm{quad}}\mathfrak k^2
 &=\mathfrak k\int_0^\infty
   \e^{-\mathfrak k\xi}\mathscr B(\xi)\,\rd\xi\,,
 \qquad \mathfrak k>0\,.
 \end{aligned}
 \label{eq:Borel}
\end{equation}
The Laplace integral converges for every real
$\mathfrak k>0$.
The nearest singularities of the Borel germ
are the logarithmic branch points
$\xi=\pm2\pi\ii/3$.
\end{proposition}
\begin{proof}
Near the origin, write
$\mathcal K(x)=\sum_{j\geq0}\kappa_{2j}x^{2j}$.
For every $0<r<2\pi/3$,
\begin{equation*}
 \sum_{n\geq1}\sum_{j\geq0}
 \frac{|\kappa_{2j}|r^{2j}}{n^{2j+2}}
 \leq\zeta(2)\sum_{j\geq0}
 |\kappa_{2j}|r^{2j}<\infty\,.
\end{equation*}
Thus the series defining $\mathscr B$ converges
locally uniformly in this disk, and the sums
may be interchanged to give
\begin{equation*}
 \mathscr B(\xi)
 =-\frac1\pi\sum_{j\geq0}
   \kappa_{2j}\zeta(2j+2)\xi^{2j}\,.
\end{equation*}
Equation~\eqref{eq:moment} identifies
$a_{2j}=-\kappa_{2j}(2j)!\zeta(2j+2)/\pi$,
so this is the Borel transform in the stated
convention. In particular,
$\mathscr B(0)=-\pi\mathcal K(0)/6=a_0$.

To continue the germ, use the factorizations
in Eq.~\eqref{eq:kernel-factorization} with
\begin{equation*}
 L_a(z)=\Log(1-\e^{-az})\,,
 \qquad \re z>0\,.
\end{equation*}
Here $|\e^{-az}|<1$, so these logarithms are
holomorphic and agree with their real values
on the positive axis.
The factorized expressions consequently define
analytic branches of the kernels in the right
half-plane, extending their Taylor germs at
the origin.
On a closed subsector
$|\arg z|\leq\pi/2-\varepsilon$,
with $0<\varepsilon<\pi/2$, one has
$\re z\geq|z|\sin\varepsilon$.
The logarithmic terms then decay exponentially
at large $|z|$, leaving the linear kernel
terms. Regularity at the origin and boundedness
on the intervening compact region give
\begin{equation*}
 |\mathcal K(z)|
 \leq A_\varepsilon+B_\varepsilon|z|
\end{equation*}
for suitable constants
$A_\varepsilon,B_\varepsilon>0$.
It follows that the sum defining $\mathscr B$
converges locally normally in the right
half-plane and satisfies
\begin{equation*}
 |\mathscr B(\xi)|
 \leq\frac{A_\varepsilon\zeta(2)}{\pi}
     +\frac{B_\varepsilon\zeta(3)}{\pi}|\xi|
\end{equation*}
on the same closed subsectors.
This supplies the analytic continuation and
growth bound required for positive-direction
Borel summation.

At $\xi=\pm2\pi\ii/3$, the $n=1$ summand
inherits the nonzero logarithmic singularities
of $\mathcal K$.
The sum over $n\geq2$ is holomorphic throughout
$|\xi|<4\pi/3$, since every argument $\xi/n$
then lies inside the kernel's Taylor disk.
Local uniform convergence follows by bounding
$\mathcal K$ on smaller disks and summing
$n^{-2}$.
The singularities of the first summand therefore
cannot cancel. There are no singularities
closer to the origin, and the continuation
constructed above has none on the positive ray.

Finally, on $x\geq0$ the kernels obey
$|\mathcal K(x)|\leq A+Bx$ for some $A,B>0$.
For every $\mathfrak k>0$,
\begin{equation*}
 \frac{\mathfrak k}{\pi}
 \sum_{n\geq1}\frac1n
 \int_0^\infty
 \e^{-n\mathfrak kx}|\mathcal K(x)|\,\rd x
 \leq
 \frac{A\zeta(2)}{\pi}
 +\frac{B\zeta(3)}{\pi\mathfrak k}<\infty\,.
\end{equation*}
This bound justifies expansion of the logarithm
under the integral. Setting $\xi=nx$ in each
term and interchanging the absolutely convergent
sum and integral gives
\begin{equation*}
 \begin{aligned}
 \frac{\mathfrak k}{\pi}\int_0^\infty
 \log(1-\e^{-\mathfrak kx})\mathcal K(x)\,\rd x
 =-\frac{\mathfrak k}{\pi}
   \sum_{n\geq1}\frac1{n^2}
   \int_0^\infty
   \e^{-\mathfrak k\xi}\mathcal K(\xi/n)\,\rd\xi
 =\mathfrak k\int_0^\infty
   \e^{-\mathfrak k\xi}\mathscr B(\xi)\,\rd\xi\,.
 \end{aligned}
\end{equation*}
This proves the exact identity
in Eq.~\eqref{eq:Borel}.
\end{proof}

As a formal series in $1/\mathfrak k$, the
expansion is Gevrey one, with constants $C,A>0$ such that
$|a_n|\leq CA^n n!$, with vanishing odd
coefficients.
When indexed instead by the powers of
$1/\mathfrak k^2$, the same coefficients
grow as $(2j-1)!$ times an exponential in $j$.
This is Gevrey-two growth, equivalently bounded
by $C'A'^j(j!)^2$.
Ordinary Borel summation in
Eq.~\eqref{eq:Borel} is performed in the
first variable. Dividing the coefficient of
$\mathfrak k^{-2j}$ by $j!$ would not give
a convergent ordinary Borel transform in
$1/\mathfrak k^2$.

Proposition~\ref{prop:Borel} identifies the Borel sums
with the integral-defined constants. Their identification
with the physical Bethe constants remains conjectural,
and Eq.~\eqref{eq:genus-all} describes the expansion of
the perturbative expressions alone.
At fixed $\lambda$, the exact actions are
\begin{equation*}
 t_{\mathrm I}
 =2\pi\sqrt{\lamh_{\mathrm I}+\frac1{2k^2}}\,,
 \qquad
 t_{\mathrm{II}}
 =2\pi\sqrt{\lamh_{\mathrm{II}}+\frac1{4N_f^2}}\,,
\end{equation*}
and hence
\begin{equation}
 q\longrightarrow
 \e^{-2\pi\sqrt{\lamh}}\ne0\,,
 \qquad
 \mathfrak k\to\infty\,,
 \quad\lambda\ \text{fixed}\,,
 \label{eq:fixed-lambda-q}
\end{equation}
where $\lamh$ is the shifted coupling of the
corresponding model.
Terms exponentially suppressed in $\sqrt N$
at fixed parameter are therefore not necessarily
beyond all orders in the fixed-$\lambda$
expansion. Their parameter-dependent prefactors
can contribute to its genus coefficients.
Determining those contributions requires a
generic-parameter completion with estimates
uniform in the simultaneous large-rank,
large-parameter limit.
The special-parameter completions at one and
three do not supply this information.

\section{M-theory localization and fixed-locus contributions}
\label{sec:localization}

\subsection{The sphere partition function and the charge ensemble}
\label{subsec:relative-localization}

Fixing the boundary value of the M-theory three-form
potential $C_3$ in an asymptotically $\mathrm{AdS}_4$
background selects a grand-canonical ensemble for M2-brane
charge~\cite[Sections~1 and~2.1]{Gautason:2025plx}.
For a squashed three-sphere $S_b^3$, with squashing
$b>0$, the relative-localization prescription of
Refs.~\cite{BenettiGenolini:2026cyc,BenettiGenolini:2026kjy}
assigns the perturbative grand potential
\begin{equation*}
 J_{\mathrm{pert}}(\mu)
 =\frac{\mathcal C}{3}\mu^3+\mathcal B\mu+\mathcal A\,.
\end{equation*}
Here $\mu$ is conjugate to Page charge, identified with
$N$ in a specified charge convention. The Airy parameters
$\mathcal C,\mathcal B,\mathcal A$ depend on the squashing
and quotient order, but not on $\mu$.
The localized action determines $\mathcal C$ and
$\mathcal B$, while the prescribed fluctuation determinant
determines $\mathcal A$. With the sphere measure and
contour, passage to fixed rank gives
\begin{equation*}
 \begin{aligned}
 Z^{\mathrm{pert}}_{S_b^3}(N)
 &=\int_{\Gamma_{\mathrm{Ai}}}\frac{\rd\mu}{2\pi\ii}
   \exp\!\left[J_{\mathrm{pert}}(\mu)-N\mu\right]\\
 &=\e^{\mathcal A}\mathcal C^{-1/3}
   \operatorname{Ai}\!\left[
      \mathcal C^{-1/3}(N-\mathcal B)\right]\,.
 \end{aligned}
\end{equation*}
For $\mathcal C>0$, $\Gamma_{\mathrm{Ai}}$ runs from
infinity at $\arg\mu=-\pi/3$ to infinity at
$\arg\mu=\pi/3$. This contour and measure are part of
the sphere prescription.

The small-squashing limit separates the two perturbative
Bethe contributions. The leading Airy exponent depends
on $N-\mathcal B$, whereas $-\mathcal A$ contributes
an additive constant to the free energy. We compare the
coefficient of $b^{-2}$ with $\W^{\mathrm{pert}}/(2\pi)$
for each model, at fixed rank and quotient order.
For the free quotient, the calculation combines an
isolated-orbifold charge with the invariant one-loop
character. For the fixed-locus quotient, the transverse
resolution supplies a localized relative index and a
curvature contribution. Its extension to the enhanced
gauge sector is an additional physical prescription,
specified below.

The common localization data are formulated on the
twelve-dimensional filling $\mathfrak W$ used to define
the M-theory Chern--Simons functional. Choose external
complex weights $b,b^{-1}$ and set
$Q_b=b+b^{-1}$. At zero flavor masses, the five ambient
quadric coordinates have weight $-Q_b/3$ and the
defining relation has weight $-2Q_b/3$.
Subtracting its normal line gives the intrinsic virtual
tangent class, whose total weight, including the external
directions, vanishes. This implements the zero-charge
condition on the preserved Killing spinor and corresponds
to the angular chemical potentials $\Delta=2\pi/3$
of Model I~\cite[Section~6]{BenettiGenolini:2026kjy}.

Write $G_4=\rd C_3$ locally and let $\Phi_G$ be its
equivariant completion. In the common-flux sector,
$\Phi_{G,0}=-\ii\ell_p^3\mu$ at every internal fixed
point, where $\ell_p$ is the eleven-dimensional Planck
length. There are no additional relative flux variables
in this sphere specialization~\cite[Eqs.~(2.15)--(2.19)]{BenettiGenolini:2026kjy}.
At a fixed point $P$, let $e_P$ be the Euler factor in
equivariant localization and let $p_{1,P},p_{2,P}$ be
the Pontryagin weight polynomials of the chosen tangent
class. Set $X_{8,P}=(p_{1,P}^2-4p_{2,P})/192$.
With the normalization of Eq.~(2.6) of that reference,
the smooth or virtual contribution is
\begin{equation}
 \begin{aligned}
 I_{\mathrm{loc}}^{\mathrm{vir}}(\mu)
 &=-\frac{\mu^3}{24\pi^2}\sum_P\frac1{e_P}
   +\mu\sum_P\frac{X_{8,P}}{e_P}\,,\\
 -I_{\mathrm{loc}}^{\mathrm{vir}}(\mu)
 &=\frac{\mathcal C^{\mathrm{vir}}}{3}\mu^3
   +\mathcal B^{\mathrm{vir}}\mu\,,\\
 \mathcal C^{\mathrm{vir}}
 &=\frac1{8\pi^2}\sum_P e_P^{-1}\,,
 \qquad
 \mathcal B^{\mathrm{vir}}
 =-\sum_P\frac{X_{8,P}}{e_P}\,.
 \end{aligned}
 \label{eq:relative-classical-coefficients}
\end{equation}
The cubic term descends from $C_3\wedge G_4\wedge G_4$,
and the linear term from the curvature coupling
$C_3\wedge X_8$. The latter belongs to the corrected
action, not to the separate one-loop determinant.
For a hypersurface, $e_P$ is the ratio of ambient and
normal Euler weights; it need not equal the top Chern
coefficient of its virtual tangent class.
The all-order use of
Eq.~\eqref{eq:relative-classical-coefficients} assumes
the proposed equivariant completion and cancellation
of relative boundary terms. Isolated quotient defects
are not included in this virtual expression.

The one-loop prescription is the relative Dolbeault
index of $T^{\mathrm{vir}}\mathfrak W-
(T^{\mathrm{vir}}\mathfrak W)^*$.
At a smooth fixed point with six additive tangent
weights $w_\alpha$, put $q_\alpha=\e^{-vw_\alpha}$,
where $v>0$ is the character parameter. The contribution
is
\begin{equation}
 \mathcal H(v)
 =\frac{\sum_{\alpha=1}^{6}(q_\alpha^{-1}-q_\alpha)}
        {\prod_{\alpha=1}^{6}(1-q_\alpha)}\,.
 \label{eq:relative-TC-character}
\end{equation}
For a hypersurface, the structure-sheaf character
includes the factor imposing the relation, and its
normal direction is subtracted from the
tangent--cotangent numerator.
The determinant is defined by
$\log Z_{\mathrm{1-loop}}=\zeta'_{\mathcal H}(0)$,
where
\begin{equation*}
 \zeta_{\mathcal H}(s)
 =\frac1{\Gamma(s)}\int_0^\infty
       v^{s-1}\mathcal H(v)\,\rd v
\end{equation*}
is continued from its convergence half-plane.
The parameter $v$ is unrelated to the instanton action
$t$ or the nomes in Section~\ref{sec:modular}.
On the noncompact filling, the localized expression
defines a relative index. Its interpretation as the
physical fluctuation determinant is the proposal of
Ref.~\cite[Section~2]{BenettiGenolini:2026kjy}; the
complete gauge-fixed supersymmetry complex and its
boundary conditions have not been constructed.

The charge sector must be fixed before the inverse
Laplace transform. Page charge is quantized but depends
on the large-gauge convention; Maxwell charge also
contains dissolved flux and curvature contributions
and need not be integral~\cite{Marolf:2000cb,Bergman:2009zh}.
A correction already included in $\mathcal B$ is not
added again to the Page variable $N$.
For a smooth eleven-dimensional spacetime $M_{11}$,
the flux condition is
\begin{equation}
 \left[\frac{G_4}{(2\pi\ell_p)^3}\right]
 -\frac14p_1(TM_{11})
 \in\operatorname{im}\!\left[
 H^4(M_{11},\mathbb Z)\longrightarrow H^4(M_{11},\mathbb R)
 \right]\,.
 \label{eq:global-flux-quantization}
\end{equation}
This is the real-cohomology form of the shifted
quantization law; torsion holonomies require its
differential refinement~\cite{Witten:1996md}.
The extension over a filling and the boundary data
are additional global requirements. The reference
sector of the free quotient and the corresponding
limitations for the transverse resolution enter the
two calculations separately.

\subsection{The classical action and the free-quotient rank shift}
\label{subsec:bulk-shift-I}

For Model I, set $u=-Q_b/3$. The intrinsic quadric
class gives
\begin{equation*}
 e_{\mathrm{vir}}=\frac{bb^{-1}u^5}{2u}
 =\frac{Q_b^4}{162}\,,
 \qquad
 p(z)=(1+b^2z)(1+b^{-2}z)
          \frac{(1+u^2z)^5}{1+4u^2z}\,,
\end{equation*}
where $z$ is a formal variable. The coefficients of
$z$ and $z^2$ are
\begin{equation*}
 p_1=\frac{10Q_b^2}{9}-2\,,\qquad
 p_2=1-\frac{2Q_b^2}{9}+\frac{5Q_b^4}{27}\,,\qquad
 \frac{X_8}{e_{\mathrm{vir}}}=\frac5{12}-\frac3{Q_b^2}\,.
\end{equation*}
Dividing the virtual contribution by the quotient
order gives
\begin{equation}
 \mathcal C_{\mathrm I}^{\mathrm{id}}(b,k)
 =\frac{81}{4\pi^2kQ_b^4}\,,
 \qquad
 \mathcal B_{\mathrm I}^{\mathrm{id}}(b,k)
 =\frac1k\left(-\frac5{12}+\frac3{Q_b^2}\right)\,.
 \label{eq:I-virtual-action}
\end{equation}
At $k=1$, these reproduce the unquotiented coefficients
of Ref.~\cite[Eq.~(6.3)]{BenettiGenolini:2026kjy}.
For $k>1$, the identity-sector expression must be
supplemented by the isolated quotient charge.

The cyclic action is preserved by the smoothing
\begin{equation}
 X_\varepsilon
 =\{z_0^2+A_1B_1+A_2B_2=\varepsilon^2\}\,,
 \qquad \varepsilon\ne0\,.
 \label{eq:I-smoothing}
\end{equation}
Every nonidentity element fixes exactly the two
points $z_0=\pm\varepsilon$, $A_a=B_a=0$.
The derivative with respect to $z_0$ is nonzero
there, so the charged coordinates give two local
$\mathbb C^4/\mathbb Z_k$ neighborhoods with weights
$(1,1,-1,-1)$~\cite[Section~5.4]{Martelli:2009ga}.
In the reference local charge sector, the gravitational
charge of each is $-(k-k^{-1})/24$, including the
noncompact boundary contribution~\cite[Section~3.1]{Bergman:2009zh}.
Their sum is
\begin{equation}
 \Delta N_{\mathrm{orb}}
 =-\frac1{12}\left(k-\frac1k\right)\,.
 \label{eq:I-orbifold-charge}
\end{equation}
Transporting this defect to the cone uses a relative
anomaly-matching prescription with the asymptotic
charge and boundary refinement held fixed.
The smoothing breaks the Reeb action and is not an
equivariant filling for the sphere calculation.
It determines the multiplicity of local defects,
while the regular equivariant contribution remains
Eq.~\eqref{eq:I-virtual-action}.
Appendix~\ref{app:free-quotient-action} supplies the
relative Euler count and explains its scope.

The defect contributes $\mu\Delta N_{\mathrm{orb}}$
to the action, hence
$\Delta\mathcal B=-\Delta N_{\mathrm{orb}}$.
This is the sign of the isolated-orbifold term in
Ref.~\cite[Eq.~(3.3)]{BenettiGenolini:2026kjy}.
The completed classical coefficients are
\begin{equation}
 \begin{aligned}
 \mathcal C_{\mathrm I}(b,k)
 &=\frac{81}{4\pi^2kQ_b^4}\,,\\
 \mathcal B_{\mathrm I}(b,k)
 &=\frac{k}{12}-\frac1{2k}+\frac3{kQ_b^2}\,,
 \qquad
 -\lim_{b\to0^+}\mathcal B_{\mathrm I}(b,k)
 =-\frac{k}{12}+\frac1{2k}=s_{\mathrm I}(k)\,.
 \end{aligned}
 \label{eq:I-classical-coefficients}
\end{equation}
Thus the relative action and isolated-charge prescription
recover the reconstructed shift without Bethe input.
The round-sphere coefficient is instead
$\mathcal B_{\mathrm I}(1,k)=k/12+1/(4k)$;
the Bethe comparison uses the small-squashing limit.

There is also a discrete choice. The free quotient has
$H^4(V^{5,2}/\mathbb Z_k,\mathbb Z)\cong\mathbb Z_{2k}$,
whose torsion sectors are associated with
$\U(N+\ell)_k\times\U(N)_{-k}$, choosing
$\ell=0,\ldots,2k-1$ as representatives~\cite[Section~3.1
and Appendix~A]{Martelli:2009ga}.
In the Stenzel charge conventions of
Ref.~\cite[Section~3.3]{Hashimoto:2011aj}, write
$m_{\mathrm{flux}}$ for the additional four-form flux
integer, denoted $q$ there. The local labels are
\begin{equation}
 \ell_{\mathrm N}=\frac{\ell+km_{\mathrm{flux}}}{2}\,,
 \qquad
 \ell_{\mathrm S}=\frac{\ell-km_{\mathrm{flux}}}{2}\,.
 \label{eq:I-local-flux-labels}
\end{equation}
Their integrality requires $\ell-km_{\mathrm{flux}}\in2\mathbb Z$.
We take $\ell=m_{\mathrm{flux}}=0$, so both local labels
vanish. This specifies the reference sector used in
Eq.~\eqref{eq:I-orbifold-charge}.
In the anomalously shifted convention of
Ref.~\cite[Eqs.~(3.11)--(3.14)]{Hashimoto:2011aj},
the corresponding dimensionless $C$-field periods are
represented by $1/2$ modulo $\mathbb Z$.
Equal ranks therefore do not imply unshifted zero periods.
A global differential $C$-field with the required relative
boundary data remains to be constructed.
Other discrete sectors can change the linear coefficient.

\subsection{The Model-I constant from the localized character}
\label{subsec:bulk-constant}

The Model-I determinant is obtained by projecting the
quadric character of Ref.~\cite[Eq.~(6.5)]{BenettiGenolini:2026kjy}
under the free action. The projection acts on both the
tangent--cotangent numerator and the structure-sheaf
denominator. Its leading small-squashing coefficient
defines
\begin{equation}
 \mathcal A_{\mathrm I}(b,k)
 =-\frac{\widehat g^{\mathrm{bulk}}_0(k)}{b^2}
   +O(|\log b|)\,,\qquad b\to0^+\,.
 \label{eq:bulk-small-b}
\end{equation}
The sign follows from the contribution $-\mathcal A_{\mathrm I}$
to the sphere free energy.

After setting $v=bu$, the finite Molien average reduces to
\begin{equation*}
 \begin{aligned}
 b^2\mathcal H_{k,b}(bu)&\longrightarrow
                 \frac{\mathscr R_k(u)}u\,,\\
 \mathscr R_k(u)
 &=\coth\frac{ku}{6}
   \left(\coth\frac u6+\coth\frac u3-\coth\frac u2\right)
   +\frac{k}{2\sinh^2(ku/6)}\,.
 \end{aligned}
\end{equation*}
Since $\mathscr R_k\to1$ at infinity, its limiting
constant is subtracted before Mellin transformation:
\begin{equation*}
 Z_k(s)=\frac1{\Gamma(s)}\int_0^\infty
 u^{s-2}[\mathscr R_k(u)-1]\,\rd u\,,\qquad \re s>3\,,
 \qquad \widehat g^{\mathrm{bulk}}_0(k)=-Z_k'(0)\,.
\end{equation*}
The derivative is taken after meromorphic continuation.
Appendix~\ref{app:bulk-constant} derives the projection
and proves that the subtracted finite-$b$ Mellin
transforms converge near $s=0$, with an error sufficient
for Eq.~\eqref{eq:bulk-small-b}. This step retains the
slow mode responsible for logarithmic terms but removes
it from the coefficient of $b^{-2}$.

The reduction uses the ABJM sphere constant
\begin{equation}
 \begin{aligned}
 A(\kappa)
 ={}\frac{2\zeta(3)}{\pi^2\kappa}
        \left(1-\frac{\kappa^3}{16}\right)
 +\frac{\kappa^2}{\pi^2}\int_0^\infty
       \frac{x\log(1-\e^{-2x})}{\e^{\kappa x}-1}\,\rd x\,,
 \qquad \kappa>0\,.
 \end{aligned}
 \label{eq:sphere-A}
\end{equation}
This is the normalization of
Ref.~\cite[Eq.~(3.13)]{Hatsuda:2014vsa}.
The plain symbol $A$ denotes this particular function,
distinct from both the sphere parameter $\mathcal A$
and the ABJM Bethe constant $C_{\mathrm B}$ in
Eq.~\eqref{eq:ABJM-constant}.
The heat character
$\mathcal H^{\mathrm B}_{\kappa}(u)
=\coth(\kappa u/4)/(2\sinh^2(u/2))$
has zeta derivative $A(\kappa)$.
Its exponential decay at infinity and odd Laurent
expansion at zero give
$\zeta_{\mathcal H^{\mathrm B}_{\kappa}}(0)=0$,
so positive rescaling of its argument leaves that
derivative unchanged.

Differentiating $\mathscr R_k$ gives the finite
combination of these characters in
Eq.~\eqref{eq:bulk-character-reduction}.
Integration by parts gives
$Z_k(s)=\zeta_{\mathscr R_k'}(s)/(1-s)$,
and the vanishing zeta values fix every scale factor.
The result is
\begin{equation}
 \begin{aligned}
 \widehat g^{\mathrm{bulk}}_0(k)
 ={}&-A(2k/3)+\frac23A(k)+\frac13A(2k)\\
 &+\frac{k}{3}\bigl[A(2/k)+A(4/k)-A(6/k)\bigr]
   -\frac{k^2\zeta(3)}{6\pi^2}\\
 ={}&\frac{\CI(k)}{2\pi}\,.
 \end{aligned}
 \label{eq:bulk-CI}
\end{equation}
The explicit quadratic term uses
$A(2)=-\zeta(3)/(2\pi^2)$.
The equality to $\CI$ is independently established
by substituting the convergent integral identity
\begin{equation*}
 D(a,b)=-\pi^2\left[\frac{A(2b/a)}b+\frac{A(2a/b)}a\right]
       +\frac{\zeta(3)}2\left(\frac a{b^2}+\frac b{a^2}\right)
\end{equation*}
into Eq.~\eqref{eq:constant-I-log-integrals}.
Its derivation in Eq.~\eqref{eq:bulk-D-A} fixes the
normalization without a determinant interpretation.
Equation~\eqref{eq:bulk-CI} is consequently an identity
of functions for every real $k>0$, although the
geometric Molien average uses integer $k$.

At unit parameter, the reduced Mellin transform is
\begin{equation*}
 Z_1(s)=\frac{3^{s-2}}{s-1}
       \bigl[20\zeta(s-2)-2L(s-1,\chi_{-3})\bigr]\,,
\end{equation*}
where
$L(s,\chi_{-3})=\sum_{m\geq1}\chi_{-3}(m)m^{-s}$
for $\re s>1$, continued elsewhere.
The character $\chi_{-3}$ was defined in
Section~\ref{subsec:finite}.
Its functional equation and that of $\zeta$ give
\begin{equation}
 \begin{aligned}
 \widehat g^{\mathrm{bulk}}_0(1)
 &=\frac{20}{9}\zeta'(-2)-\frac29L'(-1,\chi_{-3})\\
 &=-\frac{5\zeta(3)}{9\pi^2}
   -\frac{2}{9\pi}\Cltwo\left(\frac\pi3\right)
 =\frac{\CI(1)}{2\pi}=\frac{\CII(1)}{2\pi}\,.
 \end{aligned}
 \label{eq:bulk-unit-constant}
\end{equation}
Appendix~\ref{app:bulk-constant} derives the Dirichlet
reduction explicitly. This recovers the unit-quotient
result of Ref.~\cite[Eq.~(6.8)]{BenettiGenolini:2026kjy}
and extends its constant comparison to general $k$
through Eq.~\eqref{eq:bulk-CI}.

For $\Nh_{\mathrm I}>0$, the Airy argument tends to
positive infinity as $b\to0^+$.
Its leading free energy is
$\frac23\mathcal C_{\mathrm I}^{-1/2}
(N-\mathcal B_{\mathrm I})^{3/2}$.
Using $b^2Q_b^2\to1$ and
Eqs.~\eqref{eq:I-classical-coefficients} and
\eqref{eq:bulk-small-b} therefore gives
\begin{equation}
 \begin{aligned}
 \lim_{b\to0^+}b^2
 \bigl[-\log Z^{\mathrm{pert}}_{S_b^3}(N,k)\bigr]
 &=\frac{4\pi\sqrt{k}}{27}
   \left(N-\frac{k}{12}+\frac1{2k}\right)^{3/2}
   +\widehat g^{\mathrm{bulk}}_0(k)\\
 &=\frac{\WI^{\mathrm{pert}}(N,k)}{2\pi}\,.
 \end{aligned}
 \label{eq:bulk-leading-normalization}
\end{equation}
The logarithmic Airy and determinant terms vanish
after multiplication by $b^2$.
This limit of the prescribed perturbative sphere
function fixes the factor $2\pi$ in the comparison.
It uses the relative anomaly prescription for the
shift and the proposed determinant for the constant;
it neither equates the finite-squashing observables
nor determines the exponential Bethe sectors.

\subsection{The fixed-locus index from a transverse resolution}
\label{subsec:localized-index}

Write $n=N_f$ for the fixed-locus calculation.
The quotient is described by
$U=A_1^n$, $V=B_1^n$ and $T=A_1B_1$, with
$UV=T^n$ and $T=-z_0^2-A_2B_2$.
Base change of the minimal transverse $A_{n-1}$
resolution gives $n$ hypersurface charts,
\begin{equation}
 \begin{gathered}
 \xi_j\eta_j+z_0^2+A_2B_2=0\,,\\
 U=\xi_j^{j+1}\eta_j^j\,,\qquad
 V=\xi_j^{n-j-1}\eta_j^{n-j}\,,\qquad
 j=0,\ldots,n-1\,.
 \end{gathered}
 \label{eq:ADE-charts}
\end{equation}
The residue form has no exceptional zero or pole,
so this is a crepant partial resolution.
The residual quadric apices are treated with the
intrinsic virtual tangent classes of their charts.
This extends the parent prescription: those classes
are not the pullback of the intrinsic class of the
coarse hypersurface $UV=(-z_0^2-A_2B_2)^n$.
The two choices give different indices, as demonstrated
in Appendix~\ref{app:fixed-locus-character}.

The exceptional contribution can be evaluated as a
rational character before introducing a determinant.
Let $p,r$ be the normal multiplicative weights in
Eq.~\eqref{eq:relative-TC-character}.
Their chart weights are
\begin{equation}
 p_j=p^{n-j}r^{-j}\,,\qquad
 r_j=p^{-(n-j-1)}r^{j+1}\,,\qquad p_jr_j=pr\,.
 \label{eq:ADE-chart-weights}
\end{equation}
The sum of the normal structure-sheaf characters
equals the invariant Molien character. The corresponding
tangent--cotangent sum has the excess
$(n-1)[1-(pr)^{-1}]$.
Both identities follow by telescoping, without choosing
a common series-expansion chamber for the charts.
On the fixed stratum
$\mathfrak F=\mathbb C^2_{\mathrm{ext}}\times S$,
adjunction identifies $(pr)^{-1}$ with the weight of
its canonical, or dualizing, line $K_{\mathfrak F}$.
Thus
\begin{equation}
 \begin{aligned}
 \mathcal H^{\mathrm{res}}_{\mathrm{II},n,b}
 -\mathcal H^{\mathrm{inv}}_{\mathrm{II},n,b}
 &=(n-1)\mathcal H_{\mathrm{loc},b}\,,\\
 \mathcal H_{\mathrm{loc},b}
 &=\operatorname{Ind}^{\mathrm{rel}}_v\!\left[
 \bar\partial_{\mathfrak F}\otimes
 (\cO_{\mathfrak F}-K_{\mathfrak F})\right]\,.
 \end{aligned}
 \label{eq:ADE-Cartan-excess}
\end{equation}
Here ``inv'' projects the parent character, whereas
``res'' sums the intrinsic chart characters.
At the singular point of $S$, the index is evaluated
by its coherent-sheaf hypersurface character.
The factor $n-1$ is the rank of the resolved ADE sector,
and Eq.~\eqref{eq:ADE-Cartan-excess} fixes its index
per Cartan generator without reference to the Bethe
constant.

At zero masses, set
$q_1=\e^{-bv}$, $q_2=\e^{-v/b}$ and
$x=\e^{-Q_bv/3}$, so $q_1q_2=x^3$ and $p=r=x^{-1}$.
The fixed surface has Hilbert series
$H_S(x)=(1+x)/(1-x)^2$.
Converting the negative tangent weights to decaying
exponentials gives
$\operatorname{Ind}^{\mathrm{rel}}_v\bar\partial_{\mathfrak F}
=xH_S(x)/[(1-q_1)(1-q_2)]$ and
$\operatorname{ch}_vK_{\mathfrak F}=x^2$.
The localized character is therefore
\begin{equation}
 \mathcal H_{\mathrm{loc},b}(v)
 =\frac{x(1-x^2)H_S(x)}{(1-q_1)(1-q_2)}
 =\frac{x(1+x)^2}{(1-x)(1-q_1)(1-q_2)}\,.
 \label{eq:localized-finite-b-character}
\end{equation}
The canonical subtraction supplies $1-x^2$.
In particular, the nonzero multiplicative factor
beyond the fixed-surface Hilbert series is determined
by the virtual tangent--cotangent calculation.

When the exceptional curves collapse, wrapped M2-branes
supply the root states of an enhanced $\SU(n)$ gauge
sector~\cite{Anderson:2006pb}.
At an unbroken gauge background and trivial adjoint
holonomy, gauge symmetry requires the same fluctuation
operator for every adjoint generator.
We propose that the relative Cartan index extends to
this sector without an additional boundary or zero-mode
contribution at the contraction:
\begin{equation}
 \begin{aligned}
 \mathcal H^{\mathrm{ADE}}_{\mathrm{II},n,b}
 &=\mathcal H^{\mathrm{inv}}_{\mathrm{II},n,b}
   +(n^2-1)\mathcal H_{\mathrm{loc},b}\\
 &=\mathcal H^{\mathrm{res}}_{\mathrm{II},n,b}
   +n(n-1)\mathcal H_{\mathrm{loc},b}\,.
 \end{aligned}
 \label{eq:ADE-enhancement-prescription}
\end{equation}
The second expression adds the root states without
counting the Cartans twice. No central $\U(1)$ is
added, and the localized contribution vanishes at $n=1$.

The geometrically derived index in
Eq.~\eqref{eq:ADE-Cartan-excess} does not identify a
complete physical supersymmetry complex on
$\mathrm{AdS}_4\times\mathbb{RP}^3$.
The seven-dimensional vector multiplet contains a
gauge field, three real adjoint scalars and a
symplectic-Majorana gaugino; gauge fixing adds ghosts
and auxiliary fields~\cite{Anderson:2006pb}.
Their pairing must reproduce
Eq.~\eqref{eq:localized-finite-b-character} with a
supersymmetry-compatible spin structure, trivial
adjoint flat connection, and boundary conditions
realizing a global $\SU(n)$ symmetry in three dimensions.
The localization complex on compact seven-dimensional
Sasaki--Einstein spaces uses a different boundary
problem~\cite[Sections~3 and~4]{Polydorou:2017jha}.
Neither that calculation nor the rational index identity
establishes Eq.~\eqref{eq:ADE-enhancement-prescription}
for the physical spectrum.

\subsection{The localized curvature term and shifted rank}
\label{subsec:bulk-shift-II}

The same transverse geometry determines a classical
curvature contribution independently of the enhanced
one-loop index. We evaluate
Eq.~\eqref{eq:relative-classical-coefficients} with the
same $\mu$ at all $n$ charts and no additional
exceptional-flux parameters, using their intrinsic
virtual classes. Put
$\epsilon_\pm=-Q_b/3\pm\delta$ and
$\sigma=\epsilon_++\epsilon_-=-2Q_b/3$.
The chart weights are
\begin{equation}
 \omega_j^+=(n-j)\epsilon_+-j\epsilon_-\,,\qquad
 \omega_j^-=-(n-j-1)\epsilon_++(j+1)\epsilon_-\,.
 \label{eq:II-action-weights}
\end{equation}
Their sum is $\sigma$, and
$\sum_j(\omega_j^+\omega_j^-)^{-1}
=(n\epsilon_+\epsilon_-)^{-1}$.
The regulator $\delta$ is removed only after summation;
at even $n$, individual fixed-point terms are singular
at $\delta=0$.

Let $e_{\mathfrak F}$ and $p_{1,\mathfrak F}$ be the
Euler factor and first Pontryagin polynomial of the
fixed stratum, including its hypersurface relation.
At zero masses,
\begin{equation*}
 e_{\mathfrak F}=\frac{Q_b^2}{18}\,,\qquad
 p_{1,\mathfrak F}=b^2+b^{-2}-\frac{Q_b^2}{9}
                 =\frac{8Q_b^2}{9}-2\,.
\end{equation*}
The full chart Euler factor is
$e_j=e_{\mathfrak F}\omega_j^+\omega_j^-$.
After removing the regulator, the curvature sum differs
from $1/n$ times the unquotiented result by
\begin{equation}
 \sum_{j=0}^{n-1}\frac{X_{8,j}}{e_j}
 -\frac1n\left(\frac5{12}-\frac3{Q_b^2}\right)
 =\frac{n-n^{-1}}{48e_{\mathfrak F}}
      (p_{1,\mathfrak F}-\sigma^2)
 =(n-n^{-1})\left(\frac16-\frac{3}{4Q_b^2}\right)\,.
 \label{eq:II-curvature-excess}
\end{equation}
The derivation in Appendix~\ref{app:fixed-locus-character}
separates the Pontryagin expression into a term
proportional to the inverse normal Euler weight and
a term independent of the chart. Their sums determine
both the factor $n-n^{-1}$ and the normal-bundle
subtraction $-\sigma^2$.

Combining this excess with
$\mathcal B_{\mathrm I}(b,1)=-5/12+3/Q_b^2$ gives
\begin{equation}
 \begin{aligned}
 \mathcal C^{\mathrm{res}}_{\mathrm{II}}(b,n)
 &=\frac{81}{4\pi^2nQ_b^4}\,,\\
 \mathcal B^{\mathrm{res}}_{\mathrm{II}}(b,n)
 &=-\frac n6-\frac1{4n}+\frac{3(n^2+3)}{4nQ_b^2}\,,
 \qquad
 -\lim_{b\to0^+}\mathcal B^{\mathrm{res}}_{\mathrm{II}}(b,n)
 =\frac n6+\frac1{4n}=s_{\mathrm{II}}(n)\,.
 \end{aligned}
 \label{eq:II-localized-Airy}
\end{equation}
The cubic coefficient retains the quotient-volume
factor $1/n$. Both finite-squashing coefficients agree
with the independent Model-II Airy proposal of
Ref.~\cite[Eq.~(2.3) and Table~1]{Bobev:2025ltz}.
This agreement follows from the specified characteristic
classes, not from adjusting a localized contribution
to the Bethe shift.
A bulk interpretation requires a supersymmetric
relative filling and globally quantized $C$-field
realizing the common local flux data, with no further
boundary or localized term changing $\mathcal B$.
The present calculation evaluates the curvature
push-forward but does not establish those global
conditions or localized anomaly cancellation.

\subsection{The Model-II constant}
\label{subsec:bulk-constant-II}

For the action $(1,0,-1,0,0)$ on
$(A_1,A_2,B_1,B_2,z_0)$, the invariant character has
the small-squashing limit
\begin{equation}
 \begin{aligned}
 b^2\mathcal H^{\mathrm{inv}}_{\mathrm{II},n,b}(bu)
 &\longrightarrow\frac{\mathscr R^{\mathrm{inv}}_n(u)}u\,,
 &\mathscr R^{\mathrm{inv}}_n(u)
 &=f(u)\coth\frac{nu}{6}+g(u)\,,\\
 f(u)&=\frac32\coth\frac u6-\frac12\coth\frac u2\,,
 &g(u)&=\frac{x(1+x)^2}{(1-x)(1-x^3)}\,,\qquad x=\e^{-u/3}\,.
 \end{aligned}
 \label{eq:II-invariant-character}
\end{equation}
The Molien reduction includes the charged numerator
and is derived in Appendix~\ref{app:fixed-locus-character}.
Equation~\eqref{eq:localized-finite-b-character} gives
$b^2\mathcal H_{\mathrm{loc},b}(bu)\to g(u)/u$,
so the enhanced prescription becomes
\begin{equation}
 \mathscr R^{\mathrm{ADE}}_n(u)
 =f(u)\coth\frac{nu}{6}+n^2g(u)\,.
 \label{eq:II-extended-character}
\end{equation}
Both reduced characters tend to one at infinity.
The subtracted Mellin argument of
Appendix~\ref{app:bulk-constant} applies at each fixed
$n$. For either reduced character, set
\begin{equation*}
 Z_R(s)=\frac1{\Gamma(s)}\int_0^\infty
        u^{s-2}[R(u)-1]\,\rd u\,,\qquad \re s>3\,,
 \qquad \widehat g_0[R]=-Z_R'(0)\,.
\end{equation*}
The localized term is evaluated by its Dirichlet expansion,
\begin{equation}
 \begin{aligned}
 g(u)&=\frac43\sum_{m\geq1}m\e^{-mu/3}
        -\frac13\sum_{m\geq1}\chi_{-3}(m)\e^{-mu/3}\,,\\
 \alpha_{\mathrm{loc}}
 &:=\zeta'_{g'}(0)
   =\frac19\left[\frac{\zeta(3)}{\pi^2}
                  +\frac{\Cltwo(\pi/3)}{\pi}\right] >0\,.
 \end{aligned}
 \label{eq:localized-constant}
\end{equation}
Here $g'=\rd g/\rd u$.
Since $Z_R(s)=\zeta_{R'}(s)/(1-s)$ and
$\zeta_{R'}(0)=0$, a term $+g$ in $R$ contributes
$-\alpha_{\mathrm{loc}}$ to $\widehat g_0$.
Differentiating the remaining product in
Eq.~\eqref{eq:II-extended-character} and reducing it
to the sphere characters gives
\begin{equation}
 \begin{aligned}
 \widehat g^{\mathrm{ADE}}_{0,\mathrm{II}}(n)
 &=\frac12\bigl[A(2n)-A(2n/3)\bigr]
   +\frac n6\bigl[3A(2/n)-A(6/n)\bigr]
   -n^2\alpha_{\mathrm{loc}}\\
 &=\frac{\CII(n)}{2\pi}\,.
 \end{aligned}
 \label{eq:bulk-CII}
\end{equation}
The derivative decomposition is
Eq.~\eqref{eq:II-character-derivative}.
As for Model I, the second equality follows separately
from the integral-exchange identity
\eqref{eq:bulk-D-A}, now applied to
Eq.~\eqref{eq:constant-II-log-integrals}.
It holds for every real $n>0$ as an identity of functions.
The cyclic quotient and enhanced gauge sector use
integer $n$.

The invariant character alone instead gives
\begin{equation}
 \frac{\CII(n)}{2\pi}
 -\widehat g^{\mathrm{inv}}_{0,\mathrm{II}}(n)
 =-(n^2-1)\alpha_{\mathrm{loc}}\,.
 \label{eq:II-invariant-defect}
\end{equation}
The additional contribution needed at $n>1$ is thus
fixed by the derived per-Cartan index followed by the
enhancement prescription, not by solving
Eq.~\eqref{eq:II-invariant-defect} for an unknown
character. There is an independent check on
$\mathbb C^2/\mathbb Z_n\times\mathbb C^2$ at the
round-sphere superconformal point. The same geometric
index gives $\mathcal H^{\mathrm{ADHM}}_{\mathrm{loc}}
=\mathcal H^{\mathrm B}_1/2$, while the invariant
character is $(\mathcal H^{\mathrm B}_n+
\mathcal H^{\mathrm B}_1)/2$. The enhanced determinant
therefore yields
\begin{equation*}
 \mathcal A_{\mathrm{ADHM}}(n)
 =\frac12A(n)+\frac{n^2}{2}A(1)\,,
\end{equation*}
reproducing the independently proposed and tested
sphere constant in Ref.~\cite[Eq.~(3.12)]{Hatsuda:2014vsa}.
Appendix~\ref{app:fixed-locus-character} gives the
character calculation. A complementary one-loop
spectral calculation on $\mathrm{AdS}_4\times S^7/\mathbb Z_n$
reproduces the same round-sphere ADHM constant,
including the contribution of the localized seven-dimensional vector
multiplet~\cite[Appendix~H]{Kurlyand:2026cyz}.

Combining the resolved action with the enhanced
character and the sphere inverse-Laplace prescription
gives
\begin{equation}
 \begin{aligned}
 \lim_{b\to0^+}b^2
 \bigl[-\log Z^{\mathrm{res,ADE}}_{S_b^3}(N,n)\bigr]
 &=\frac{4\pi\sqrt n}{27}
    \left(N+\frac n6+\frac1{4n}\right)^{3/2}
   +\widehat g^{\mathrm{ADE}}_{0,\mathrm{II}}(n)\\
 &=\frac{\WII^{\mathrm{pert}}(N,n)}{2\pi}\,.
 \end{aligned}
 \label{eq:II-small-b-normalization}
\end{equation}
Here $Z^{\mathrm{res,ADE}}$ denotes the Airy expression
with $\mathcal C^{\mathrm{res}}_{\mathrm{II}}$,
$\mathcal B^{\mathrm{res}}_{\mathrm{II}}$ and
$\mathcal A^{\mathrm{ADE}}_{\mathrm{II}}
=\zeta'_{\mathcal H^{\mathrm{ADE}}_{\mathrm{II},n,b}}(0)$.
Its small-$b$ limit exists at every fixed $N,n\geq1$
by the same large-argument Airy calculation as
Eq.~\eqref{eq:bulk-leading-normalization}.
The prescribed classical and one-loop contributions
therefore reproduce both perturbative Bethe inputs.
Their physical identification requires the relative
bulk action, the localized-sector extension and the
charge and boundary data specified above. None of
these perturbative calculations derives the divisor
laws or a bulk realization of their Eisenstein structure.

\section{Discussion}
\label{sec:discussion}

The two $V^{5,2}$ quotients separate the geometric data
that determine the leading holographic answer from those
that govern its quantum corrections. At equal nontrivial
quotient order, their volumes and leading Bethe critical values
agree, although only Model II has a gauge sector localized
on the link. The reconstructed critical values distinguish
them through the shifted ranks, constants and exponential
amplitudes. At quotient order three, matching the shifted
powers and instanton actions by a rank translation still
leaves a difference in the leading exponential amplitude.
Thus the finite-rank comparison probes information beyond
both the volume and the leading instanton action.

The perturbative terms acquire a geometric interpretation
through the small-squashing limit of the sphere partition
function. Fixing the M-theory three-form potential places
the bulk calculation in the ensemble conjugate to
M2-brane Page charge~\cite{Gautason:2025plx}. Within the
relative-localization prescription of
Ref.~\cite{BenettiGenolini:2026kjy}, the cubic and linear
terms of the action generate the shifted power after
conversion to fixed rank, while the one-loop determinant
supplies its additive constant. For Model I, the virtual
curvature contribution and the isolated-orbifold charge
defect account for the rank shift, and the projected
character reduces to the independently reconstructed
constant integral. Model II requires a different
curvature contribution and additional localized fields.
The agreement in both cases concerns the full coefficient
of $b^{-2}$ in the prescribed perturbative free energy,
not an identification of the sphere and Bethe observables
at finite squashing.

For $N_f>1$, the relative tangent--cotangent index on the
transverse $A_{N_f-1}$ resolution differs from the invariant
parent character by $N_f-1$ copies of the relative Dolbeault
index of $\cO_{\mathfrak F}-K_{\mathfrak F}$ on the fixed
stratum $\mathfrak F$. This contribution is determined by
the exceptional geometry before any determinant is evaluated.
Extending it from the Cartan multiplets to the enhanced
$\SU(N_f)$ sector gives the small-squashing coefficient
$\CII/(2\pi)$, whereas the invariant projection alone
leaves the nonzero discrepancy in
Eq.~\eqref{eq:II-invariant-defect}.
At the round-sphere point ($b=1$) of the ADHM orbifold,
the same extension reproduces the independent constant
proposal~\cite[Eq.~(3.12)]{Hatsuda:2014vsa}.
The extension includes localized contributions absent
from the invariant projection of the parent fields.
The curvature calculation uses the same resolution but
does not require this nonabelian extension, so the shift
and constant give distinct tests of the bulk description.

The round-sphere coefficient
$\mathcal B_{\mathrm I}(1,k)=k/12+1/(4k)$ also fixes the
coefficient left unspecified in the Model-I Wilson-loop
Airy ratio of Ref.~\cite[Eq.~(4.49)]{Gautason:2025bft}.
For matched quotient orders greater than two, the selected
Wilson-loop M2-branes have the same classical action and
one-loop determinant in the two geometries. The different
sphere Airy coefficients give distinct perturbative rank
dependence for the corresponding expectation values.
This comparison uses the proposed Wilson-loop completion
and is a further prediction of the localized action.

For the localized index to describe physical fluctuations,
the supersymmetry complex on
$\mathrm{AdS}_4\times\mathbb{RP}^3$ must reproduce it
from the gauge, scalar, gaugino and ghost fields, with
boundary conditions that realize the flavor symmetry
of Model II. The enhancement assumes that
contraction of the exceptional curves introduces no
unpaired boundary or zero-mode contribution beyond the
added root multiplets. For both quotients, the local flux
data must also extend to a globally quantized differential
$C$-field with the prescribed boundary data. The smoothing
in the free-quotient charge calculation is used for relative
anomaly matching and does not replace the equivariant
filling. The shifted holonomies of the equal-rank Model-I
sector~\cite{Hashimoto:2011aj} illustrate why gauge
ranks alone do not specify that extension.

These perturbative calculations leave the arithmetic of
the exponential sectors unexplained. The reconstructed
completions for $V^{5,2}$ show that the weight-four
organization found in ABJM and ADHM is not confined to
their enhanced supersymmetry or flat-space quotient cones.
The two $V^{5,2}$ theories share the unit-parameter function
but select different divisor laws at quotient order three.
The recurring structure is a restriction on the
coefficient laws, not a universal sequence of amplitudes.
The instanton nome is assembled from the rank and coupling;
it is not the complex structure of a microscopic torus.
The translation defect in Eq.~\eqref{eq:correction-monodromy}
excludes a scalar modular transformation law for the
completed correction on the groups considered here.
A microscopic explanation must account for the arithmetic
without assuming such a symmetry of the observable.

A derivation from the Bethe equations must establish
both the $t+1/m$ prefactors and the restriction of the
Fourier generators to the span of
$E_4(Q)$, $E_4(Q^3)$ and $E_4(Q^9)$. Within this span,
three exact Fourier coefficients determine the generator;
the prescribed primitive and first-order operator then
determine every sector.
The finite-rank comparisons test this continuation far
beyond the fitted orders, but do not establish these
microscopic restrictions. Holomorphic-block difference
equations offer possible additional constraints, since
they follow from supersymmetric line-operator
identities~\cite[Section~2.4]{Beem:2012mb}.
They act on flavor fugacities with deformation parameter
$\mathfrak q$, however, and do not directly constrain
the dependence on the rank-dependent nome $q$.
Extracting such a constraint requires following the
selected Bethe vacuum as the rank changes.

At parameter two, the unresolved problem already includes
the action dependence within each exponential sector.
The reconstructed Laurent prefactors could terminate or
could belong to longer inverse-action expansions. An
additional inverse power multiplying $\e^{-mt}$ would
ultimately dominate the next exponential sector, so
specifying instanton actions alone cannot resolve this
question. In the physical type-IIA limit,
$\lambda=N/\mathfrak k$ is held fixed and the nome tends
to $\e^{-2\pi\sqrt{\lamh}}\ne0$, so these sectors can
contribute to ordinary genus coefficients.
Borel summation recovers the constant integral from
its inverse-parameter expansion but supplies no missing
rank-dependent exponential sectors.
A complete fixed-$\lambda$ expansion requires the
exponential prefactors at general parameter and estimates
uniform in the joint limit, which the special-parameter
completions do not determine.

The reconstructed sectors contribute to the imaginary
part of the selected semiclassical block action.
Equation~\eqref{eq:intro-block-asymptotics} places the
critical twisted superpotential in the coefficient
of $\hbar^{-1}$, so corrections exponentially
suppressed at large $N$ still modify the leading
block exponent as $\hbar\to0$ at fixed rank.
Recovering the superconformal index from this action
also requires the block fluctuations and the gluing
sum over vacua. These additional data can carry
nonperturbative information of their own, as the
ABJM topologically twisted index demonstrates:
its Hessian, effective-dilaton and flux factors
produce arithmetic beyond that of the critical
value~\cite{Hosseini:2026lvj}.
A bulk instanton calculation must therefore isolate
the contribution to the same selected exponent.
When the amplitude is computed at fixed M2-brane
chemical potential, this comparison also requires
conversion to fixed rank with the measure and
contour appropriate to the
observable~\cite{Bobev:2026gir}.

Wrapped Euclidean M2-branes could account for the
reconstructed exponential amplitudes. For suitable
Reeb-invariant three-cycles in Sasaki--Einstein
manifolds, equivariant fluctuation complexes organize
the one-loop calculation~\cite{Kurlyand:2026yke},
while M2-brane localization has reproduced an ABJM
instanton contribution in the grand-canonical
ensemble~\cite{Gautason:2025per}.
Applying these methods requires identifying which
supersymmetric configurations contribute to the
chosen observable. Fermionic zero modes are absent
from the quadratic action, so integration over
them vanishes unless interactions or operator
insertions supply the corresponding Grassmann
factors. A nonzero determinant of the remaining
fluctuations therefore does not by itself establish
a nonzero instanton amplitude.
Once these zero-mode integrals are understood,
the higher exponential orders require an analysis
of possible multiple covers and configurations
of several branes, whose contributions are not
determined by the single-brane calculation.
For the two $V^{5,2}$ quotients, the resulting
amplitudes must then be converted to the
fixed-rank small-squashing exponent and compared
with the reconstructed coefficients, including
their signs, divisibility dependence and
action-dependent prefactors.
The localized relative indices and reconstructed
Bethe amplitudes furnish quantitative tests for
extending M-theory localization to ADE gauge
sectors and wrapped M2-branes on non-toric
backgrounds.

\section*{Acknowledgments}
This research was supported by UK Research and Innovation (UKRI) under the UK government's Horizon Europe funding guarantee (Grant No.~EP/Y027604/1).

\textbf{Statement on use of AI.}
The ideas, arguments, and writing in this paper came from the author. An OpenAI model provided useful assistance with code development, computational checks, and copy editing.

\section*{Data and code availability}
The data and code supporting the findings of this manuscript are not publicly available and may be obtained from the author upon reasonable request.

\appendix
\section{A logarithmic integral at integer arguments}\label{app:clausen}

For positive integers $a,b$, consider the holomorphic function
\begin{equation}
 F(z)=\int_0^z\frac{\log(1-w^a)\log(1-w^b)}w\,\rd w
 =\sum_{r,s\geq1}\frac{z^{ar+bs}}{rs(ar+bs)}\,.
 \label{eq:F-disk}
\end{equation}
The sum is absolutely and uniformly convergent on the closed unit disk:
its sum of positive coefficients is $D(a,b)<\infty$.
It follows that $\re F(1)=D(a,b)$ and the boundary mean of $F$ is
zero. For $0<\vartheta<2\pi$, away from the finitely many zeros, write
\begin{equation}
 \log(1-\e^{\ii a\vartheta})=\Lambda_a(\vartheta)
                              +\ii\alpha_a(\vartheta)\,,
 \quad
 \Lambda_a=\log|1-\e^{\ii a\vartheta}|\,,
 \quad
 \alpha_a=\frac{a\vartheta}{2}-\pi\left\lfloor\frac{a\vartheta}{2\pi}\right\rfloor
                            -\frac\pi2\,.
 \label{eq:boundary-log}
\end{equation}
Thus $(\re F)'=-\alpha_a\Lambda_b-\alpha_b\Lambda_a$. Integrating this
identity against $(2\pi-\vartheta)/(2\pi)$ and using the zero boundary mean
gives
\begin{equation}
 D(a,b)=J(a,b)+J(b,a)\,,\qquad
 J(a,b)=\frac1{2\pi}\int_0^{2\pi}
        (2\pi-\vartheta)\alpha_a(\vartheta)\Lambda_b(\vartheta)\,\rd \vartheta\,.
 \label{eq:J}
\end{equation}
The logarithmic singularities are integrable, so the interval can be split
at their locations before integrating by parts.

Let $f_a(\vartheta)=(2\pi-\vartheta)\alpha_a(\vartheta)$,
$c_b(\vartheta)=\Cltwo(b\vartheta)$ and
$r_b(\vartheta)=\re\Li_3(\e^{\ii b\vartheta})$.
Then $c_b'=-b\Lambda_b$ and $r_b'=-bc_b$. On each continuity interval,
$f_a''=-a$. At $\vartheta_j=2\pi j/a$, $1\leq j<a$, the jumps are
\begin{equation}
 [f_a]_j=-\pi(2\pi-\vartheta_j)\,,\qquad [f_a']_j=\pi\,.
 \label{eq:jumps}
\end{equation}
The endpoint derivatives obey $f_a'(0^+)=a\pi+\pi/2$ and
$f_a'(2\pi^-)=-\pi/2$. Two integrations by parts, including these endpoints
and all the jumps, therefore give
\begin{equation}
 J(a,b)=\frac\pi{ab}\sum_{j=1}^{a-1}j\Cltwo\left(\frac{2\pi bj}{a}\right)
 +\frac{a\zeta(3)}{2b^2}
 +\frac1{2b^2}\sum_{j=0}^{a-1}
             \re\Li_3(\e^{2\pi\ii bj/a})\,.
 \label{eq:J-reduced}
\end{equation}
Here the constant-weight Clausen sum vanishes. The root-of-unity identities
\begin{equation}
 \sum_{j=0}^{a-1}\Cltwo(2\pi bj/a)=0\,,
 \qquad
 \sum_{j=0}^{a-1}\re\Li_3(\e^{2\pi\ii bj/a})
                       =\frac{\gcd(a,b)^3}{a^2}\zeta(3)\,,
 \label{eq:Clausen-filter}
\end{equation}
follow by retaining Fourier indices divisible by $a/\gcd(a,b)$.
Substituting into~\eqref{eq:J-reduced} and symmetrizing proves
\eqref{eq:D-finite}. This proof uses boundary limits of convergent
polylogarithm series and keeps the sawtooth discontinuities explicitly.

\section{Numerical implementation}\label{app:numerics}

\subsection{Continuation and precision controls}

The root calculation uses analytic reduced Jacobians and midpoint Newton
steps in arbitrary-precision real and complex ball arithmetic to locate a
candidate, followed by a separate enclosure test for the nearby zero. The precision
schedule in Section~\ref{subsec:numerics} ends at $920$ working decimal digits
for $P=800$ and at $320$ digits for $P=200$. At each intermediate precision the
Newton iteration is stopped when the largest reduced residual is below
$10^{-(d-12)}$, before increasing $d$ and refining the same solution.

At fixed $k$, the Model-I initialization at $N=2$ uses the continuum uniform
profile, with $y=\sqrt N\,\xi$,
$|\xi|\leq2\pi/(3\sqrt k)$ and $p=-\sqrt k\,\xi/2$.
The Model-II initialization uses quantiles of the triangular continuum density
with endpoint $|y|=2\pi\sqrt{N/N_f}/3$. For each subsequent rank, the ordered
solution at $N-1$ is linearly interpolated at the new rank quantiles. The
imaginary coordinates are multiplied by $\sqrt{N/(N-1)}$. Reflection is
imposed exactly in the reduced variables. In Model I the common parity center
is removed during interpolation, and the real coordinates are continued in
the strip of~\eqref{eq:symmetry-I}. The reduced Jacobian includes the
dependence of the reflected coordinates on the solved half of the roots.

Backtracking rejects steps that violate ordering or the Model-I real-profile
sign and strip conditions. A further Model-I diagnostic compares the largest
predictor correction with the minimum imaginary-root spacing. A ratio at
most five is used as an empirical continuity check, not as an existence or
uniqueness criterion. Reverse continuation from rank $N+1$ was also performed
at representative ranks and levels $k=1,2,3,10,30$, including ranks across the
change of precision target. These runs requested sixty additional decimal
digits. A common local enclosure for the forward and reverse candidates
checks that they determine the same nearby zero. Each fixed integer
parameter has its own ascending-rank sequence.

The final decimal roots are imported again for the residual and enclosure
checks. For Model I, verification at $P+150$ digits uses the sine-logarithm
kernel~\eqref{eq:sine-kernel} and evaluates the full $2N$ physical gradients.
For Model II, verification at $P+130$ digits uses the real arctangent
equations~\eqref{eq:real-II}. These evaluations differ from the stable
decaying-exponential expressions used during iteration. The multiplicative
equations~\eqref{eq:exp-I} and~\eqref{eq:exp-II} are checked separately.
The residual bounds include the rounding of the serialized centers.

\subsection{Root neighborhoods and critical-value radii}

In Model I, let $g$ be the real system comprising the real and imaginary
parts of~\eqref{eq:reduced-I} on the positive half. Its dimension is
$2\lfloor N/2\rfloor$. Write $J(x)=\partial g/\partial x$ for its
Jacobian. Let $x_0$ be the exported center and $C$ a binary
midpoint approximation to $J(x_0)^{-1}$. For the cube
$X=\{x:\|x-x_0\|_\infty\leq r\}$, the enclosure calculation bounds
\begin{equation}
 \eta\geq\|Cg(x_0)\|_\infty\,,\qquad
 \rho\geq\sup_{x\in X}\|I-CJ(x)\|_\infty\,.
 \label{eq:contraction-data}
\end{equation}
The radius is chosen as $r=4\eta$, with a positive fallback when the computed
$\eta$ vanishes. The entire cube must remain in the analytic strip and obey
\begin{equation}
 \rho<1\,,\qquad \eta+\rho r<r\,.
 \label{eq:contraction-test}
\end{equation}
The map $x\mapsto x-Cg(x)$ then maps $X$ strictly into itself and is a
contraction. Since $CJ$ is invertible under the first bound, its fixed point
is the unique zero in this cube. A radius $r$ in each real coordinate gives
a complex-root radius at most $\sqrt2r$ before the reflection and parity
center are restored.

For Model II, a direct enclosure uses monotonicity. Suppose the exported
center $\widehat y$ obeys $\max_i|B_i(\widehat y)|\leq\varepsilon$ and
$\max_i|\widehat y_i|\leq M$. Set
\begin{equation}
 m_0=\frac{N_f\sqrt3/2}{\cosh(M+1)-1/2}\,,\qquad
 \delta=\frac{2\varepsilon}{m_0}<1\,.
 \label{eq:monotone-radius}
\end{equation}
On the positive $i$th face of the cube of radius $\delta$, all differences
$y_i-y_j$ are at least their center values. The pair forces are monotone,
while the fundamental force increases by at least $m_0\delta$. Thus
$B_i\geq\varepsilon$ there; on the negative face $B_i\leq-\varepsilon$.
For $\varepsilon>0$, the outward signs of the gradient
exclude a boundary minimum of $\mathcal Y_{N_f}$ on the
closed cube. Its minimum is interior and is therefore a
zero of the full real system. Theorem~\ref{thm:branch}
identifies it uniquely. An exactly vanishing residual
already specifies the solution. This estimate is independent of the Newton inverse and
uses the full real system.

Accepted neighborhoods must preserve strict ordering. Absolute and relative
root radii are recorded, with relative radius bounded by
$r/(|a|-r)$ for a center $a$ when $|a|>r$. An odd-rank central zero is excluded
from a relative-error requirement. Both logarithmic and multiplicative
residual bounds must meet the target $10^{-P}$. The root neighborhoods are
then propagated through the stable critical-value formulas at $P+150$
digits. The decimal scalar center $\bar w$ and its exported absolute radius $r$
include root uncertainty, special-function evaluation and decimal rounding.
The fits below use $\bar w$, while enclosure comparisons also use $r$.

\subsection{Parent and exponential fits}

For Model I, Eq.~\eqref{eq:shift-fit} is solved for
$s_{\mathrm{eff}}(N,k)$ at $N=50,\ldots,299$, using
secant starting values $(0,1)$.
The estimates at $(N,k)=(299,1)$ and $(299,2)$ determine
the two coefficients within the assumed form
$s=\alpha k+\beta/k$.
Rational reconstruction uses denominator bound
$1000$. Free fits of the leading coefficient, shift and constant use the
triples $(100,150,200)$, $(150,200,250)$ and $(200,250,300)$, at $180$ and
$220$ working digits. The retrospective integral-kernel reconstruction fixes its three
coefficients from the $N=300$ constants at $k=1,2,3$, at $140$ digits. This
is a reconstruction within the specified three-dimensional kernel family.
It is not a uniqueness statement among all convergent integrals.

For Model II, the shifted rank is taken from
\cite{Bobev:2023lkx,Bobev:2024mqw}. Constant estimates are formed at
$N=50,100,150,200,250,300$. Positive-axis quadrature at $75$ and $110$
working digits is compared with the finite Clausen evaluations. Every
constant estimate extracted at finite rank retains an exponentially small
bias. The integral functions used in subsequent subtractions are evaluated
directly, without replacing them by a rounded fitted plateau.

The sector fits use $850$ working decimal digits and QR least
squares~\cite[Section~5.3]{GolubVanLoan:2013}.
Let $q_0$ be the nome at the lowest training rank. After dividing the
normalized remainder $2\pi\W^{\mathrm{np}}/\mathfrak k^2$ by $q_N$,
the columns for sector $m$ are
\begin{equation}
 t_N(q_N/q_0)^{m-1}\,,\qquad (q_N/q_0)^{m-1}\,.
 \label{eq:QR-columns}
\end{equation}
The solved coefficients are divided by $q_0^{m-1}$ to recover $A_m,B_m$ in
$\sum_m(A_mt+B_m)q^m$. Both are fitted freely. Dividing the data and model by
$q_N$ weights the squared residuals of
Eq.~\eqref{eq:free-fit} by $q_N^{-2}$. The subsequent
column normalization leaves this weighted least-squares
problem unchanged in exact arithmetic.

For Model I the two final reconstructions use every rank in $10\leq N\leq54$
and $56\leq N\leq100$, with sixteen sectors in each fit. For Model II the
fits use even ranks $20\leq N\leq100$ with twelve sectors and even ranks
$10\leq N\leq100$ with sixteen sectors. Rational approximants have denominator
at most $10^5$ and are formed separately for $A_m$ and $B_m$.
In Model I the reconstruction is reported through twelve sectors. The
rational candidates must agree across the disjoint windows; their numerical
deviations and the relation $mB_m=A_m$ are then recorded and checked.
The reporting cutoff does not assert failure of all higher fitted sectors.
In Model II a pair is accepted only if both rational candidates agree across
the windows, satisfy $mB_m=A_m$, and differ from the fitted values by less
than $10^{-8}$. Eleven sectors pass this rule. Model I does not impose this
additional numerical-deviation cutoff. Neither procedure gives an interval
error bound on a physical all-order coefficient. The reported ranges are
all displayed in
Table~\ref{tab:coefficients}. The divisor ansatz extrapolates these finite
sequences and is subsequently tested against the critical values, including
ranks outside the reconstruction windows.

The parameter-two calculations use a different fit family at $500$ digits.
For Model I, after subtracting $\frac43(t+2)q$, the second sector is fitted
to $q^2t^j$ for $-1\leq j\leq3$, together with nuisance third-sector columns
$q^3t^j$ for $-3\leq j\leq5$. The training ranks are even integers in
$[100,198]$ and $[200,298]$, and the odd ranks between them are retained for
prediction. The variables $q$ and $t$ are rescaled at the first rank of each
window. Candidate second-sector coefficients are tested against the basis
$(1,\pi^{-2})$ by the PSLQ integer-relation algorithm~\cite{ferguson1999analysis}
with tolerance $10^{-25}$,
coefficient bound $10^4$ and at most $1000$ steps. The nuisance third-sector
coefficients are not interpreted as a reconstructed physical sector.

For Model II at $N_f=2$, the reconstructed leading term is subtracted, the second
sector is fitted in the basis $(t,1,t^{-1})$, and the third in
$(1,t^{-1},t^{-2},t^{-3})$ after subtraction of the proposed second sector.
The training ranks are $200,210,\ldots,300$, and the prediction ranks are
$201,211,\ldots,291$. The fitted second-sector coefficients are compared with
$(-23/3,-19/6,\pi^2)$. For the third-sector coefficients,
the script tests rational multiples of $\pi^j$,
$j=0,\ldots,4$, with denominator at most $10^4$. No infinite parameter-two completion is inferred from these fits.

\subsection{Infinite-series comparison}

For the special-parameter series, the interval calculation uses $1000$
working decimal digits through $N=100$ and $400$ above that rank. Given the
scalar radius $r$, the target tail is $10^{-30}r$. An initial cutoff
$\max(1,\lceil-\log(10^{-30}r)/t\rceil+5)$ is estimated in machine arithmetic;
the cutoff is then increased until the outward-rounded analytic tail bound
in~\eqref{eq:tail} is below the target. Machine arithmetic selects only the
initial trial cutoff. It does not determine the accepted error bound.
The finite sum is evaluated by reverse Horner recursion, followed by the
factor $\mathfrak k^2/(2\pi)$. Adding the symmetric tail interval gives a prediction
interval $J$. The comparison requires
\begin{equation}
 \sup_{x\in J}|x-\bar w|<r\,.
 \label{eq:scalar-compatibility}
\end{equation}
This proves containment in the recorded scalar neighborhood, not equality of
two exact real numbers. The analytic tail estimate controls the infinite
series, while the scalar neighborhood controls the numerical Bethe value.

\section{Fixed-rank large-parameter expansions}
\label{app:fixed-rank}

We derive Eqs.~\eqref{eq:fixed-rank-II} and~\eqref{eq:fixed-rank-I}
from the finite Bethe equations near vanishing centred
holonomies, holding the integer $N$ fixed. The error
constants may depend on $N$. The argument uses local
analyticity of the branch equations and involves no
continuum approximation or proposed constant function.

\subsection{The flavored branch}

For Model II, put $s_i=2i-N-1$. Multiplying
Eq.~\eqref{eq:real-II} by $\varepsilon=1/N_f$ gives
\begin{equation*}
 f_\varphi(y_i)+3\varepsilon\sum_j f_\theta(y_i-y_j)
 -\pi\varepsilon s_i=0\,.
\end{equation*}
At $(\boldsymbol y,\varepsilon)=(0,0)$ the Jacobian
with respect to $\boldsymbol y$ is $\sqrt3 I$, because
$f_\varphi'(0)=\sqrt3$.
The real-analytic implicit-function theorem supplies a
solution $\boldsymbol y(\varepsilon)=O(\varepsilon)$.
For positive $\varepsilon$, Theorem~\ref{thm:branch}
identifies it with the unique real branch and imposes
$\sum_i y_i=0$.

Since $f_\theta'(0)=1/\sqrt3$, expanding the even interaction
terms of $\mathcal Y_{N_f}$ and retaining its linear lift gives
\begin{equation}
 \mathcal Y_{N_f}^{(2)}(\boldsymbol y)
 =\frac{\sqrt3(N_f+N)}2\sum_i y_i^2
  -\pi\sum_i s_i y_i\,.
 \label{eq:quadratic-II}
\end{equation}
Here $\sum_{i<j}(y_i-y_j)^2=N\sum_i y_i^2$ on the
zero-sum subspace. The linearized equations are therefore
$\sqrt3(N_f+N)y_i=\pi s_i$.
The first omitted forces are cubic, of size
$O(N_f\|\boldsymbol y\|^3)=O(N_f^{-2})$;
the inverse Jacobian is $O(N_f^{-1})$ at fixed $N$.
Consequently,
\begin{equation}
 y_i=\frac{\pi s_i}{\sqrt3(N_f+N)}+O(N_f^{-3})\,.
 \label{eq:large-flavor-roots}
\end{equation}
Completing the square in Eq.~\eqref{eq:quadratic-II}
gives its minimum
$-\pi^2\sum_i s_i^2/[2\sqrt3(N_f+N)]$.
The exact value relation~\eqref{eq:value-minimum},
$\sum_i s_i^2=N(N^2-1)/3$ and
$\Cltwo(\theta)=2\Cltwo(\varphi)/3$ then give
\begin{equation}
 \WII(N,N_f)
 =2N(N_f+N)\Cltwo\left(\frac\pi3\right)
 -\frac{\pi^2N(N^2-1)}{6\sqrt3(N_f+N)}
 +O(N_f^{-3})\,.
 \label{eq:fixed-II-derivation}
\end{equation}
The omitted interactions are quartic and contribute
$O(N_f\|\boldsymbol y\|^4)=O(N_f^{-3})$.
The displacement from the quadratic root is $O(N_f^{-3})$;
since the quadratic Hessian is $O(N_f)$, its contribution
to the stationary value is $O(N_f^{-5})$.
The denominator $N_f+N$ is retained exactly within the
quadratic calculation.

\subsection{The local large-level branch}

For Model I, remove the parity center as in
Eq.~\eqref{eq:symmetry-I}, and write $L_i=N+1-2i$.
The analytic kernel has the expansion
$K_\theta(z)=-\ii z/\sqrt3+O(z^3)$.
On the reflection-symmetric subspace,
$\sum_i z_i=\sum_i\overline z_i=0$, so the reduced
equations become
\begin{equation}
 -(k+\ii\sqrt3N)z_i+\pi L_i+O(\|\boldsymbol z\|^3)=0\,.
 \label{eq:linear-I}
\end{equation}
To specify the local branch, set $z_i=\varepsilon w_i$
and $\varepsilon=1/k$. At $\varepsilon=0$, the reduced
real system for the real and imaginary parts of
$\boldsymbol w$ has the solution $w_i=\pi L_i$ and
Jacobian $-I$. The real-analytic implicit-function theorem
gives a unique nearby branch for sufficiently small
$|\varepsilon|$. The omitted cubic force and the
$O(k^{-1})$ inverse Jacobian imply that its roots differ
by $O(k^{-4})$ from the quadratic solution
\begin{equation}
 p_i^{(2)}=\frac{\pi kL_i}{k^2+3N^2}\,,
 \qquad
 y_i^{(2)}=-\frac{\pi\sqrt3N L_i}{k^2+3N^2}\,.
 \label{eq:large-level-roots}
\end{equation}
In particular, $p_i=O(k^{-1})$ and $y_i=O(k^{-2})$.
The branch lies within the analytic strip near the
origin; the argument does not establish continuation
to every smaller positive level.

Expanding the imaginary part of the canonical twisted
superpotential on this slice, before imposing its
stationary equations, gives
\begin{equation}
 \WI^{(2)}
 =4N^2\Cltwo\left(\frac\pi3\right)
 -2k\sum_i p_i y_i+2\pi\sum_i L_i y_i
 -\sqrt3N\sum_i p_i^2+\sqrt3N\sum_i y_i^2\,.
 \label{eq:quadratic-I}
\end{equation}
The equations obtained by varying $p_i$ and $y_i$ are
$k y_i=-\sqrt3N p_i$ and
$k p_i-\sqrt3N y_i=\pi L_i$, respectively, reproducing
Eq.~\eqref{eq:large-level-roots}.
At their solution the homogeneous quadratic term is
minus one half of the linear term, so the correction
to the constant in Eq.~\eqref{eq:quadratic-I} is
$\pi\sum_iL_i y_i^{(2)}$. Therefore
\begin{equation}
 \WI(N,k)
 =4N^2\Cltwo\left(\frac\pi3\right)
 -\frac{\pi^2N^2(N^2-1)}{\sqrt3(k^2+3N^2)}
 +O(k^{-4})\,.
 \label{eq:fixed-I-derivation}
\end{equation}
The paired chiral terms are even at the origin, so the
omitted interactions contribute $O(\|\boldsymbol z\|^4)
=O(k^{-4})$. The displacement from the quadratic saddle
is $O(k^{-4})$ and its quadratic Hessian is $O(k)$,
giving a stationary-value correction of $O(k^{-7})$.
The denominator in Eq.~\eqref{eq:fixed-I-derivation}
therefore retains only the exact quadratic contribution;
the full $k^{-4}$ coefficient also receives quartic
interaction terms. At $N=1$, both quadratic corrections vanish,
consistently with the exact benchmarks in
Eq.~\eqref{eq:rank-one}.

\section{Reduction and regularization of the bulk characters}
\label{app:bulk-constant}

The projected characters are first combined at finite
squashing. Their determinants use
\begin{equation}
 \zeta_{\mathcal H}(s)
 =\frac1{\Gamma(s)}\int_0^\infty
       v^{s-1}\mathcal H(v)\,\rd v\,,
 \qquad \log Z_{\mathrm{1-loop}}=\zeta'_{\mathcal H}(0)\,.
 \label{eq:bulk-zeta-definition}
\end{equation}
The integral converges for $\re s>3$ for the characters
below. Meromorphic continuation defines the derivative
at zero. The small-squashing argument is given separately
from the rational-character reduction, since a pointwise
heat-kernel limit alone would not determine this derivative.

\subsection{Free projection and the small-squashing limit}

At zero masses, put $q_1=\e^{-bv}$,
$q_2=\e^{-v/b}$ and $x=\e^{-Q_bv/3}$, so
$q_1q_2=x^3$.
The intrinsic quadric character of
Ref.~\cite[Eq.~(6.5) and Appendix~E]{BenettiGenolini:2026kjy}
is
\begin{equation}
 \mathcal H_{1,b}(v)
 =\frac{x^3(1+x)}{(1-q_1)(1-q_2)(1-x)^4}
 \bigl[q_1^{-1}-q_1+q_2^{-1}-q_2
       +5(x-x^{-1})+x^{-2}-x^2\bigr]\,.
 \label{eq:bulk-character}
\end{equation}
The factor $1-x^2$ of the defining relation has already
cancelled one ambient factor $1-x$ in this expression.
For the free quotient, the weights on
$(A_1,A_2,B_1,B_2,z_0)$ are $(1,1,-1,-1,0)$.
Projecting both numerator and denominator, then setting
$v=bu$ and taking $b\to0^+$ at fixed $u>0$, with
$x=\e^{-u/3}$ in the limiting expression, gives
\begin{equation}
 \begin{aligned}
 b^2\mathcal H_{k,b}(bu)&\longrightarrow
                         \frac{\mathscr R_k(u)}u\,,\\
 \mathscr R_k(u)
 &=\frac{x^3(1+x)}{1-x^3}\frac1k
   \sum_{\zeta^k=1}
   \frac{x^{-3}-x^3+(x-x^{-1})[2(\zeta+\zeta^{-1})+1]
                   +x^{-2}-x^2}
        {(1-\zeta x)^2(1-\zeta^{-1}x)^2}\,.
 \end{aligned}
 \label{eq:bulk-Molien}
\end{equation}
The elementary filter is
\begin{equation}
 U_k(x):=\frac1k\sum_{\zeta^k=1}
 \frac1{(1-\zeta x)(1-\zeta^{-1}x)}
 =\frac{1+x^k}{(1-x^2)(1-x^k)}\,,\qquad |x|<1\,.
 \label{eq:bulk-basic-filter}
\end{equation}
Indeed, the phase average retains $x^{r+s}$ with
$k\mid(r-s)$. The diagonal sum is $(1-x^2)^{-1}$,
and each off-diagonal sum is
$x^k/[(1-x^2)(1-x^k)]$.
For $S_\zeta=\zeta+\zeta^{-1}$ and
$G_\zeta(x)=(1-S_\zeta x+x^2)^{-1}$,
\begin{equation*}
 G_\zeta+xG_\zeta'=(1-x^2)G_\zeta^2\,,\qquad
 G_\zeta'=S_\zeta G_\zeta^2-2xG_\zeta^2\,.
\end{equation*}
Thus the two squared-denominator averages in
Eq.~\eqref{eq:bulk-Molien} are
$(U_k+xU_k')/(1-x^2)$ and
$U_k'+2x(U_k+xU_k')/(1-x^2)$.
With $T_k(x)=(1+x^k)/(1-x^k)$, substitution gives
\begin{equation}
 \begin{aligned}
 \mathscr R_k(u)
 &=xT_k'(x)
   +\left[\frac{1+x}{1-x}+\frac{1+x^2}{1-x^2}
                       -\frac{1+x^3}{1-x^3}\right]T_k(x)\\
 &=\coth\frac{ku}{6}
   \left(\coth\frac u6+\coth\frac u3-\coth\frac u2\right)
   +\frac{k}{2\sinh^2(ku/6)}\,.
 \end{aligned}
 \label{eq:bulk-R}
\end{equation}
In particular, $\mathscr R_k$ is even in $u$,
$\mathscr R_k=O(u^{-2})$ at zero, and
$\mathscr R_k-1$ decays exponentially at infinity.
The reduced Mellin transform is
\begin{equation}
 Z_k(s)=\frac1{\Gamma(s)}\int_0^\infty
 u^{s-2}[\mathscr R_k(u)-1]\,\rd u\,,
 \qquad \re s>3\,.
 \label{eq:bulk-reduced-zeta}
\end{equation}

\paragraph{The Mellin limit.}
\label{subsec:bulk-Mellin-limit}

The limiting constant $\mathscr R_k(\infty)=1$ comes
from the slow finite-$b$ mode
$S_b(v)=(\e^{bv}-1)^{-1}$.
Its zeta function is $b^{-s}\zeta(s)$, so its
zeta derivative is
$\frac12\log b-\frac12\log(2\pi)$.
It cannot contribute to the coefficient of $b^{-2}$,
but must be removed before taking the limiting Mellin
integral. The following argument applies to both
projected characters and to the enhanced Model-II
character at each fixed quotient order.

Let $\mathcal H_b$ denote one of these characters at fixed
quotient order. Set $\beta=b^2$ and define the subtracted
scaled character
\begin{equation*}
 F_\beta(u)=\beta\bigl[\mathcal H_b(bu)-S_b(bu)\bigr]\,.
\end{equation*}
For a finite cyclic projection, the character has the form
\begin{equation*}
 \mathcal H_b(bu)
 =\frac{(a+d)P(x)+V(x)}{(1-a)(1-d)}\,,\qquad
 a=\e^{-\beta u}\,,\quad d=\e^{-u}\,,\quad
 x=\e^{-(1+\beta)u/3}\,.
\end{equation*}
Here $P,V$ are rational functions regular for $0\leq x<1$,
with $P(0)=1$ and $V(0)=0$.
This form follows from $ad=x^3$ and
$x^3(a^{-1}-a+d^{-1}-d)=(a+d)(1-x^3)$ in the
full projected numerator.
Adding the localized character changes only $V$ by a
multiple of $x(1+x)^2/(1-x)$ and preserves these
properties. The exact subtraction gives
\begin{equation*}
 F_\beta(u)
 =\frac{\beta}{1-a}
   \frac{a[P(x)-1]+ad+dP(x)+V(x)}{1-d}\,.
\end{equation*}
For $u\geq1$ and $0\leq\beta\leq\beta_0$, every
numerator term decays exponentially in $u$.
The factor $\beta/(1-\e^{-\beta u})$ extends smoothly
to $1/u$ at $\beta=0$, obeys
$\beta/(1-\e^{-\beta u})\leq\beta+1/u$, and has
bounded first $\beta$ derivative. Differentiating the
remaining factors adds at most polynomial powers of
$u$. Hence $F_\beta$ and $\partial_\beta F_\beta$
are bounded by a polynomial times $\e^{-cu}$, for
some $c>0$ independent of sufficiently small $\beta$.
In particular,
$F_\beta-F_0=O(\beta)$ in the exponentially weighted
tail integral.

Near $u=0$, the projected characters are meromorphic,
with poles of order at most three. After multiplication
by $\beta$, their Laurent coefficients are analytic
at $\beta=0$. In a fixed sufficiently small disk,
subtracting the terms through the constant coefficient
leaves an $O(u)$ remainder, with the same bound for
its $\beta$ derivative. Splitting the Mellin integral
inside that disk and integrating the finitely many
Laurent monomials explicitly therefore gives
\begin{equation*}
 \zeta_{F_\beta}(s)=Z_R(s)+O(\beta)
 \quad\text{near }s=0\,,\qquad
 Z_R(s)=\frac1{\Gamma(s)}\int_0^\infty
 u^{s-2}[R(u)-1]\,\rd u\,.
\end{equation*}
The statement holds after meromorphic continuation,
uniformly on a small neighborhood of zero, and also
for the first $s$ derivative.
Here $F_0(u)=[R(u)-1]/u$ and $R=\mathscr R_k$ or
one of the reduced Model-II characters.
Evenness of $R$ excludes a constant Laurent term in
$F_0$, so $Z_R(0)=0$.

At finite $b$, the zero total equivariant weight makes
$\mathcal H_b$ odd in $v$. For a cyclic projection,
inversion of the group element is included in the
average; the localized character is odd as well.
Since
$S_b(bu)=(\beta u)^{-1}-1/2+O(u)$, the constant
Laurent coefficient of $F_\beta$ is $\beta/2$ and
$\zeta_{F_\beta}(0)=\beta/2$.
Restoring the scale in the original determinant gives
\begin{equation}
 \begin{aligned}
 \zeta_{\mathcal H_b}(s)
 &=b^{-s}\zeta(s)+b^{s-2}\zeta_{F_\beta}(s)\,,\\
 \zeta'_{\mathcal H_b}(0)
 &=\frac{Z_R'(0)}{b^2}+O(|\log b|)\,.
 \end{aligned}
 \label{eq:bulk-Mellin-limit}
\end{equation}
This proves $\widehat g_0[R]=-Z_R'(0)$ with the
claimed remainder. The vanishing limiting value
$Z_R(0)$ excludes a $b^{-2}\log b$ term.
The estimates are at fixed quotient order; they do
not supply a uniform simultaneous large-order limit.

For an independent integral evaluation of the finite
part, write $R(u)=r_{-2}u^{-2}+r_0+O(u^2)$ at zero.
The preceding continuation also gives
\begin{equation}
 \begin{aligned}
 Z_R'(0)
 ={}&\int_0^1\frac{R(u)-r_{-2}u^{-2}-r_0}{u^2}\,\rd u
    +\int_1^\infty\frac{R(u)-1}{u^2}\,\rd u\\
 &-\frac{r_{-2}}3-(r_0-1)\,.
 \end{aligned}
 \label{eq:bulk-Mellin-finite-part}
\end{equation}
Both integrals converge absolutely. The final terms
are the analytic integrals of the subtracted monomials.
This representation fixes the regularization without
introducing an independent finite subtraction.

\subsection{Sphere-character reduction and the constant integral}

The ABJM sphere character is
\begin{equation}
 \mathcal H^{\mathrm B}_\kappa(u)
 =\frac{\coth(\kappa u/4)}{2\sinh^2(u/2)}\,.
 \label{eq:ABJM-heat-character}
\end{equation}
In the determinant convention of
Ref.~\cite[Section~3]{BenettiGenolini:2026kjy},
$\zeta'_{\mathcal H^{\mathrm B}_\kappa}(0)=A(\kappa)$
with $A$ in Eq.~\eqref{eq:sphere-A}.
The character decays exponentially at infinity and
has only odd Laurent powers at zero. Since
$1/\Gamma(s)=s+O(s^2)$, its Mellin transform has
zero value at $s=0$.
Under $u\mapsto cu$, $c>0$, the zeta function becomes
$c^{-s}\zeta_{\mathcal H}(s)$. Its derivative thus
changes by $-\log c\,\zeta_{\mathcal H}(0)$, which
vanishes here.

Differentiation of Eq.~\eqref{eq:bulk-R} gives
\begin{equation}
 \begin{aligned}
 \mathscr R_k'(u)={}&
 \mathcal H^{\mathrm B}_{2k/3}(u)
 -\frac23\mathcal H^{\mathrm B}_k(2u/3)
 -\frac13\mathcal H^{\mathrm B}_{2k}(u/3)\\
 &-\frac{k}{3}\left[
 \mathcal H^{\mathrm B}_{2/k}(ku/3)
 +\mathcal H^{\mathrm B}_{4/k}(ku/3)
 -\mathcal H^{\mathrm B}_{6/k}(ku/3)\right]
 -\frac{k^2}{3}\mathcal H^{\mathrm B}_2(ku/3)\,.
 \end{aligned}
 \label{eq:bulk-character-reduction}
\end{equation}
The two sets of terms arise by differentiating the
bracket and $\coth(ku/6)$, respectively; differentiating
the final $\sinh^{-2}$ term supplies the last character.
The boundary term in integration by parts vanishes
for $\re s>3$, giving
$Z_k(s)=\zeta_{\mathscr R_k'}(s)/(1-s)$.
Every character on the right of
Eq.~\eqref{eq:bulk-character-reduction} has zero zeta
value at the origin. Therefore
$Z_k'(0)=\zeta'_{\mathscr R_k'}(0)$, and its negative
is the first expression in Eq.~\eqref{eq:bulk-CI}.

The reduction to $\CI$ can be proved independently
by convergent integrals. For $a,b>0$, define
$\mathcal J(a,b)=\int_0^\infty xL_a(x)/(\e^{bx}-1)\,\rd x$.
Rescaling the integral in Eq.~\eqref{eq:sphere-A}
gives $\mathcal J(a,b)=4a^{-2}\mathcal J(2,2b/a)$.
Solving that equation for its integral, then integrating
the derivative of $xL_a(x)L_b(x)$, yields
\begin{equation}
 \begin{aligned}
 \mathcal J(a,b)
 &=\frac{\pi^2}{b^2}A(2b/a)
   -\frac{a\zeta(3)}{b^3}+\frac{\zeta(3)}{2a^2}\,,\\
 D(a,b)
 &=-b\mathcal J(a,b)-a\mathcal J(b,a)\\
 &=-\pi^2\left[\frac{A(2b/a)}b+\frac{A(2a/b)}a\right]
   +\frac{\zeta(3)}2\left(\frac a{b^2}+\frac b{a^2}\right)\,.
 \end{aligned}
 \label{eq:bulk-D-A}
\end{equation}
The boundary term is $O(x\log^2x)$ at zero and
exponentially small at infinity, and all integrals
converge absolutely. This is the exchange method of
Ref.~\cite[Appendix~B.1]{Hong:2026zul}.
Substitution into Eq.~\eqref{eq:constant-I-log-integrals}
proves the second equality of Eq.~\eqref{eq:bulk-CI}
for every real $k>0$.

\subsection{The unit-parameter Dirichlet evaluation}

At $k=1$, the reduced character is the rational
function of Ref.~\cite[Eq.~(E.4)]{BenettiGenolini:2026kjy}:
\begin{equation}
 \begin{aligned}
 \mathscr R_1(u)
 &=\frac{(1+x)^2(x^2+3x+1)}{(1-x)^2(1+x+x^2)}\\
 &=1+\frac{20x}{3(1-x)^2}-\frac{2x}{3(1+x+x^2)}\\
 &=1+\frac{20}{3}\sum_{m\geq1}m x^m
      -\frac23\sum_{m\geq1}\chi_{-3}(m)x^m\,,
 \qquad x=\e^{-u/3}\,.
 \end{aligned}
 \label{eq:bulk-unit-series}
\end{equation}
The two series follow from
$x/(1-x)^2=\sum m x^m$ and
$x/(1+x+x^2)=(x-x^2)/(1-x^3)$.
Their absolute Mellin integrals converge for $\re s>3$,
so termwise integration gives
\begin{equation}
 Z_1(s)=\frac{3^{s-2}}{s-1}
        [20\zeta(s-2)-2L(s-1,\chi_{-3})]\,.
 \label{eq:bulk-unit-Dirichlet}
\end{equation}
Both terms in brackets vanish at $s=0$.
The Riemann functional equation gives
$\zeta'(-2)=-\zeta(3)/(4\pi^2)$.
For the primitive odd character $\chi_{-3}$, whose Gauss sum is
$\mathrm{i}\sqrt3$, the Dirichlet functional equation, combined
with the gamma reflection and duplication
formulas~\cite[Eqs.~(25.15.5)--(25.15.6), (5.5.3)
and~(5.5.5)]{NIST:DLMF}, gives
\begin{equation*}
 \Lambda(s,\chi_{-3})
 =\left(\frac3\pi\right)^{(s+1)/2}
  \Gamma\left(\frac{s+1}{2}\right)L(s,\chi_{-3})
 =\Lambda(1-s,\chi_{-3})\,.
\end{equation*}
At $s=-1$, the gamma pole is cancelled by the simple
trivial zero of $L$, giving
$\Lambda(-1,\chi_{-3})=2L'(-1,\chi_{-3})$.
At $s=2$, the completed value is
$3\sqrt3 L(2,\chi_{-3})/(2\pi)$.
The sine series gives
$\Cltwo(2\pi/3)=\sqrt3 L(2,\chi_{-3})/2$,
and $\Cltwo(\pi/3)=3\Cltwo(2\pi/3)/2$.
Consequently $L'(-1,\chi_{-3})=\Cltwo(\pi/3)/\pi$,
and $-Z_1'(0)$ is exactly
Eq.~\eqref{eq:bulk-unit-constant}.

\section{The transverse ADE contribution}
\label{app:fixed-locus-character}

The transverse-resolution prescription uses the intrinsic
virtual classes of the charts in Eq.~\eqref{eq:ADE-charts}.
The following calculations determine their relative index
and curvature push-forward. The physical enhancement in
Eq.~\eqref{eq:ADE-enhancement-prescription} is separate
from these identities.

\subsection{Normal characters and the localized virtual class}

Let $\mathrm{vol}_3=\rd z_0\wedge\rd A_2\wedge\rd B_2$.
On the coarse hypersurface, a residue generator on the
patch $U\ne0$ is $\rd U/U\wedge\mathrm{vol}_3$.
Its pullback to chart $j$ is
\begin{equation*}
 \frac{\rd U}{U}\wedge\mathrm{vol}_3
 =\left[(j+1)\frac{\rd\xi_j}{\xi_j}
        +j\frac{\rd\eta_j}{\eta_j}\right]\wedge\mathrm{vol}_3
 =\frac{\rd\xi_j}{\xi_j}\wedge\mathrm{vol}_3\,.
\end{equation*}
The last equality uses $\xi_j\eta_j=T$ and
$\rd T\wedge\mathrm{vol}_3=0$.
The resulting form is the residue generator of
$\xi_j\eta_j+z_0^2+A_2B_2=0$ and extends across its
smooth locus without an exceptional zero or pole.
This proves crepancy of the partial resolution.
Its residual ordinary double points retain the
intrinsic hypersurface prescription.

For generic nonzero normal weights $p,r$, define
\begin{equation*}
 H(p,r)=\frac1{(1-p)(1-r)}\,,\qquad
 J(p,r)=(p^{-1}-p+r^{-1}-r)H(p,r)\,.
\end{equation*}
These are the normal structure-sheaf and
tangent--cotangent rational characters.
Put $\eta=pr$ and use the weights in
Eq.~\eqref{eq:ADE-chart-weights}.
Since $p_{j+1}=p_j/\eta=1/r_j$,
\begin{equation*}
 H(p_j,r_j)=\frac1{1-\eta}
       \left(\frac1{1-p_j}-\frac1{1-p_{j+1}}\right)\,.
\end{equation*}
The endpoints $p_0=p^n$ and $p_n=r^{-n}$ give
\begin{equation}
 \sum_{j=0}^{n-1}H(p_j,r_j)
 =\frac{1-(pr)^n}{(1-p^n)(1-r^n)(1-pr)}
 =\frac1n\sum_{\zeta^n=1}H(\zeta p,\zeta^{-1}r)\,.
 \label{eq:normal-Hilbert-identity}
\end{equation}
The last expression is the Molien character of the
invariant ring generated by $A_1^n,B_1^n,A_1B_1$
with relation $UV=T^n$.
This calculation takes place in rational equivariant
characters. For $n>1$, consecutive charts obey
$p_{j+1}r_j=1$, so the inequalities
$|p_j|,|r_j|<1$ cannot hold in every chart.
The invariant Molien series can be expanded
for $|p|,|r|<1$, but the chart sum is combined as a
rational function before any specialization or Mellin
transform.

For a pair with product $\eta$, direct algebra gives
\begin{equation*}
 J(p,r)=(\eta^{-1}-\eta)H(p,r)+1-\eta^{-1}\,.
\end{equation*}
The first term cancels between the chart sum and
Molien average by Eq.~\eqref{eq:normal-Hilbert-identity}.
The second appears $n$ times in the chart sum and
once in the normalized average, yielding
\begin{equation}
 \sum_{j=0}^{n-1}J(p_j,r_j)
 -\frac1n\sum_{\zeta^n=1}J(\zeta p,\zeta^{-1}r)
 =(n-1)(1-\eta^{-1})\,.
 \label{eq:normal-TC-excess}
\end{equation}
The tangential contribution on $\mathfrak F$ multiplies
the same normal structure-sheaf character in both
expressions and cancels as well. The total canonical
weight of $\mathfrak W$ is zero. Equivariant adjunction
therefore gives $K_{\mathfrak F}$ weight $\eta^{-1}$,
and multiplication by the structure-sheaf index of
$\mathfrak F$ proves Eq.~\eqref{eq:ADE-Cartan-excess}.

At zero masses, the three fixed-surface ambient weights
are $x^{-1}$ and the relation weight is $x^{-2}$.
Thus
\begin{equation*}
 \operatorname{Ind}^{\mathrm{rel}}_v\bar\partial_{\mathfrak F}
 =\frac{1-x^{-2}}{(1-q_1)(1-q_2)(1-x^{-1})^3}
 =\frac{x(1+x)}{(1-q_1)(1-q_2)(1-x)^2}\,.
\end{equation*}
Together with $\eta^{-1}=x^2$, this gives the localized
character in Eq.~\eqref{eq:localized-finite-b-character}.
On the smooth stratum, its virtual Dolbeault complex is
$\Omega^{0,\bullet}(\mathfrak F,\cO_{\mathfrak F})-
\Omega^{0,\bullet}(\mathfrak F,K_{\mathfrak F})$.
The $A_1$ point of $S$ can be evaluated on its crepant
resolution: rationality preserves the structure-sheaf
index and crepancy preserves the canonical line.
This identifies an alternating virtual index, not the
individual physical bosonic, fermionic or ghost modes.

The choice of class can also be tested directly.
The coarse hypersurface has normal virtual weights
$p^n,r^n,\eta$ with the relation $\eta^n$ subtracted.
Its structure-sheaf character is still the middle
expression of Eq.~\eqref{eq:normal-Hilbert-identity},
but its tangent--cotangent numerator is
\begin{equation*}
 p^{-n}-p^n+r^{-n}-r^n+\eta^{-1}-\eta
                         -\eta^{-n}+\eta^n\,.
\end{equation*}
At $p=r=x^{-1}$, using
$\sum_{j=0}^{n-1}x^{n-1-2j}=(x^n-x^{-n})/(x-x^{-1})$
gives the full difference
\begin{equation}
 \mathcal H^{\mathrm{coarse}}_{\mathrm{II},n,b}(v)
 -\mathcal H^{\mathrm{inv}}_{\mathrm{II},n,b}(v)
 =\mathcal H_{\mathrm{loc},b}(v)
   \bigg[\bigg(\sum_{j=0}^{n-1}x^{n-1-2j}\bigg)^2-1\bigg]\,.
 \label{eq:coarse-index-difference}
\end{equation}
Consequently the difference of the scaled characters
$b^2\mathcal H(bu)$ tends to
$g(u)/u$ times the bracket, with $x=\e^{-u/3}$.
The difference of their reduced numerators is $g(u)$
times that bracket. These are distinct from
$(n^2-1)g(u)$.
For $n\geq2$, the expression in
Eq.~\eqref{eq:coarse-index-difference} grows at large
positive $v$, starting at $x^{3-2n}$.
It therefore has no Mellin convergence half-plane
on the same positive heat ray without a further
spectral prescription. The calculation distinguishes
the virtual classes; it does not define a second
physical determinant with the regularization used here.

\subsection{The curvature push-forward}

Let $\sigma=\epsilon_++\epsilon_-$ and set
$a_j=\omega_j^+=n\epsilon_+-j\sigma$.
Then $\omega_j^-=-a_{j+1}$, and
\begin{equation*}
 \frac1{\omega_j^+\omega_j^-}
 =\frac1\sigma\left(\frac1{a_j}-\frac1{a_{j+1}}\right)\,.
\end{equation*}
The endpoints are $a_0=n\epsilon_+$ and
$a_n=-n\epsilon_-$, giving
\begin{equation}
 \sum_{j=0}^{n-1}\frac1{\omega_j^+\omega_j^-}
 =\frac1{n\epsilon_+\epsilon_-}\,.
 \label{eq:normal-Euler-sum}
\end{equation}
The identity is first established for generic weights
and then specialized. In particular, vanishing
individual weights at even $n$ do not obstruct the
combined expression.

Write $E=\omega_j^+\omega_j^-$ for the normal Euler
weight in one chart. With fixed-stratum Pontryagin
polynomials $p_{1,\mathfrak F},p_{2,\mathfrak F}$,
\begin{equation*}
 \begin{aligned}
 p_1&=p_{1,\mathfrak F}+\sigma^2-2E\,,\\
 p_2&=p_{2,\mathfrak F}
       +p_{1,\mathfrak F}(\sigma^2-2E)+E^2\,.
 \end{aligned}
\end{equation*}
Thus the curvature expression separates as
\begin{equation}
 \frac{X_8}{e_{\mathfrak F}E}
 =\frac{(p_{1,\mathfrak F}-\sigma^2)^2
             -4p_{2,\mathfrak F}}{192e_{\mathfrak F}E}
  +\frac{p_{1,\mathfrak F}-\sigma^2}{48e_{\mathfrak F}}\,.
 \label{eq:normal-X8-polynomial}
\end{equation}
The inverse-$E$ term sums to $1/n$ of the unquotiented
term by Eq.~\eqref{eq:normal-Euler-sum}. The remaining
term occurs $n$ times on the resolution and $1/n$
times in that comparison. Their difference proves
Eq.~\eqref{eq:II-curvature-excess}.
The normal contribution $-\sigma^2$ follows from the
full Pontryagin polynomial and cannot be dropped.

On $\mathfrak F$, the ambient internal weights are
three copies of $u=-Q_b/3$ and the relation has
weight $2u$. Including the external weights gives
\begin{equation*}
 e_{\mathfrak F}=\frac{bb^{-1}u^3}{2u}
                =\frac{Q_b^2}{18}\,,\qquad
 p_{1,\mathfrak F}=b^2+b^{-2}-u^2\,,\qquad
 \sigma^2=4u^2\,.
\end{equation*}
Their substitution yields
$(n-n^{-1})(1/6-3/(4Q_b^2))$.
The sign of the shift then follows from
$\mathcal B=-\sum X_8/e$.
This push-forward computes the specified relative
curvature contribution. It does not derive a
complete localized anomaly-inflow action or establish
which global $C$-field and boundary data realize the
chosen fixed-point fluxes.

\subsection{Invariant projection and Mellin reduction}

For the Model-II action, projecting the full finite-$b$
character gives
\begin{equation}
 \begin{aligned}
 \mathcal H^{\mathrm{inv}}_{\mathrm{II},n,b}(v)
 ={}&\frac{x^3(1+x)}{(1-q_1)(1-q_2)(1-x)^2}
 \frac1n\sum_{\zeta^n=1}
 \frac{1}{(1-\zeta x)(1-\zeta^{-1}x)}\\
 &\quad\times\bigl[
 q_1^{-1}-q_1+q_2^{-1}-q_2
 +(x-x^{-1})(3+\zeta+\zeta^{-1})+x^{-2}-x^2\bigr]\,.
 \end{aligned}
 \label{eq:II-full-Molien}
\end{equation}
Here $q_1=\e^{-bv}$, $q_2=\e^{-v/b}$ and
$x=\e^{-Q_bv/3}$.
Along $v=bu$, $b\to0^+$, one has
$x\to\e^{-u/3}$, $b^2/(1-q_1)\to1/u$ and the
numerator tends to
$x^{-3}-x^3+(x-x^{-1})(3+\zeta+\zeta^{-1})+x^{-2}-x^2$.
The two required averages are $U_n(x)$ from
Eq.~\eqref{eq:bulk-basic-filter} and
\begin{equation*}
 \frac1n\sum_{\zeta^n=1}
 \frac{\zeta+\zeta^{-1}}{(1-\zeta x)(1-\zeta^{-1}x)}
 =\frac{(1+x^2)U_n(x)-1}{x}\,.
\end{equation*}
This follows by solving the denominator for
$\zeta+\zeta^{-1}$.
Substitution proves Eq.~\eqref{eq:II-invariant-character};
at $n=1$ it agrees with Eq.~\eqref{eq:bulk-R}, and
at $n=2$ it reduces to $\coth^2(u/6)$.
The localized finite-$b$ character decays exponentially
and adds no slow mode. The common subtracted-Mellin
argument in Appendix~\ref{subsec:bulk-Mellin-limit}
therefore applies to the enhanced character as well.

For $g(u)$ in Eq.~\eqref{eq:II-invariant-character},
partial fractions give
\begin{equation}
 \begin{aligned}
 g(u)&=\frac{4x}{3(1-x)^2}-\frac{x}{3(1+x+x^2)}\,,
 \qquad x=\e^{-u/3}\,,\\
 \zeta_{g'}(s)
 &=-3^{s-1}\left[\frac43\zeta(s-2)
                   -\frac13L(s-1,\chi_{-3})\right]\,.
 \end{aligned}
 \label{eq:II-localized-Dirichlet}
\end{equation}
The derivative contributes $-m/3$ to each Fourier
mode; Mellin integration then gives the displayed
power $3^{s-1}$.
The bracket vanishes at $s=0$.
Using the functional-equation evaluations in
Appendix~\ref{app:bulk-constant} proves
$\zeta_{g'}(0)=0$ and Eq.~\eqref{eq:localized-constant}.

Differentiating the product in the enhanced numerator
gives
\begin{equation}
 \begin{aligned}
 (\mathscr R_n^{\mathrm{ADE}})'(u)
 ={}&\frac12\mathcal H^{\mathrm B}_{2n/3}(u)
    -\frac12\mathcal H^{\mathrm B}_{2n}(u/3)\\
 &-\frac n2\mathcal H^{\mathrm B}_{2/n}(nu/3)
   +\frac n6\mathcal H^{\mathrm B}_{6/n}(nu/3)
   +n^2g'(u)\,.
 \end{aligned}
 \label{eq:II-character-derivative}
\end{equation}
Every term has vanishing zeta value at zero, so
integration by parts and positive rescaling give
the first line of Eq.~\eqref{eq:bulk-CII} with no
extra logarithmic normalization.
The independent integral calculation uses
Eq.~\eqref{eq:bulk-D-A} in
Eq.~\eqref{eq:constant-II-log-integrals} and gives
\begin{equation*}
 \frac{\CII(n)}{2\pi}
 =\frac12[A(2n)-A(2n/3)]
  +\frac n6[3A(2/n)-A(6/n)]
  -\frac{n^2}{9}\left[
       \frac{\zeta(3)}{\pi^2}+\frac{\Cltwo(\pi/3)}\pi\right]\,.
\end{equation*}
All exchanges in this last calculation concern
absolutely convergent integrals for $n>0$.

\subsection{A round-sphere ADHM check}

For $\mathbb C^2/\mathbb Z_n\times\mathbb C^2$ at the
round superconformal point, the four internal weights
are $-1/2$ and the external weights are $1,1$.
Put $x=\e^{-v/2}$. The fixed-stratum structure character
is $x^2/[(1-x)^2(1-x^2)^2]$, and its canonical weight
is $x^2$. The excess in
Eq.~\eqref{eq:normal-TC-excess} therefore gives
\begin{equation*}
 \mathcal H_{\mathrm{loc}}^{\mathrm{ADHM}}(v)
 =\frac{x^2(1+x)}{(1-x)(1-x^2)^2}
 =\frac12\mathcal H^{\mathrm B}_1(v)\,.
\end{equation*}
The normal Molien filter gives
$\mathcal H_n^{\mathrm{ADHM,inv}}
=(\mathcal H_n^{\mathrm B}+\mathcal H_1^{\mathrm B})/2$.
Extending the per-Cartan contribution to $n^2-1$
generators consequently yields
$\mathcal A_{\mathrm{ADHM}}=[A(n)+n^2A(1)]/2$.
This is the independently proposed and numerically
tested sphere constant of
Ref.~\cite[Eq.~(3.12)]{Hatsuda:2014vsa}.
The check uses no $V^{5,2}$ constant, but does retain
the enhancement prescription. It does not determine
the fluctuation boundary conditions on the present
$\mathrm{AdS}_4\times\mathbb{RP}^3$ locus.

\section{The virtual action and isolated free-quotient defect}
\label{app:free-quotient-action}

The Model-I action uses two different constructions.
The virtual characteristic number is evaluated on the
Reeb-equivariant cone. The isolated-orbifold charge is
assigned by relative excision in the circle-preserving
smoothing. The latter does not preserve the Reeb action
and cannot replace the former localization problem.

Set $u=-Q_b/3$ and let $z$ be a formal variable.
The internal virtual Chern polynomial is
\begin{equation}
 c(T^{\mathrm{vir}}X;z)
 =\frac{(1+uz)^5}{1+2uz}
 =1+3uz+4u^2z^2+2u^3z^3+u^4z^4+O(z^5)\,.
 \label{eq:I-virtual-Chern}
\end{equation}
The localization Euler factor is instead
$e_X=u^5/(2u)=u^4/2$.
It is the ratio of ambient and normal Euler weights,
not the coefficient $c_4(T^{\mathrm{vir}}X)=u^4$.
Multiplying the Chern polynomial by
$(1+bz)(1+b^{-1}z)=1+Q_bz+z^2$
includes the external directions and makes the total
first Chern class vanish. The identities
$p_1=c_1^2-2c_2$ and
$p_2=c_2^2-2c_1c_3+2c_4$ then give
$p_1^2-4p_2=-8c_4$.
Consequently,
\begin{equation}
 \mathcal B_{\mathrm I}^{\mathrm{id}}
 =\frac{c_4(X)+Q_bc_3(X)+c_2(X)}{24k e_X}
 =\frac1{12k}-\frac1{2k}+\frac3{kQ_b^2}\,.
 \label{eq:I-Chern-decomposition}
\end{equation}
Here $c_j(X)$ are the virtual coefficients in
Eq.~\eqref{eq:I-virtual-Chern}.
This gives an independent Chern-class check of the
Pontryagin calculation in Section~\ref{subsec:bulk-shift-I}.
It separates the internal contribution from the mixed
internal--external terms.

The smoothing $X_\varepsilon$ is diffeomorphic to
$T^*S^4$ and has Euler characteristic two.
For $k>1$, every nontrivial subgroup of the cyclic
action fixes the two points $P_\pm$.
The commuting-pair orbifold Euler count is therefore
\begin{equation}
 \chi_{\mathrm{orb}}(X_\varepsilon/\mathbb Z_k)
 =\frac1k\sum_{g,h\in\mathbb Z_k}\chi(X_\varepsilon^{g,h})
 =\frac{2+2(k^2-1)}{k}=2k\,.
 \label{eq:I-orbifold-Euler}
\end{equation}
The superscript denotes the common fixed set of $g,h$.
Subtracting the identity contribution $2/k$ leaves
$2(k-k^{-1})$, including the vanishing excess at $k=1$.
This is an orbifold index count, not the ordinary
Euler characteristic of the quotient cone.

The standard local gravitational charge of
$\mathbb C^4/\mathbb Z_k$ uses the bulk Euler number
$k-k^{-1}$, with $k^{-1}$ supplied by the boundary
at infinity~\cite[Section~3.1]{Bergman:2009zh}.
Applying that prescription to both local neighborhoods
gives $-2(k-k^{-1})/24$, as in
Eq.~\eqref{eq:I-orbifold-charge}.
The relative Euler count confirms the multiplicity
and quotient dependence. It does not derive the
anomaly prescription or its transport to the original
cone. This use holds the asymptotic charge sector and
boundary refinement fixed; a finite contribution from
changing those data would have to be included separately.

With that prescription, $\Delta\mathcal B
=(k-k^{-1})/12$ changes the first term in
Eq.~\eqref{eq:I-Chern-decomposition} from $1/(12k)$
to $k/12$, leaving the mixed terms unchanged.
This is the origin of the $k$-linear term in
Eq.~\eqref{eq:I-classical-coefficients}.
No further shift of the Page variable is then made.
The calculation determines the stated anomaly
contribution but not a global $C$-field quadratic
refinement, an independent boundary counterterm,
or the full supersymmetric saddle.

\bibliographystyle{ytamsalpha}
\def\arxivfont{\rm}
\begingroup
\small
\bibliography{references}
\endgroup
\end{document}